\documentclass[a4paper,11pt]{article}
\usepackage{jheppub} 
\usepackage{graphicx,color,rotating}
\usepackage{hyperref}
\usepackage{epsfig,color}
\usepackage{slashed}
\usepackage{amsfonts}
\usepackage{ulem}
\usepackage{color}

\graphicspath{{D:/Desktop/draft}}
\usepackage{tabularx}
\usepackage{graphicx}
\usepackage{adjustbox}
\usepackage{tabularx}
\usepackage{amsmath}

\usepackage{float}
\usepackage{subfig}
\usepackage{amsmath,mathtools}
\begin{document}
	\author{
	Po-Yan Tseng$^{1,2}$ and Yu-Min Yeh$^{1}$}
	\affiliation{
        $^1$ Department of Physics, National Tsing Hua University,
		101 Kuang-Fu Rd., Hsinchu 300044, Taiwan \\
        $^2$ Physics Division, National Center for Theoretical Sciences,
		Taipei 106319, Taiwan \\
	}
	
	\date{\today}

\abstract{The neutrino-matter interaction cause the final flavor compositions deviating from those expected in vacuum. In this work we consider neutrinos interact with ultra-light scalar dark matters $\phi$ ($m_\phi\ll 1\,{\rm eV}$). When the neutrinos emitted from a distant source and propagate through the dark matter medium, the MSW potential in the Hamiltonian is replaced by the hypothetical $\nu\phi$ effective potential. Two types of neutrino sources are adopted in our calculations. The first is point-like neutrino source, the Active Galactic Nuclei (AGN), which produce $\mathcal{O}({\rm TeV})$ neutrinos primarily through charged pion decay, yielding an initial flavor ratio of ($\nu_e,\nu_\mu,\nu_\tau$)=($1:2:0$). We focus on two specific AGNs, NGC 1068 and TXS 0506+056 in our analysis, whose distances from Earth are much larger than the  neutrino oscillation lengths. 
We consider the range of $\nu\phi$ coupling constant such that the adiabatic condition and mean free path can be satisfied.
Hence, by assuming neutrinos are produced randomly from their corona regions, we compute the averaged flavor ratio at Earth with $m_\phi=10,1,0.1\,{\rm peV}$ and $m_\phi=1,0.1,0.01\,{\rm feV}$, generate the ternary plots of neutrino flavors, and compare the results with the IceCube present and future sensitivities. The second source considered is the Diffuse Supernova Neutrino Background (DSNB), it is an isotropic neutrino source with energy $\mathcal{O}({\rm MeV})$. Applying the core-collapse simulation neutrino temperatures to the DSNB flux, we repeat the similar analysis for the DSNB with $m_\phi=10^{-22}\,{\rm eV}$, and we estimate different neutrino flavors lies in the HK/DUNE/JUNO combined sensitivities limit. }

\title{AGN and DSNB Neutrino Oscillation in Dark Matter Background}
\maketitle
\section{Introduction}
Neutrinos are known to oscillate between different flavor eigenstates while propagating in vacuum due to their tiny but non-negligible mass differences. It is also known that neutrinos interact with electrons in the medium, and the Hamiltonian of neutrinos must be modified by an additional ``effective potential.'' This is known as the Mikheyev–Smirnov–Wolfenstein (MSW) effect. The neutrino-matter interaction will change the averaged flavor ratio between different flavor eigenstates in the vacuum. In this work, we consider neutrinos interacting with ultra-light scalar dark matter (DM) $\phi$ 
which modify the Hamiltonian through MSW effect. 
Active Galactic Nuclei (AGNs) are point-like sources of neutrinos with energy scale $\mathcal{O}({\rm TeV})$, and the DM density around the central supermassive black hole has been proposed to be a spiked-like distribution~\cite{Gondolo:1999ef, Herrera:2023nww, Cline:2023tkp}. The IceCube collaboration has identified high energy neutrinos from two distinct AGNs, the NGC 1068 and TXS 0506+056, with energy $E\in [1.5,15]\,{\rm TeV}$ and $[40,4000]\,{\rm TeV}$, respectively~\cite{IceCube:2022der, IceCube:2018cha}. We assumed the flavor-dependent interaction between neutrinos and complex dark scalar $\phi$~\cite{Tseng:2024akh}, and computed the flavor ratio by adding the effective potential to the Hamiltonian using the spiked density profile of NGC 1068 and TXS 0506+056.

The Medium Energy Starting Event (MESE), which is based on the 11.4 years of IceCube data, measures whole sky neutrino events with energies of at least 1 TeV and provides a three-flavor composition of astrophysical neutrinos~\cite{IceCube:2025uyt}. Therefore, we compared our predictions for AGN-produced neutrinos, produced from pion decay and the initial flavor ratio $(\nu_e:\nu_\mu:\nu_\tau)=(1:2:0)$, with the IceCube MESE data. The future IceCube-Gen2~\cite{IceCube-Gen2:2020qha, IceCube-Gen2:2023rds}
predicts a more restricted flavor ratio constraint.
In addition to IceCube observations, KM3NeT \cite{KM3Net:2016zxf}, Baikal-GVD \cite{Baikal-GVD:2019fko}, P-ONE \cite{P-ONE:2020ljt}, and TAMBO \cite{Romero-Wolf:2020pzh} have the potential to detect high energy neutrinos and contribute to the flavor ratio constraints. Combining IceCube MESE, IceCube-Gen2, and additional contributions from future neutrino telescopes~\cite{Agarwalla:2023sng},
we are able to distinguish the flavor ratio signal from different DM mass 
and from different production sources. Furthermore, the couplings of the neutrino-DM interactions would be constrained by these experimental limits.

The Diffuse Supernova Neutrino Background (DSNB) is an alternative neutrino source considered in this work, with energy scale $\mathcal{O}(10\,{\rm MeV})$. The neutrino fluxes with certain flavor structure are produced by core-collapse supernovae that have occurred throughout the cosmic history and are nearly isotropic. Detectors such as Super-Kamiokande/Hyper-Kamiokande (SK/HK) \cite{Super-Kamiokande:2021jaq, Hyper-Kamiokande:2022smq}, Deep Underground Neutrino Experiment (DUNE) \cite{DUNE:2020lwj, DUNE:2020ypp}, and Jiangmen Underground Neutrino Observatory (JUNO)~\cite{JUNO:2015zny} are aimed to detect the DSNB flux and their flavor ratio. The detection of DSNB from these experiments relies on different neutrino interactions. For example, the DUNE depends on the charge current interaction on liquid Argon to identify electron neutrino; the SK/HK and JUNO depend on the Inverse Beta Decay (IBD), which is sensitive to $\bar{\nu}_{e}$. On the other hand, the muon and tau neutrinos can be detected from neutrino-electron elastic scattering. Applying the $\nu\phi$ interactions into the Hamiltonian would alter the neutrino flavors from DSNB, and we compared the final flavor ratio with the experimental limits.

The paper is organized as follow: In Section \ref{nu_in_DM}, we introduce the model of $\nu\phi$ interaction, the spiked DM density around AGN, the DSNB flux from core-collapse supernovae, and modified the vacuum neutrino Hamiltonian by adding the effective potential. Then we compute the average flavor ratio by taking the oscillation length and the adiabatic condition into account. We set our analysis upper limit on coupling constant with the neutrino mean free path. In Section \ref{analy}, we generate and analyze the ternary plot of flavor ratio distributions for AGN neutrinos with different set of DM mass, the event averaged deviation of final flavor ratio from neutrinos propagate in the vacuum with initial flavor ratio $(1:2:0)$, and show the constraint on coupling constant with normal and inverse mass ordering. We repeat the similar analysis for DSNB neutrinos but with the flavor component $(\nu_e,\nu_\mu,\nu_\tau)$ replaced by $(\bar{\nu}_e,\nu_x,\nu_e)$, and the DM mass is fixed at $10^{-22}\,{\rm eV}$. We summarize our work in Section \ref{sum}.

\bigskip

\section{Neutrino Oscillation in Dark Matter}\label{nu_in_DM}
\begin{figure}
    \centering
    \includegraphics[width=0.25\linewidth]{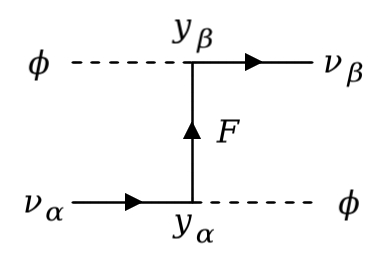}
    \caption{The process leads to effective potential $V_{\alpha\beta}$.}
    \label{feynman}
\end{figure}

We assume neutrinos are Majorana fermions and couple to hypothetical dark scalar $\phi$ through a fermionic mediator $F$. The Lagrangian is given by \cite{Tseng:2024akh}
    \begin{equation}\label{lagrange}
    \mathcal{L}_{\rm int}=\sum_{\alpha}y_\alpha(\phi\overline{P_L\nu_{\alpha}}F_R+\phi^\ast\overline{F_R}P_L\nu_\alpha),
    \end{equation}
where the Greek alphabet $\alpha=e,\mu,\tau$ is the neutrino flavor label, $y_\alpha$ is coupling constant. In the flavor basis, the Hamiltonian is
    \begin{equation}\label{flavor_totH}
    H=\frac{1}{2E}U\begin{pmatrix}
    0&0&0\\
    0&\Delta m^2_{21}&0\\
    0&0&\Delta m^2_{31}
    \end{pmatrix}U^\dagger+V,
    \end{equation}
where $U$ is the PMNS matrix, $V$ is the effective potential matrix and it depends on the $\nu\phi$ interaction. The differences of mass square for Normal Ordering (Inverse Ordering) are given by~\cite{ParticleDataGroup:2024cfk}
\begin{equation}\label{massorder}
    \begin{aligned}
        &m_1<m_2<m_3\,({\rm NO}):\quad \Delta m^2_{21}=7.50\times10^{-5}\,{\rm eV}^2,\quad \Delta m^2_{32}=2.45\times10^{-3}\,{\rm eV}^2,\\
        &m_3<m_1<m_2\,({\rm IO}):\quad \,\ \Delta m^2_{21}=7.50\times10^{-5}\,{\rm eV}^2,\quad \Delta m^2_{32}=-2.53\times10^{-3}\,{\rm eV}^2.\\
    \end{aligned}
\end{equation} 
From (\ref{lagrange}), we can write down the components of $V$:
    \begin{equation}\label{effV}
    V_{\alpha\beta}=G_{\alpha\beta}\frac{\rho_{\rm DM}}{m_\phi},\quad G_{\alpha\beta}=\frac{y_\alpha y_\beta}{m^2_F}.
    \end{equation}
The Feynman diagram of $\nu\phi$ scattering is shown in Fig. \ref{feynman}.
The effective mass squared is given by
    \begin{equation}
    H_m=U^{\prime\dagger}HU'=\frac{1}{2E}\begin{pmatrix}
    0&0&0\\
    0&\Delta m^2_{21,\rm eff}&0\\
    0&0&\Delta m^2_{31,\rm eff}
    \end{pmatrix},
    \end{equation}
where $U'$ is the mixing matrix that diagonalized the flavor basis total Hamiltonian Eq.(\ref{flavor_totH}). Then oscillation length is 
    \begin{equation}
    L_{ij}=\frac{4\pi E}{\Delta m^2_{ij,\rm eff}},
    \end{equation}
where the Latin alphabets $i,j$ are the mass eigenstate label.

We denote the initial flavor ratio by $f_0=(f_{0e},f_{0\mu},f_{0\tau})$, then after propagating a distance that is far longer than the oscillation length, the final average flavor ratio is given by \cite{deSalas:2016svi}
    \begin{equation}\label{final_ratio}
    f_{1\beta}(r_1)=\sum_{\alpha=e,\mu,\tau}\left(\sum^3_{i=1}|U'_{\beta i}(r_0) U^{\ast\prime}_{\alpha i}(r_0)|^2f_{0\alpha}\right).
    \end{equation}
where $r_0$ is the distance to the source. By iterating Eq.(\ref{final_ratio}) $n+1$ times, we obtain the final average ratio at Earth
    \begin{equation}\label{ratio_at_earth}
    f^\oplus_{\beta}=\sum_{\alpha=e,\mu,\tau}\left(\sum^3_{i=1}|U'_{\beta i}(r_n) U^{\ast\prime}_{\alpha i}(r_n)|^2f_{n\alpha}\right).
    \end{equation}
If the DM density changes slowly along the path, the neutrino propagates adiabatically. The adiabatic parameter is defined by 
    \begin{equation}
    \eta=\frac{\langle|U^{\prime\dagger}\dot{U}'|\rangle}{\langle|H_m-({\rm Tr}(H_m)/3)I|\rangle},
    \end{equation}
where $I$ is the $3\times 3$ identity matrix. The $\langle...\rangle$ denotes the average over all the matrix elements. The propagation of neutrinos is adiabatic if $\eta\ll1$.

\bigskip

\subsection{Active Galactic Nuclei}
The neutrino spectrum from AGN can be expressed by the power law $\Phi(E)\propto E^{-\gamma}$ \cite{IceCube:2022der, Doring:2023vmk}. Assuming $N_{\rm tot}$ the total event number measured by IceCube, the differential event contributed from certain energy $E$ is written as
    \begin{equation}
    \frac{dN}{N_{\rm tot}}=\frac{(-\gamma+1)E^{-\gamma}dE}{E_{\rm max}^{-\gamma+1}-E_{\rm min}^{-\gamma+1}},
    \end{equation}
where $E_{\rm max(\rm min)}$ is the energy upper (lower) limit of neutrino from AGN.
The DM density around AGN is given by the spike-profile,
    \begin{equation}
    \rho_{\rm DM}(r)=\begin{cases}
    0&,r\leq 4R_s\\
    \frac{\rho_{\rm sp}(r)\rho_{\rm sat}}{\rho_{\rm sp}(r)+\rho_{\rm sat}}&, 4R_s\leq r\leq R_{\rm sp}\\
    \frac{\rho_{\rm NFW}(r)\rho_{\rm sat}}{\rho_{\rm NFW}(r)+\rho_{\rm sat}}&, r\geq R_{\rm sp}
    \end{cases}
    \end{equation}
where $R_s$ is the Schwarzschild radius. The saturation density $\rho_{\rm sat}$ depends on the $\phi\phi^\ast$ annihilation, and the cross section is proportional to $y_{\alpha}^2y^2_\beta$ \cite{Tseng:2024akh}. 
The total contributions to annihilation cross section is $\sum_{\alpha,\beta}\sigma_{\phi\phi^\ast\rightarrow\nu_\alpha\bar{\nu}_\beta}$. $\rho_{\rm sp }$ is the spike density around the central supermassive black hole and is given by \cite{Herrera:2023nww}
    \begin{equation}
    \rho_{\rm sp}(r)=\rho_R g_\gamma(r)\left(\frac{R_{\rm sp}}{r}\right)^{\gamma_{\rm sp}},
    \end{equation}
where $R_{\rm sp}$ indicates the size of the spike. For $r\geq R_{\rm sp}$ the spike density is replace by the Navarro-Frenk-White
(NFW) profile.

For NGC 1068, the oscillation length averaged over energy and coupling constant is around $10^{-6}\,{\rm pc}$, which is much shorter than the distance from NGC 1068 to Earth. The maximum of adiabatic parameter $\eta$ is below $10^{-10}$, which means neutrinos propagates extremely adiabatically.
Therefore, we can simply consider the DM density at the source location and around the Earth. We assume neutrinos from NGC are produced within the corona, which possess  radius $R\in [10,10^2]R_s$\footnote{NGC's corona radius is estimated by $R\in [3,100]R_s$~\cite{Blanco:2023dfp}. Ref.~\cite{Fiorillo:2025cgm} states that the TXS's neutrino flux from corona isn't sufficient enough to produce IceCube observed neutrino.}. 
The DM density around the Earth is taken to be $\rho^\oplus_{\rm DM}=0.4\,{\rm GeV/cm^3}$. The relevant parameters for TXS 0506+056 are $\eta\simeq 6\times10^{-11}$ and averaged $L_{ij} \sim 10^{-4}\,{\rm pc}$.

The $\nu\phi$ scattering may affect the direction of propagation. For detection on Earth, we require that the neutrinos be forward-scattered through their path. So the mean free path should be longer than the distance between AGN and Earth. The mean free path $l$ is related to the scattering cross section and the number density of DM 
    \begin{equation}l=\left(\sigma_{\nu\phi}\frac{\rho_{\rm DM}}{m_\phi}\right)^{-1}.
    \end{equation}
The cross section is given by $\sigma_{\nu\phi}=y^2_\alpha y^2_\beta m_\phi E/(32\pi m_F^4)$ under the heavy mediator limit, i.e. $m_F\geq 1\,{\rm TeV}$ and $m_F^2\gg m_\phi E$. 
The $\rho_{\rm DM}$ is nearly independent of $m_\phi$ when $m_\phi\ll m_F$, so $l$ depends mainly on the coupling constant and neutrino energy. For the analysis of AGNs, $m_F$ is fixed to $1\,{\rm TeV}$. For a given $y_{\alpha,\beta}$, $l$ reaches its minimum when (i) $\rho_{\rm DM}$ is maximum, i.e., the peaked spike density, and (ii) $E=E_{\rm max}$. Concerning this extreme case, we find $l_{\rm min}\sim 1.81\times10^{-4}\,{\rm pc}$ ($l_{\rm min}\sim 1.85\times10^{-4}\,{\rm pc}$) for NGC 1068 (TXS 0506+056) when $\rho_{\rm DM}$ is evaluated at $10R_s$ and $y_{\alpha,\beta}=10$. In terms of Schwarzschild radius, $l_{\rm min}/R_s\sim 190$ ($l_{\rm min}/R_s\sim 6$) for NGC 1068 (TXS 0506+056). As neutrinos propagate outward, $l$ increases because $\rho_{\rm DM}$ gets lower. Therefore, neutrinos produced in the direction forward to Earth are rarely deflected by DM.

\bigskip

\subsection{Diffuse Supernova Neutrino Background}
The neutrino spectrum emitted from a supernova is given by \cite{Beacom:2010kk}
	\begin{equation}
	F(E)=\frac{E_{\rm tot}}{6}\frac{120}{7\pi^4}\frac{E^2}{T^4}\frac{1}{\exp(E/T)+1},
	\end{equation}
where $E_{\rm tot}=3\times 10^{53}\,{\rm erg}$ 
is the total emitted energy, the factor $1/6$ factor stands for energy of one of $\nu_e$, $\nu_{\bar{e}}$, $\nu_\mu$, $\nu_{\bar{\mu}}$, $\nu_\tau$, $\nu_{\bar{\tau}}$, and $T$ is the neutrino temperature.
The diffuse differential neutrino flux is given by
	\begin{equation}\label{DSNBflux}
	\frac{d\Phi_\nu}{dE}=\int_{0}^{z_{\rm max}}{\frac{R_{\rm CCSN}(z)F(E,z)}{H(z)}dz}.
	\end{equation}
The core-collapse supernova rate $R_{\rm CCSN}$ and the star-formation
rate are parameterized as~\cite{Horiuchi:2008jz, Yuksel:2008cu}
\begin{subequations}\label{rccsn}
	\begin{align}
	R_{\rm CCSN}(z)&=\dot{\rho}_\ast(z)\frac{\int_{8M_\odot}^{50M_\odot}{\psi(M)dM}}{\int_{0.1M_\odot}^{100M_\odot}{M\psi(M)dM}},\label{rccsn}\\
	\dot{\rho}_\ast(z)&=\dot{\rho}_0\left[
	(1+z)^{-10\alpha}+\left(\frac{1+z}{B}\right)^{-10\beta}+\left(\frac{1+z}{C}\right)^{-10\gamma}
	\right]^{-1/10}\label{sfr},
	\end{align}
	\end{subequations}
where $B=2^{1-\alpha/\beta}$, $C=2^{(\beta-\alpha)/\gamma}\cdot 5^{1-\beta/\gamma}$, $\dot{\rho}_0=0.0178^{+0.0035}_{-0.0036}\,{\rm M}_\odot\,{\rm yr}^{-1}\,{\rm Mpc}^{-3}$, and $\alpha=3.4\pm 0.2$, $\beta=-0.3\pm 0.2$, $\gamma=-3.5\pm 1$. The initial mass function $\psi(M)$ is proportional to $M^{-2.35}$ \cite{Salpeter1955}.
We fixed $z_{\rm max}=5$ for which there is a reasonable amount of star formation. The neutrino temperature can be obtained by the core-collapse simulation of neutrino emissions \cite{Horiuchi:2008jz, Mathews:2014qba}. The simulations predict the temperature hierarchy $T_{\nu_e}<T_{\bar{\nu}_e}<T_{\nu_x}$. We adopt $T_{\nu_e}=5\,{\rm MeV}$, $T_{\bar{\nu}_e}=6\,{\rm MeV}$, and $T_{\nu_x}=7\,{\rm MeV}$ in our calculations.

The initial flavor ratio is given by \cite{Tabrizi:2020vmo} 
\begin{subequations}\label{ini_fDSNB}
    \begin{equation}
    f_{0\alpha}=\frac{\Phi_\alpha}{\sum_{\beta=\nu_e,\bar{\nu}_e,\nu_x}\Phi_\beta} ,
    \end{equation}
    \begin{equation}
    f_{0\alpha}(E)=\frac{\frac{d\Phi_\alpha}{dE}(E)\Delta E}{\sum_{\beta=\nu_e,\bar{\nu}_e,\nu_x}\frac{d\Phi_\beta}{dE}(E)\Delta E},
    \end{equation}    
\end{subequations}
where $\nu_x$ is the collective notation of $\nu_\mu,\bar{\nu}_\mu,\nu_\tau,\bar{\nu}_\tau$. Eq.(\ref{ini_fDSNB}a) indicates the initial flavor ratio of the DSNB flux, while Eq.(\ref{ini_fDSNB}b) is the initial ratio for each energy bin. 
Since the neutrinos are propagating in the extragalactic medium, the dark matter density is far smaller than the AGN spike density. Furthermore, the neutrino energy from DSNB scales around $\mathcal{O}({\rm MeV})$, hence the vacuum Hamiltonian is several order larger than those of AGN neutrinos. We must therefore increase the value of effective potential $V_{\alpha\beta}$. The theoretical lower limit of ultra-light dark matter mass is approximately $10^{-22}\,{\rm eV}$. We fix $m_\phi$ to this value and $m_F$ to $0.1\,{\rm GeV}$, then use the exact $\sigma_{\nu\phi}$ \cite{Tseng:2024akh} to compute the neutrino mean free path. It turns out $l\sim 400\,{\rm kpc}$ around Earth (the maximum DM density is $\rho^\oplus_{\rm DM}=0.4\,{\rm GeV/cm^3}$) and $l\sim 130\,{\rm Gpc}$ in extragalactic medium.
The maximum average oscillation length is $\sim 10^{-5}\,{\rm pc}$. We assume the  DM density between CCSN and Earth varies as a linear function of $z$, denoted by $\rho_{\rm DM}(z)$. When $z=0$, $\rho_{\rm DM}(0)=\rho^\oplus_{\rm DM}$. Suppose a single CCSN locates at redshift $z_0$, then $\rho_{\rm DM}(z_0)=1.27\,{\rm GeV/m^3}$ is the average extragalactic DM density. Then the adiabatic parameter $\eta\sim 10^{-19}\sim10^{-13}$ for $z=5\sim 10^{-5}$. The contribution to the number of events from certain energy $E$ is
    \begin{equation}\label{DeltaN_N}
    \frac{\Delta N}{N_{\rm tot}}=\frac{\sum_{\beta=\nu_e,\bar{\nu}_e,\nu_x}\frac{d\Phi_\beta}{dE}\Delta E}{\sum_{\beta=\nu_e,\bar{\nu}_e,\nu_x}\Phi_\beta}.
    \end{equation}

\begin{figure}
    \renewcommand{\thefigure}{2-1}
    \captionsetup[subfloat]{labelformat=empty}
    \centering
    \subfloat[\hspace{0.75cm}(a1)\hspace{3.1cm} (b1)\hspace{3.3cm}(c1)\hspace{3.1cm}(d1)]{\includegraphics[height=0.26\linewidth]{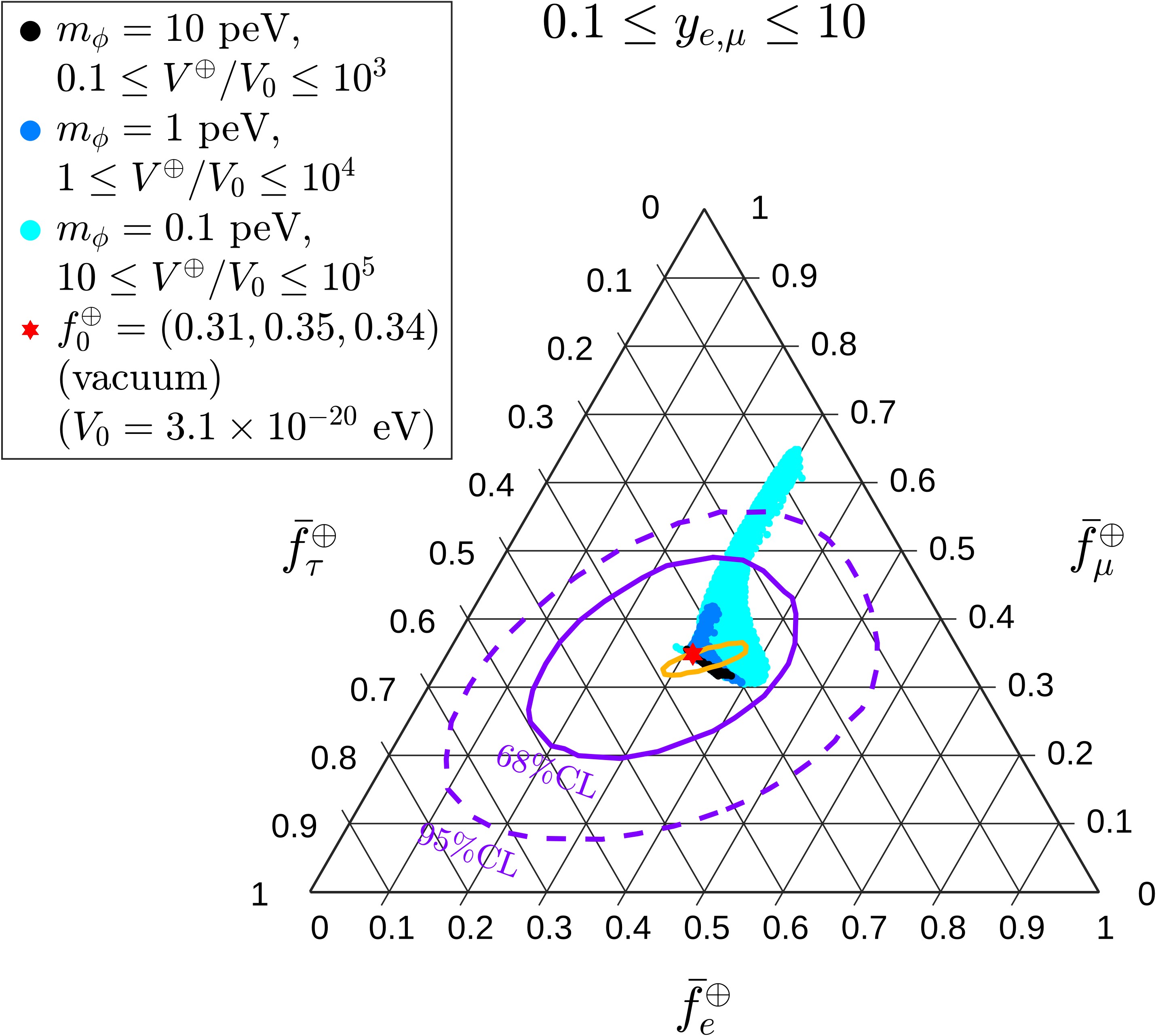}\quad
    \includegraphics[height=0.26\linewidth]{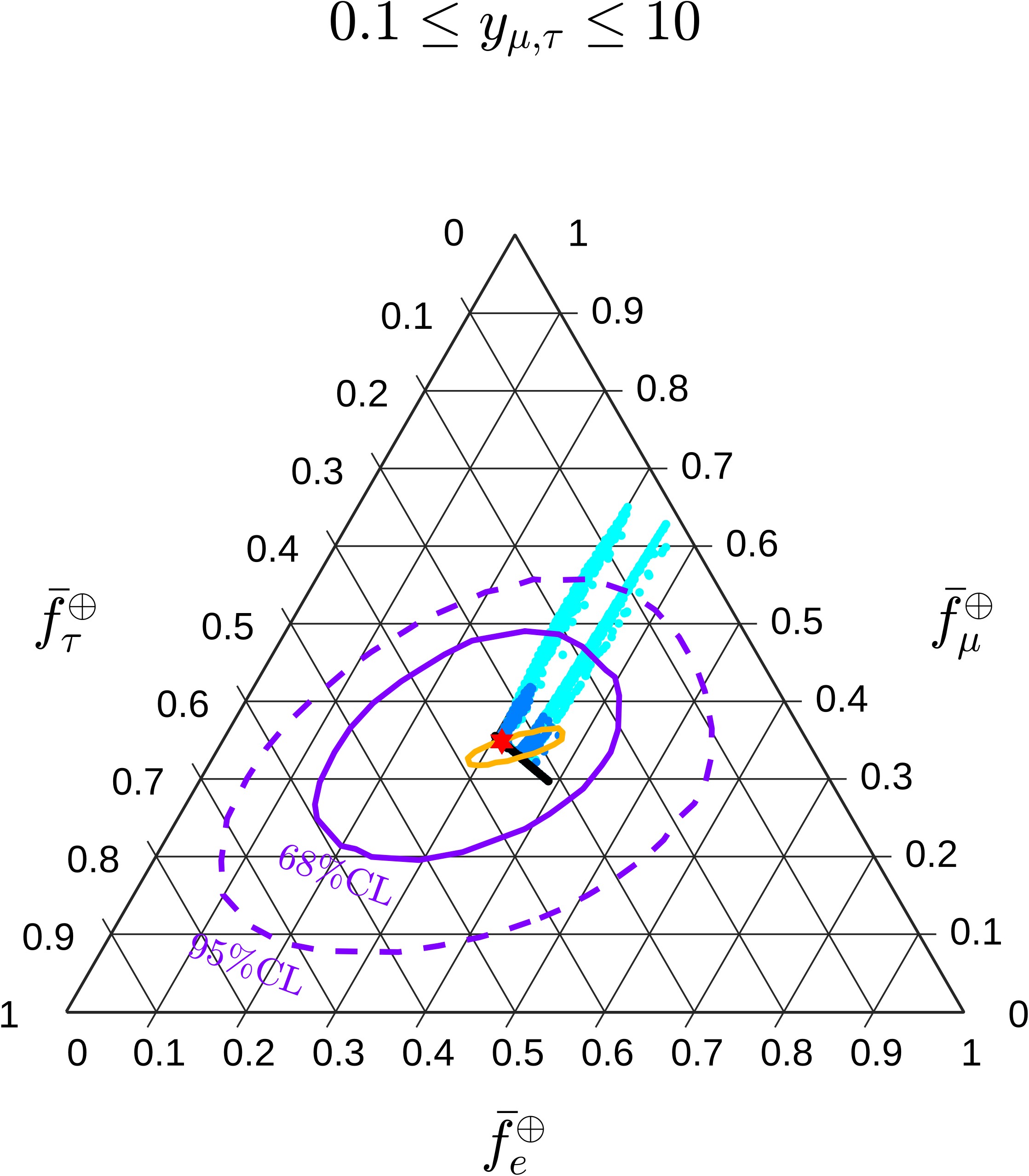}\quad
    \includegraphics[height=0.26\linewidth]{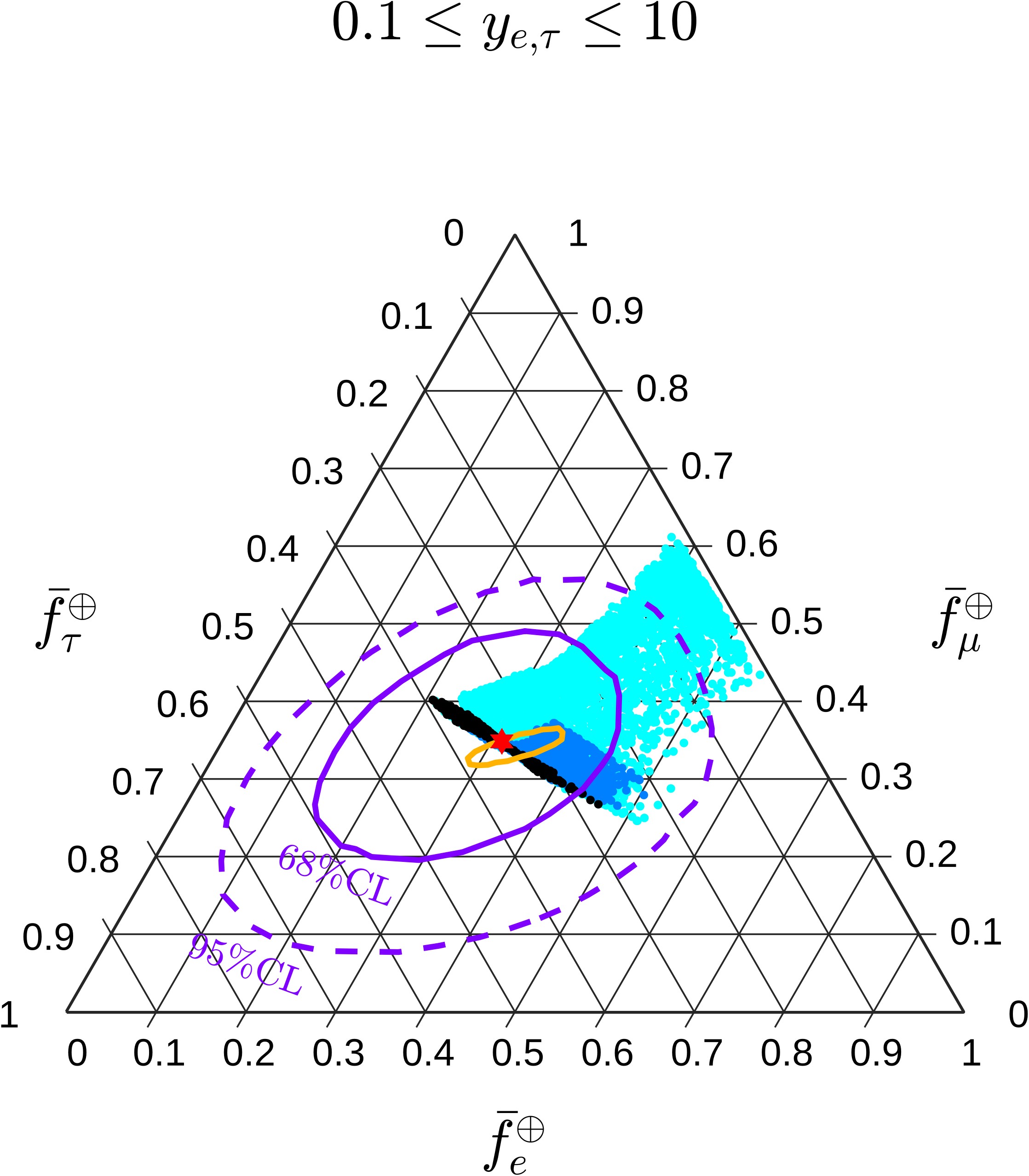}\quad
    \includegraphics[height=0.26\linewidth]{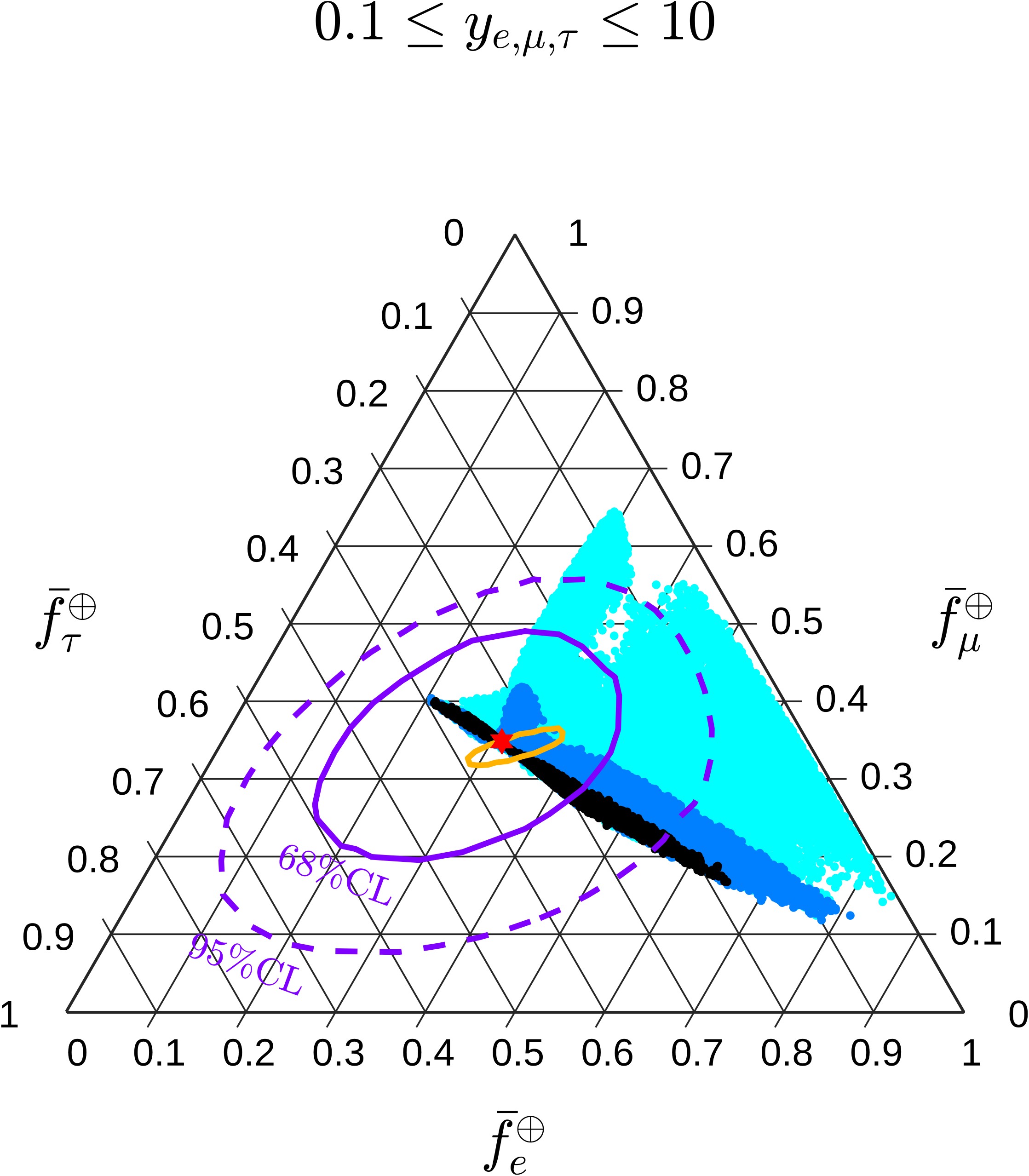}
    }
    
    \subfloat[\hspace{0.75cm}(a2)\hspace{3.1cm} (b2)\hspace{3.3cm}(c2)\hspace{3.1cm}(d2)]{\includegraphics[height=0.26\linewidth]{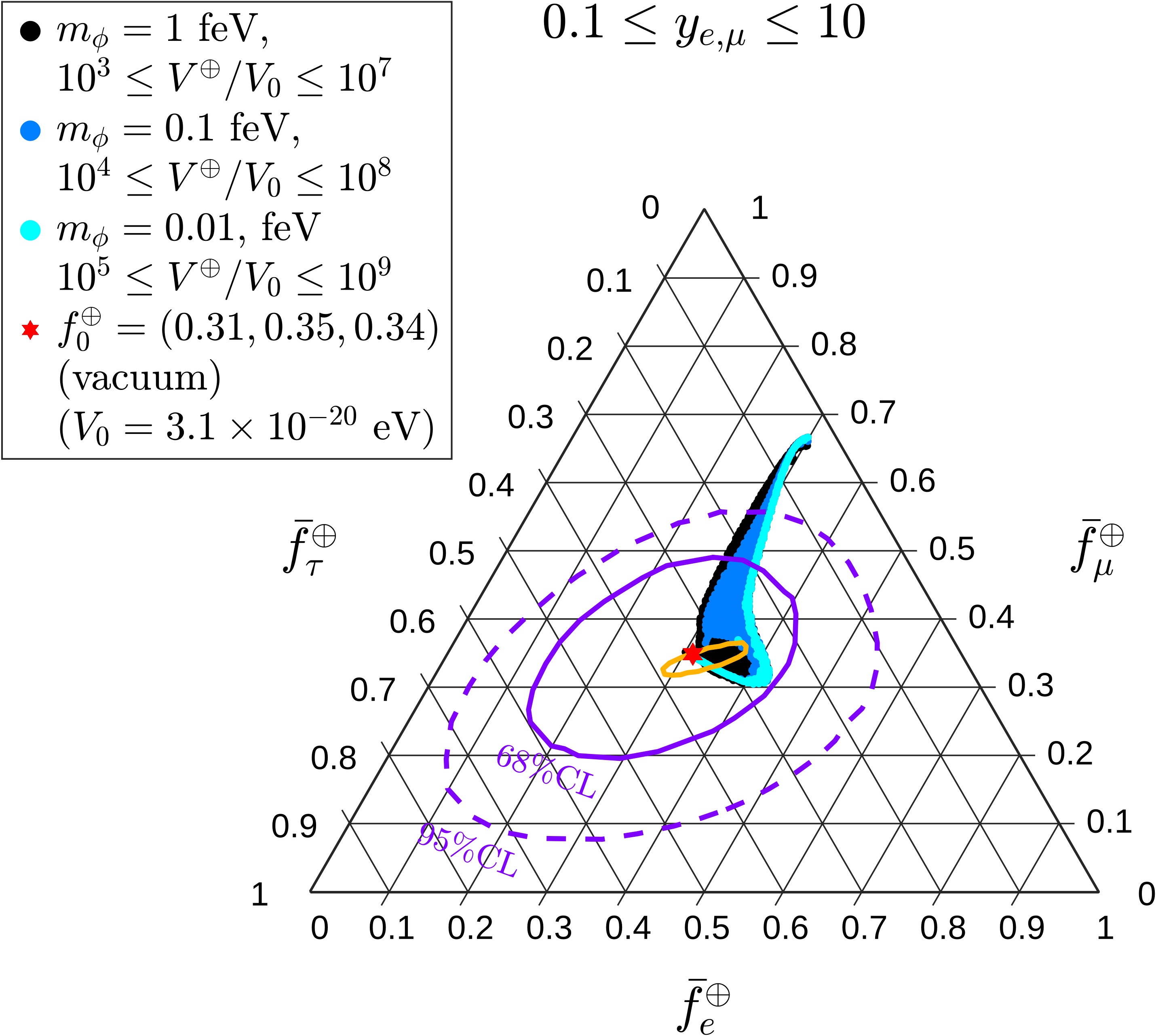}\quad
    \includegraphics[height=0.26\linewidth]{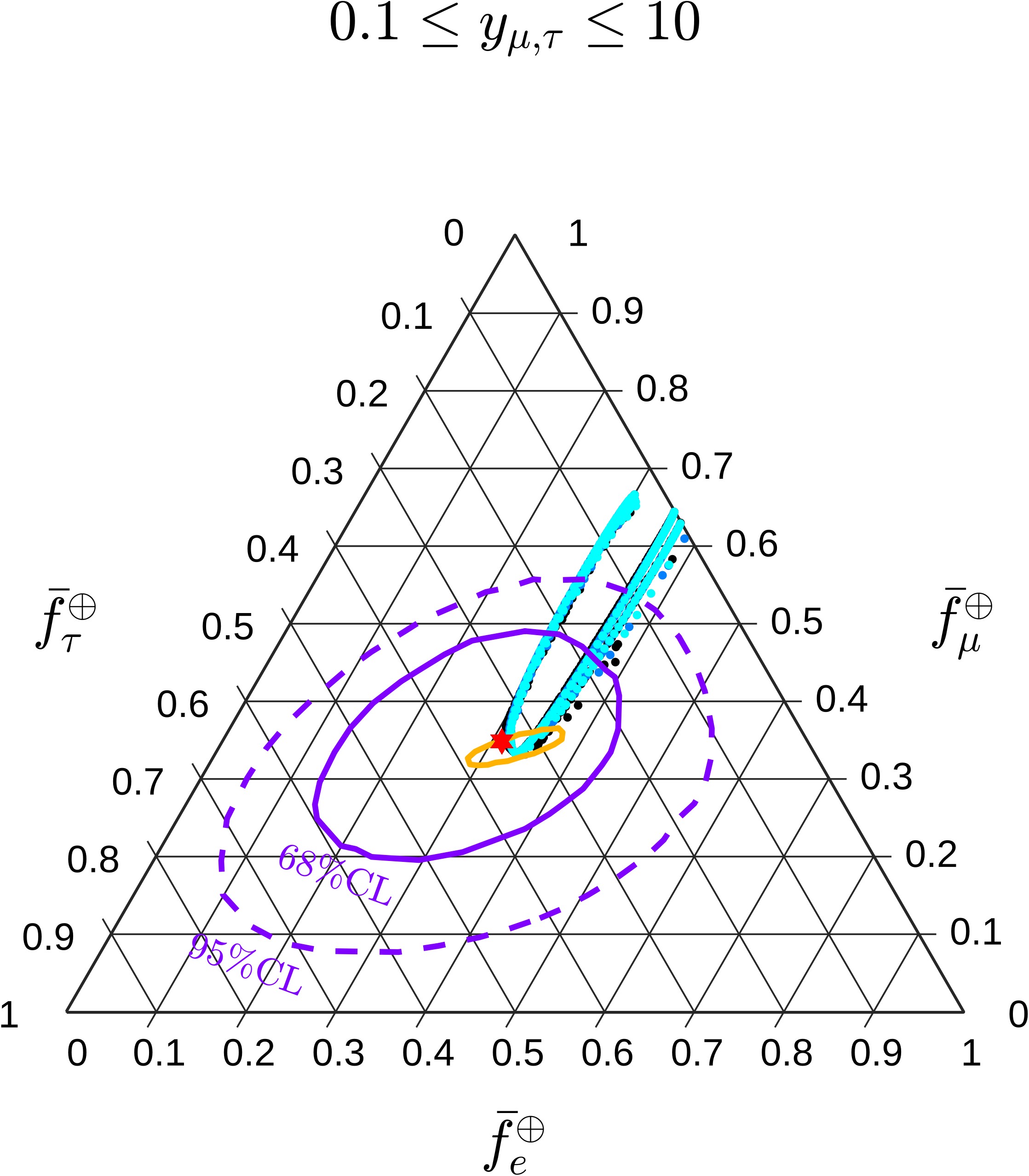}\quad
    \includegraphics[height=0.26\linewidth]{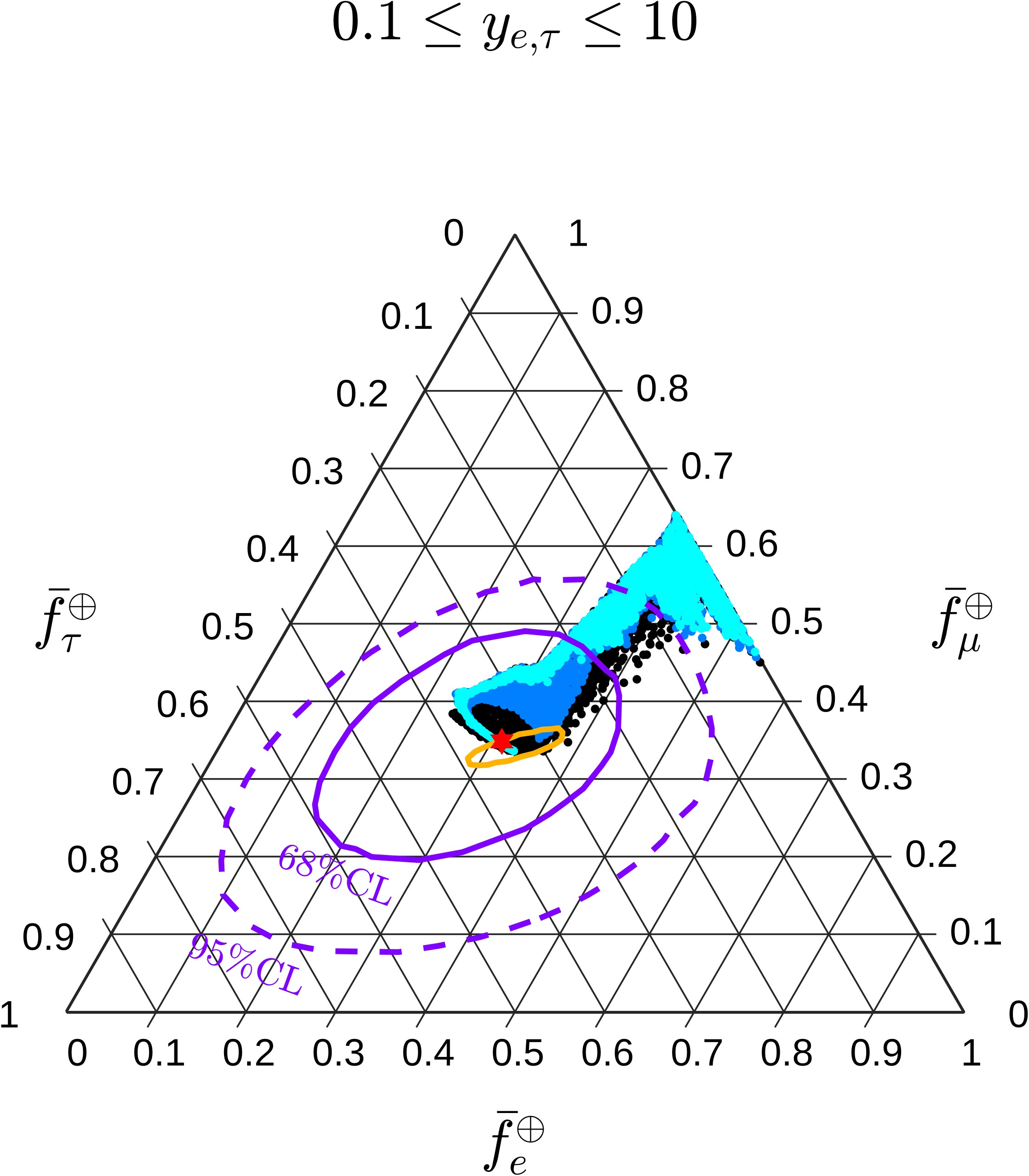}\quad
    \includegraphics[height=0.26\linewidth]{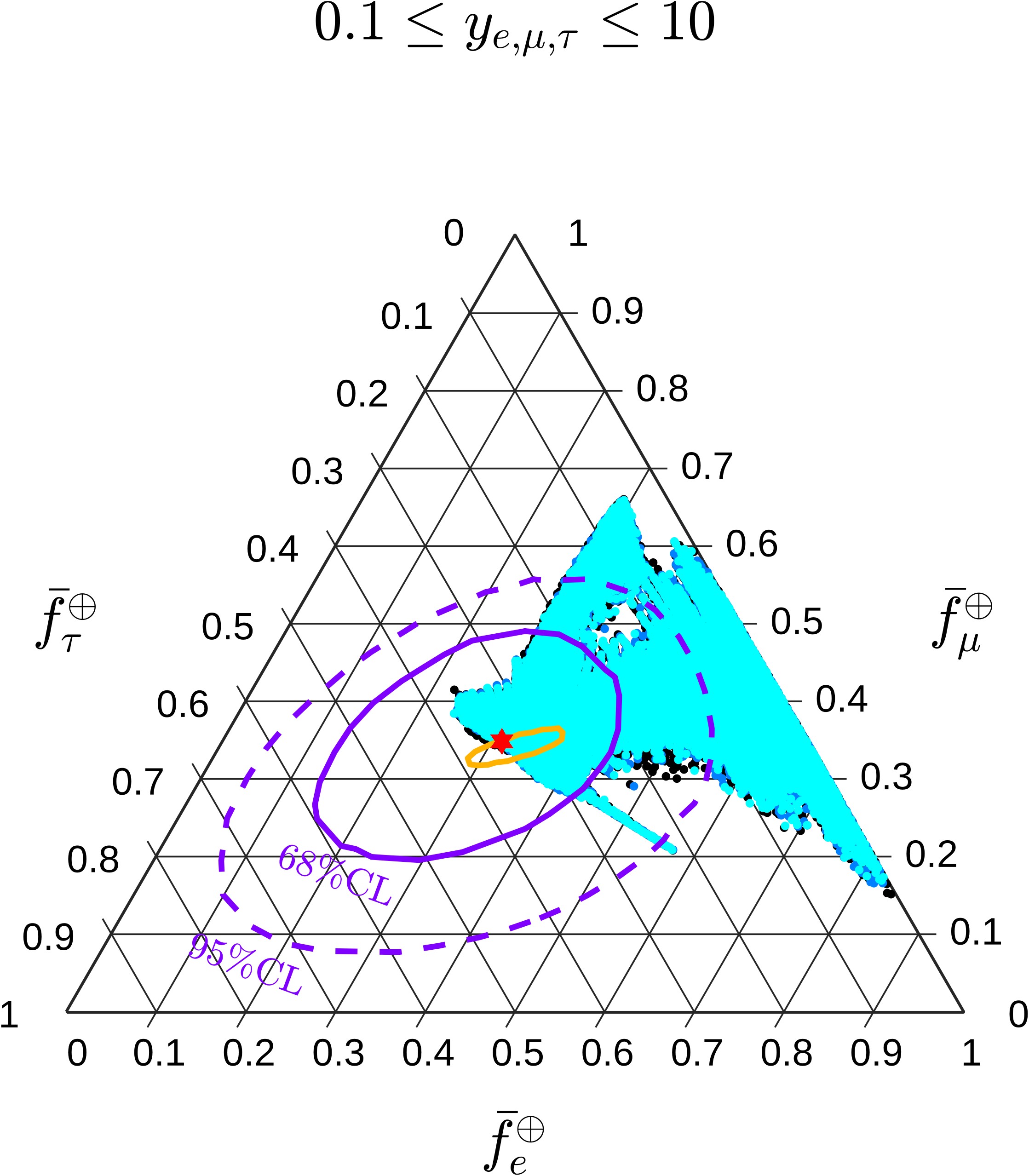}
    }

    \caption{The ternary plots of flavor ratio emitted from NGC 1068 and detected on Earth for Case-(a,b,c,d). In each plot the effective potential around NGC 1068 is far above $V^{31}_r$. In the first row, $V^\oplus_{\alpha\beta}$ approaches $V^{31}_r$ from below as we lower the value of $m_\phi$ to 0.1 peV; while in the second row, $V^\oplus_{\alpha\beta}$ moves away from $V^{31}_r$ as we further decrease value of $m_\phi$ to 0.01 feV.
    The neutrinos from NGC 1068 are assumed to be produced within $[10,10^2]R_s$. The purple solid (dashed) contours are the 68\% (95\%) C.L. of SPL best fitting of IceCube MESE data. The orange contour is the combined limit using 15 years of IceCube plus 10 years of IceCube-Gen2, and using the additional contribution of future neutrino telescopes.
    }
    \label{fig1}
\end{figure}

\begin{figure}
    \renewcommand{\thefigure}{2-2}
    \captionsetup[subfloat]{labelformat=empty}
    \centering
    \subfloat[(a1)\hspace{4.8cm} (a2)\hspace{4.8cm}(a3)]{\includegraphics[width=0.33\linewidth]{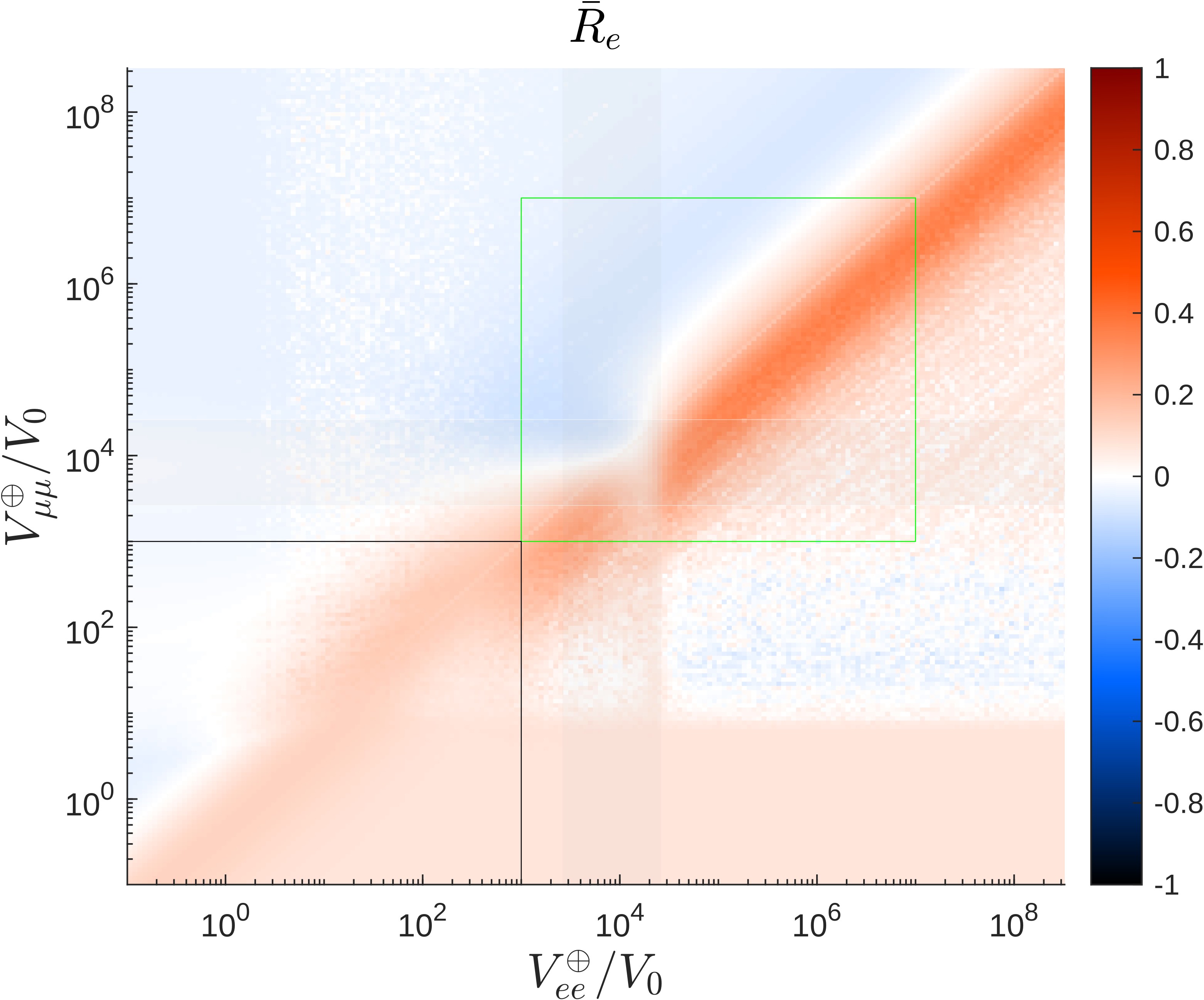}\quad
    \includegraphics[width=0.33\linewidth]{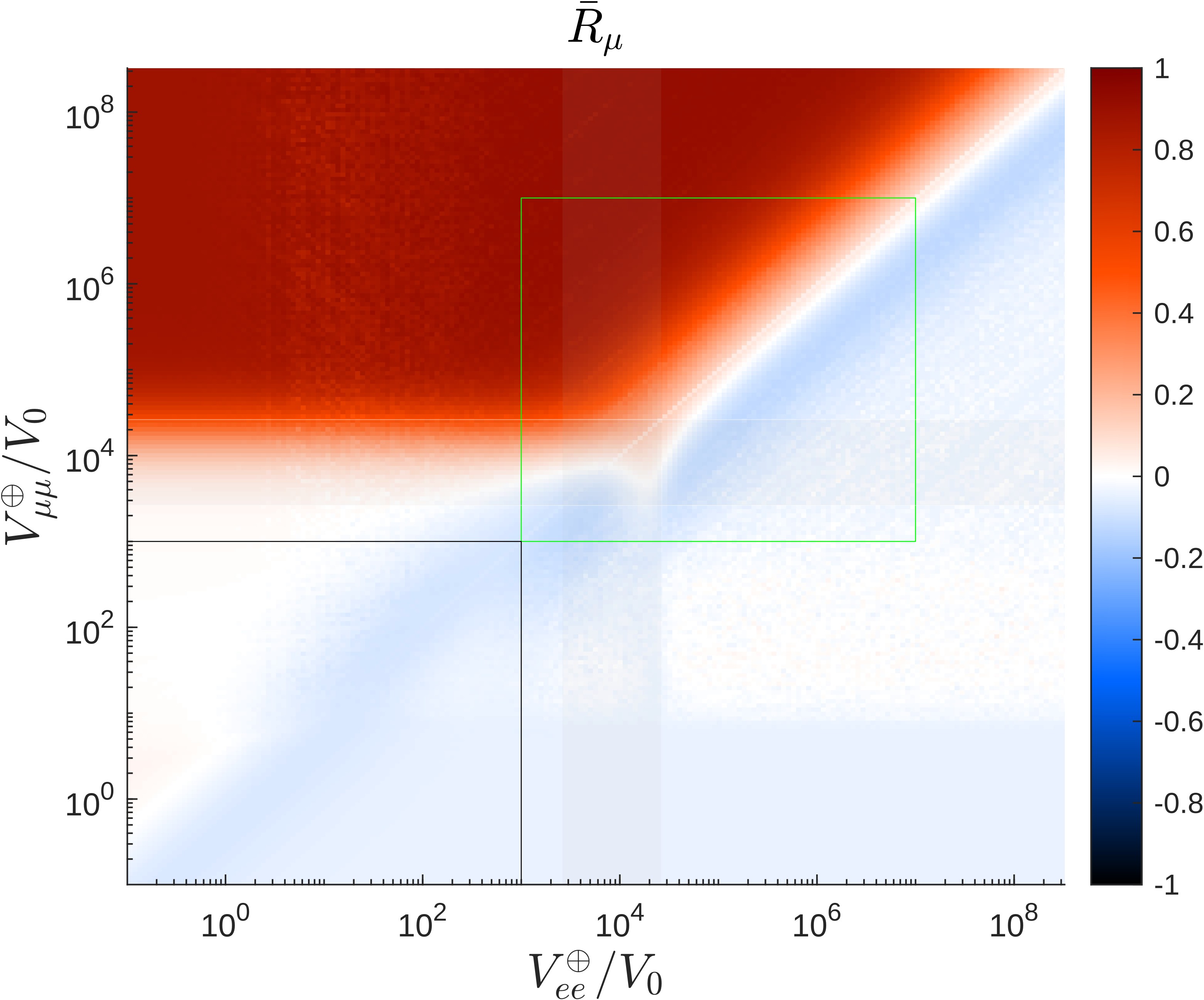}\quad
    \includegraphics[width=0.33\linewidth]{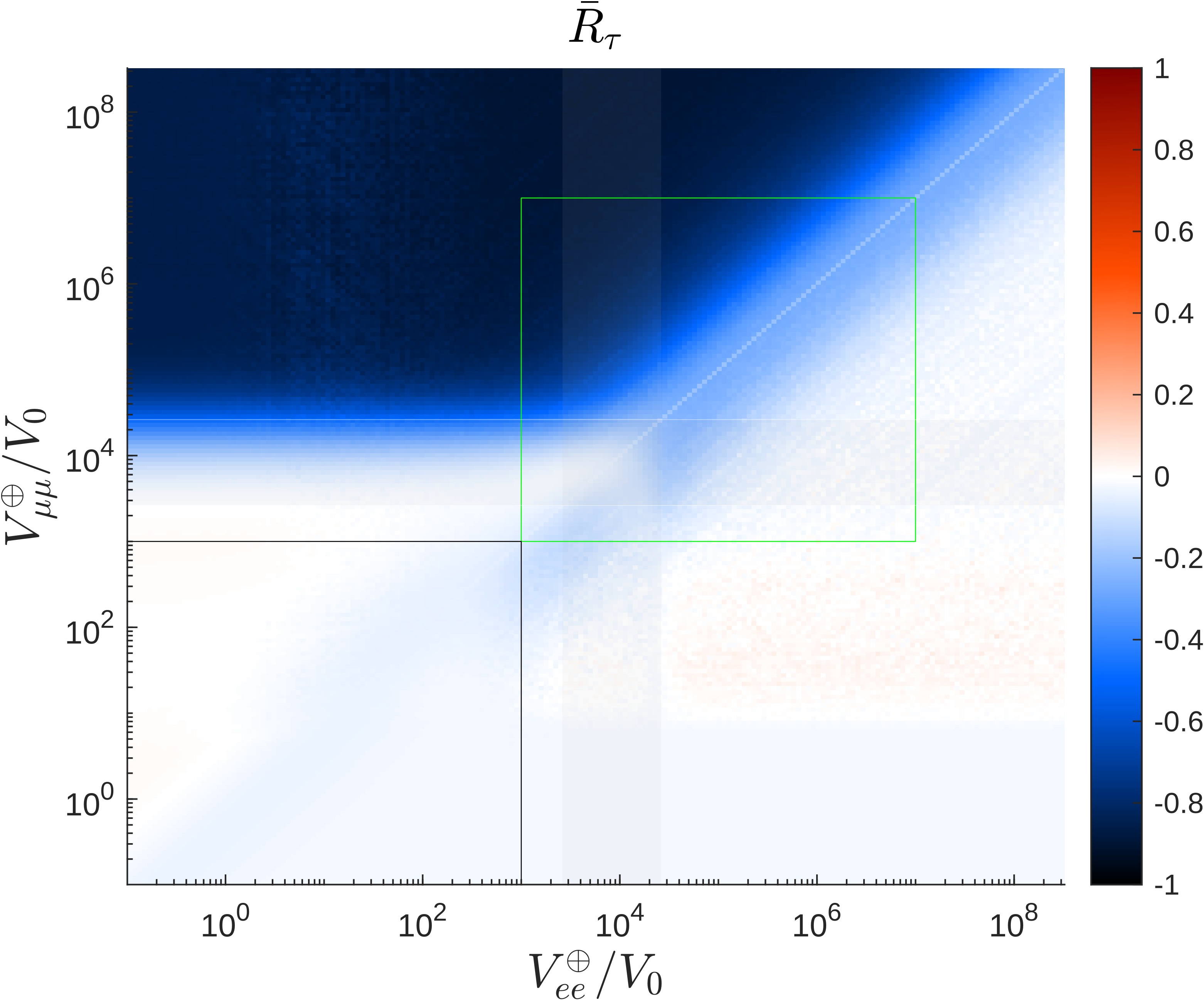}}

    \subfloat[(b1)\hspace{4.8cm} (b2)\hspace{4.8cm}(b3)]{\includegraphics[width=0.33\linewidth]{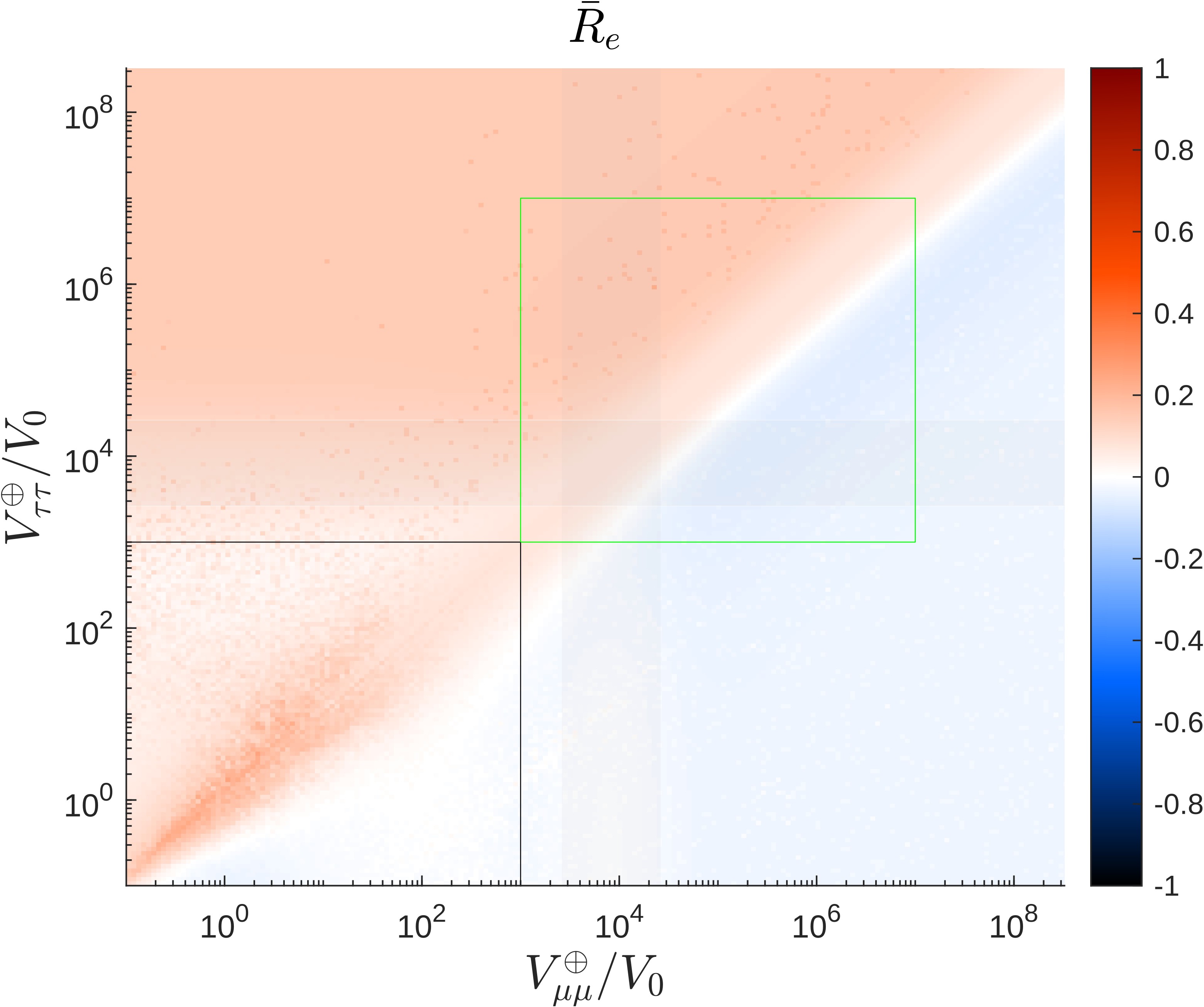}\quad
    \includegraphics[width=0.33\linewidth]{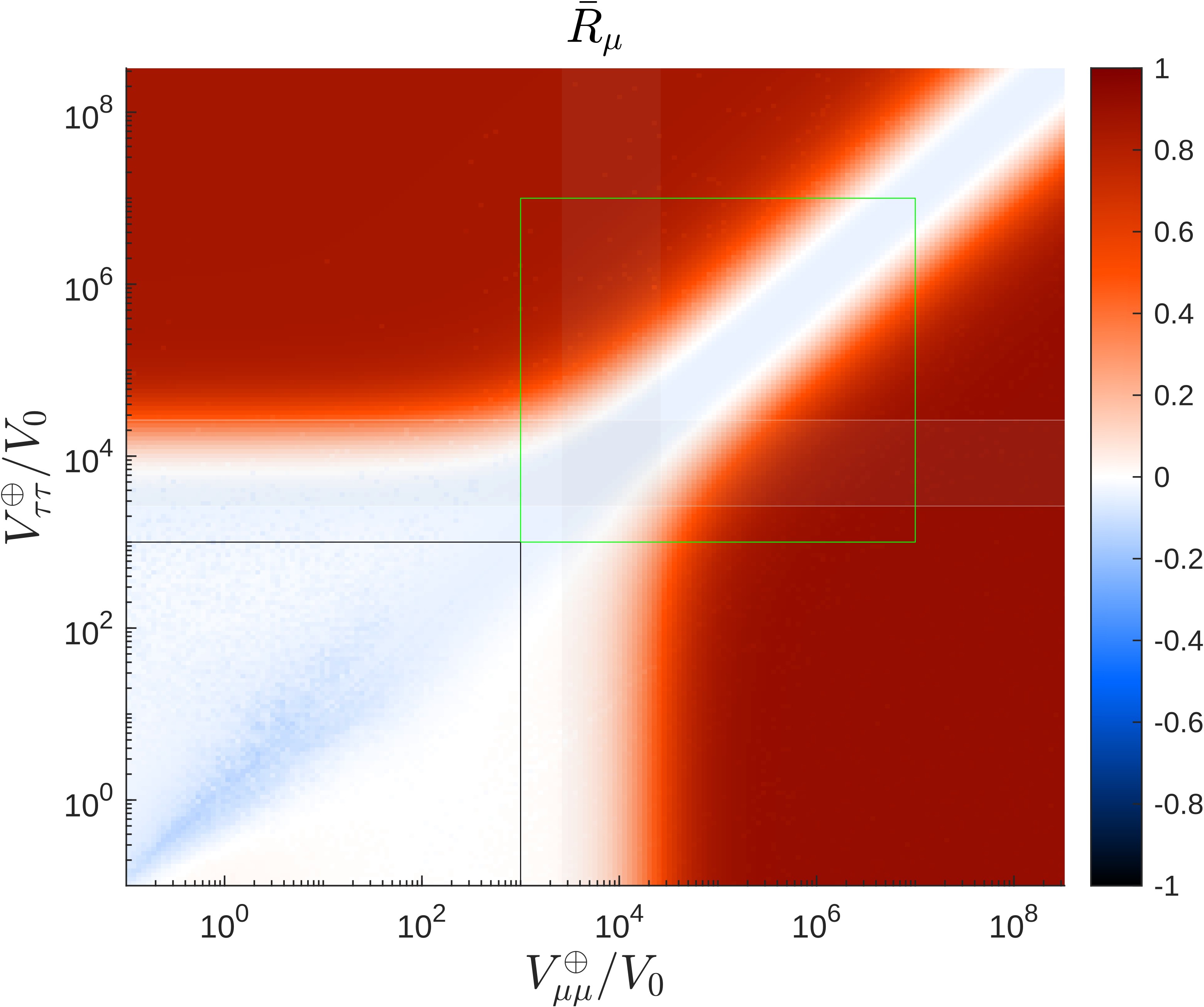}\quad
    \includegraphics[width=0.33\linewidth]{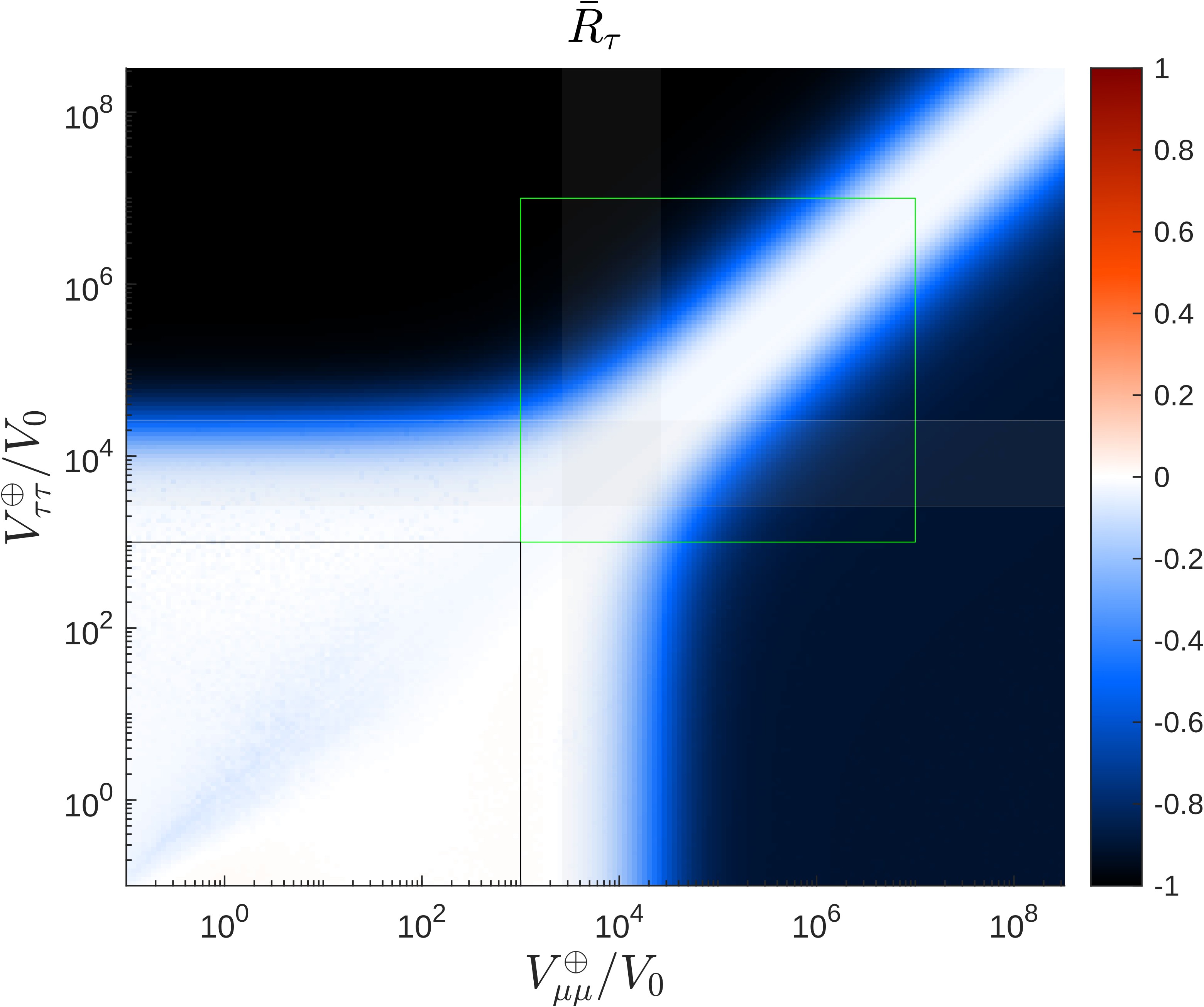}}

    \subfloat[(c1)\hspace{4.8cm} (c2)\hspace{4.8cm}(c3)]{\includegraphics[width=0.33\linewidth]{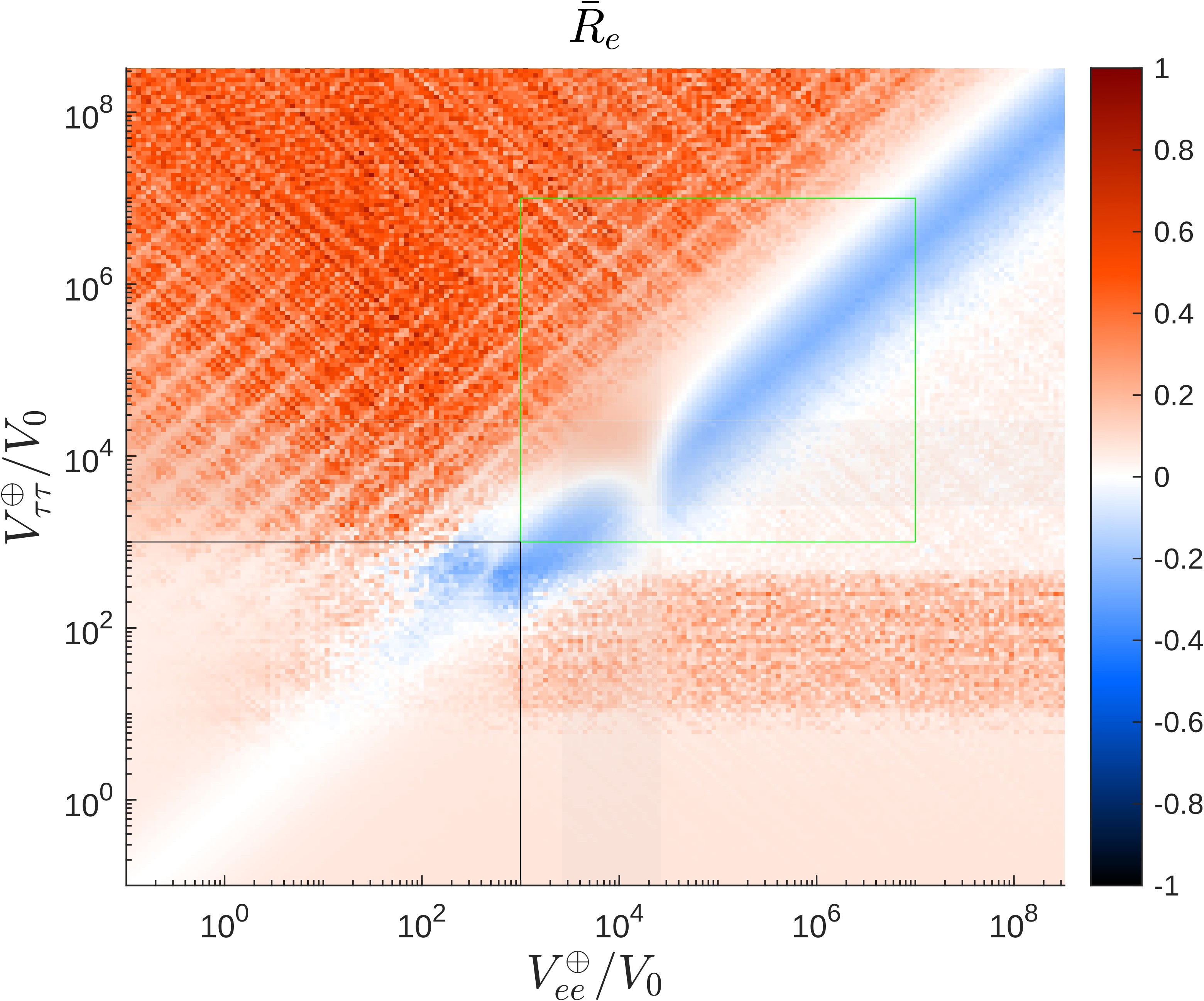}\quad
    \includegraphics[width=0.33\linewidth]{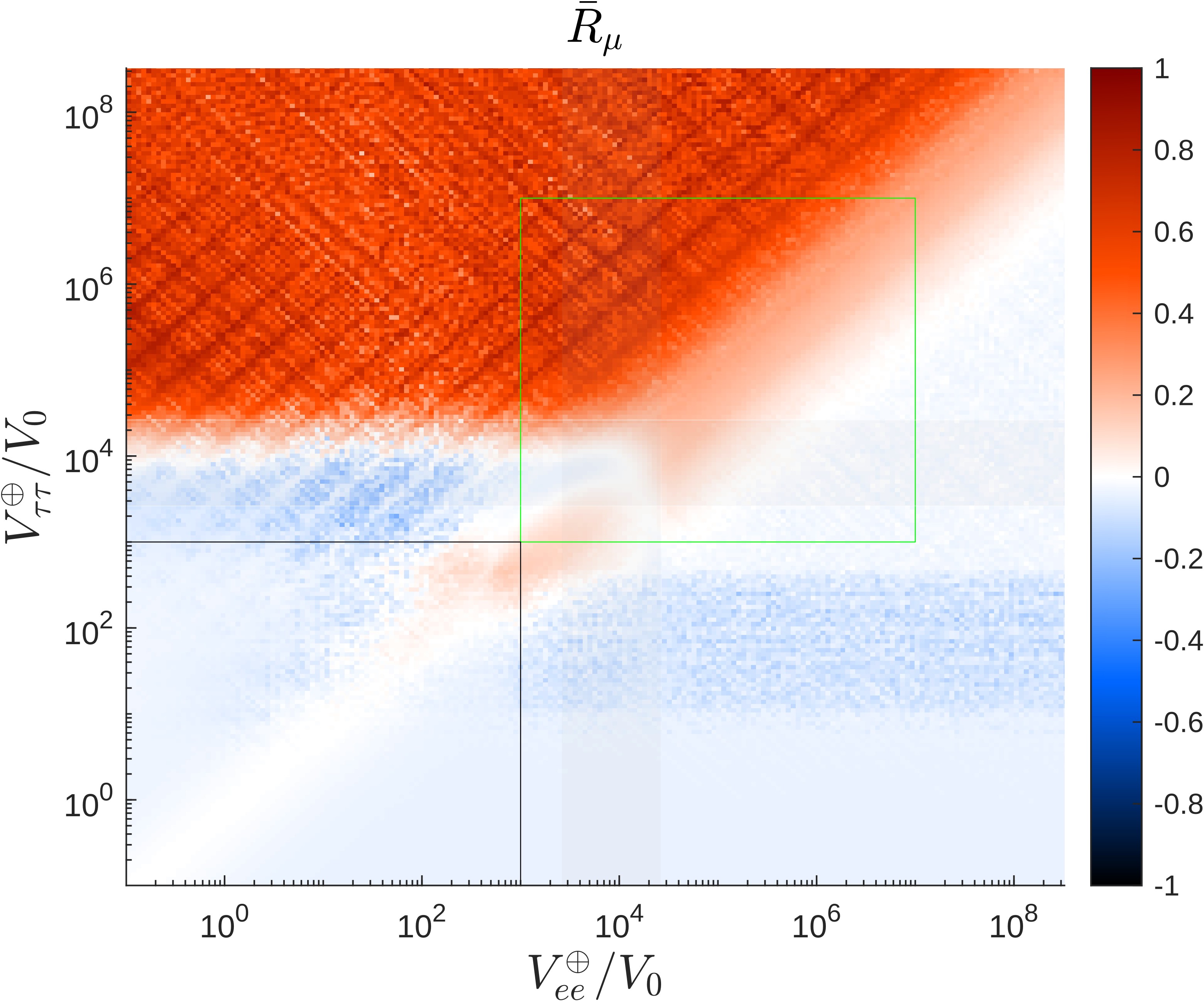}\quad
    \includegraphics[width=0.33\linewidth]{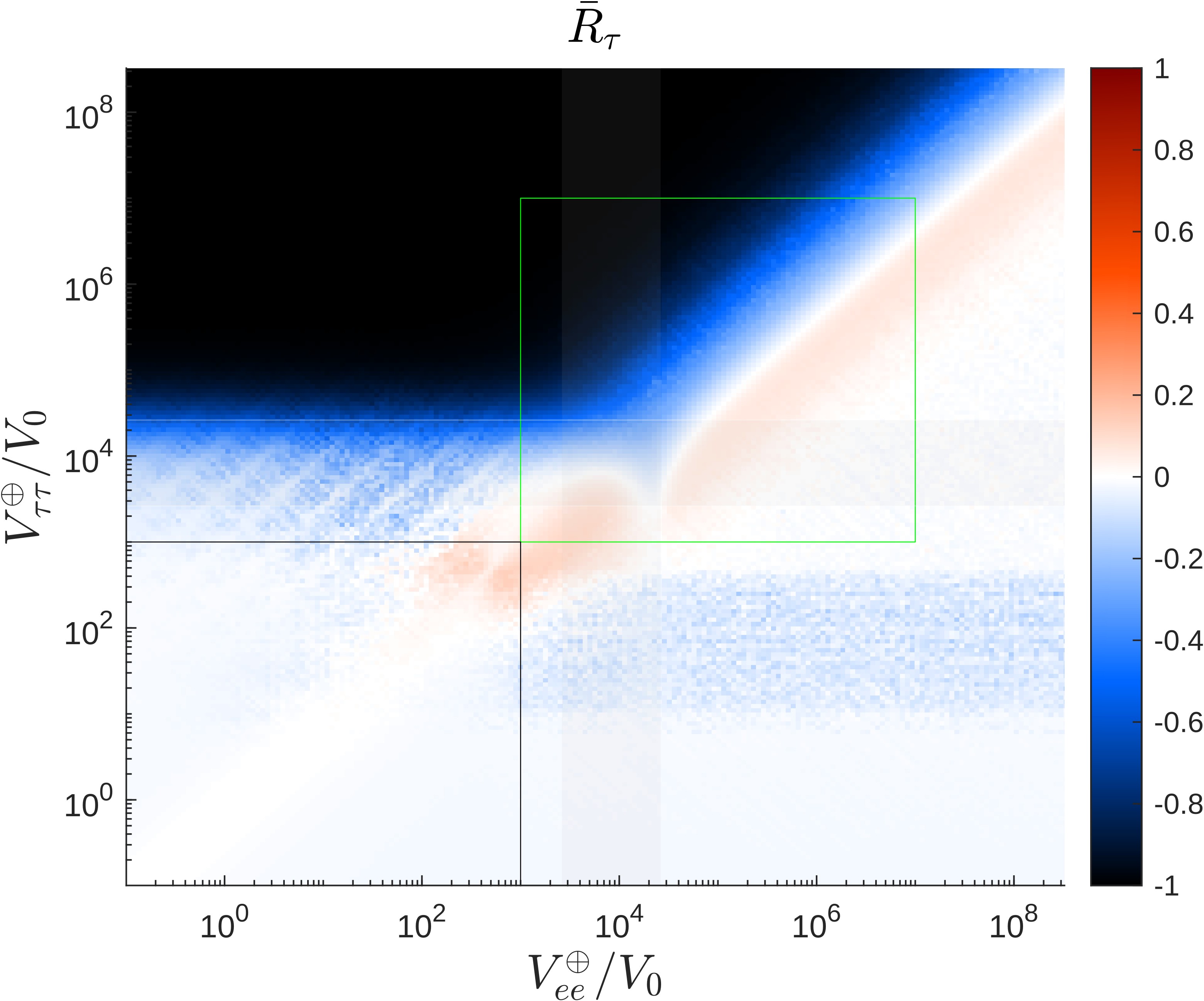}}
    \caption{The distributions of event-averaged deviation $\bar{R}_{e,\mu,\tau}$ of Fig.~\ref{fig1} for Case-(a,b,c).   
     The green square in the middle of each panel is the range of $V^\oplus$  for $m_\phi=1\,{\rm feV}$ ($10^3\leq V^\oplus/V_0\leq 10^7$), and the black square at the bottom-left corner is for $m_\phi=10\,{\rm peV}$ ($0.1\leq V^\oplus/V_0\leq 10^3$), where $V_0=3.1\times10^{-20}\,{\rm eV}$.}
    \label{fig1_2}
\end{figure}

\bigskip

\section{Result and Analysis}\label{analy}
\subsection{AGN Neutrino Flavor Ratio in IceCube}
The neutrinos from AGN are mainly produced from the charged pion decay, which yields the initial flavor ratio $f_0=(1,2,0)/3$. Due to the adiabatic condition, we only need to consider the average flavor ratio at the production location and the Earth. Fig.~\ref{fig1} shows ternary plots of the average flavor ratio emitted from NGC 1068 and detected on Earth. 
Following Ref.~\cite{deSalas:2016svi}, we obtain the average flavor ratio 
by
    \begin{equation}\label{barR}
    \bar{R}_\beta(y_e,y_\mu,y_\tau)=\frac{1}{N}\sum_N\frac{f^\oplus_\beta-f^\oplus_{0\beta}}{f^\oplus_{0\beta}}, \quad\bar{f}^\oplus_\beta=\frac{1}{N}\sum_N f^\oplus_\beta
    \end{equation}
where $f^\oplus_0=(0.34,0.31,0.35)$ is the flavor ratio detected at Earth when neutrinos propagate in vacuum without DM interactions, referring to the red point in each ternary plot.
$N$ is the total numbers of event. 
For each event, the neutrinos are assumed to be produced within the range $[10,10^2]R_s$ randomly in logarithmic scale.
The coupling constants are chosen below 10 by the perturbative condition, and the lower limit of $y_{\alpha}$ is fixed at $0.1$, which is not too low for DMs to affect the neutrino flavor.
Throughout this work, we examine four classified coupling combinations:
\begin{itemize}
\item Case-(a): $(y_e,y_\mu,0)$\,,
\item Case-(b): $(0,y_\mu,y_\tau)$\,,
\item Case-(c): $(y_e,0 ,y_\tau)$\,,
\item Case-(d): $(y_e,y_\mu,y_\tau)$\,,
\end{itemize}
as shown in each figure with corresponding labels
(a), (b), (c), and (d).


The MSW resonance effective potential is given by
    \begin{equation}
    V^{r}_{ij}=\frac{\Delta m^2_{ij}\cos(2\theta_{ij})}{2E},
    \end{equation}
Although this resonance potential is obtained by considering only $V_{ee}$, we included it in our plots as a reference and see whether the effective potential within these reference values deviates the flavor ratio. The ranges of resonance effective potential for NGC 1068 and TXS 0506+056 are given by
    \begin{equation}
    \label{eq_Vr}
    \begin{aligned}
    V^{r}_{21}&\in [9.65\times10^{-19},9.65\times10^{-18}]\,{\rm eV},\quad 
    V^{r}_{31}\in [8.06\times10^{-17},8.06\times10^{-16}]\,{\rm eV},\quad\text{(NGC)}\\
    V^{r}_{21}&\in [3.62\times10^{-21},3.62\times10^{-19}]\,{\rm eV},\quad 
    V^{r}_{31}\in [3.02\times10^{-19},3.02\times10^{-17}]\,{\rm eV}.\quad\text{(TXS)}
    \end{aligned}
    \end{equation}

In the first row of Fig.~\ref{fig1}, i.e. panels (a1, b1, c1, d1), the $m_\phi$ is chosen to be 0.1 peV (cyan), 1 peV (blue), and 10 peV (black). For $m_\phi=10$ peV, the effective potentials at Earth lie within the range $[3.1\times10^{-21}\,, 3.1\times10^{-17}]$ eV, which is less than $V_{31}^r$. Since $V^\oplus_{\alpha\beta}=G_{\alpha\beta}\rho^\oplus_{\rm DM}/m_\phi$  is proportional to $m^{-1}_\phi$, it gradually covers $V_{31}^r$ as $m_\phi$ decreases to $1$ and $0.1$ peV. The distribution of the flavor ratio spreads significantly when $V^\oplus$ (The matrix element of $V^\oplus$ is $V^\oplus_{\alpha\beta}$) covers the whole range of $V_{31}^r$. 
In the second row of Fig. \ref{fig1} we consider lighter DM masses, $m_\phi=$0.01 feV (cyan), 0.1 feV (blue), and 1 feV (black). In this case, only the effective potential of $m_\phi=1\,{\rm feV}$ case covers $V_{31}^r$, and most of the values of $V^\oplus$ for $m_\phi=0.1\,{\rm feV}$ and $0.01\,{\rm feV}$ are larger than $V_{31}^r$. Therefore, the Hamiltonian is dominated by $V^\oplus$ and the final flavor ratio depends on the relative strength between each $V^\oplus_{\alpha\beta}$. 
These distributions exhibit some common features: (i) for $m_\phi=10$ peV, the electron flavor ratio $\bar{f}^\oplus_e$ tends to increase rather than decrease relative to the red point, while $\bar{f}^\oplus_{\mu,\tau}$ tends to change in the opposite direction;
(ii) for $m_\phi \leq 0.1$ peV, the electron flavor ratio $\bar{f}^\oplus_e$ mildly increase, while $\bar{f}^\oplus_\tau$ ($\bar{f}^\oplus_\mu$) tends to substantially decrease (increase) relative to the red point $f^\oplus_0=(0.34,0.31,0.35)$.
We included the Single-Power Law (SPL) best fitting of IceCube MESE data with 68\% and 95\% confidence level \cite{IceCube:2025uyt} and the combined limit using 15 years of IceCube plus 10 years of IceCube-Gen2 and the future neutrino telescopes \cite{Agarwalla:2023sng} in our plots. Our simulations are partially overlap with the IceCube observations.

In Fig.~\ref{fig1_2}, we show $\bar{R}_{e,\mu,\tau}$ for the three coupling combinations, Case-(a,b,c), with the corresponding labels (a), (b), (c).
The color bar on the right of each panel indicates the deviation from $f^\oplus_0$. The darker red (darker blue) regions correspond to larger (smaller) values of $\bar{f}^\oplus_{e,\mu,\tau}$ compared with $f^\oplus_{0 e,0\mu,0\tau}$. In addition, the transparent vertical and horizontal bands in each panel indicate the resonance range of $V^{31}_r$ from Eq.(\ref{eq_Vr}), and the color gradient of $\bar{R}_{e,\mu,\tau}$ guides us to read the flavor ratio distributions. We illustrate this point by taking Fig.~\ref{fig1_2}(a) as an example. The bottom-left black and upper-right green squares in each panel of Fig.~\ref{fig1_2}(a1,a2,a3) are the range of $V^\oplus$ for $m_\phi=10\,{\rm peV}$ and $m_\phi=1\,{\rm feV}$ respectively, corresponding to the black points in Fig. \ref{fig1}(a1) and (a2).
The $m_\phi=10\,{\rm peV}$ black squares in $\bar{R}_{e,\mu,\tau}$ containing both light red and light blue portions are consistent with the distribution of the black points that aggregate around the red point in Fig. \ref{fig1}(a1), where their values are close to $f^\oplus_0$.
In particular, the bottom-right corner in this square of $\bar{R}_e$ is covered by light red indicating the mild enhancement of $\bar{f}^\oplus_e$ from $\bar{f}^\oplus_{0e}$, meanwhile the corresponding squares for $\bar{R}_{\mu,\tau}$ show light blue in the bottom-right corner implying the decrements of $\bar{f}^\oplus_{\mu,\tau}$. Therefore, the black points in Fig.~\ref{fig1}(a1) form a strip extending 
to lower-right direction.
Similarly, from the  $m_\phi=1\,{\rm feV}$ green squares in $\bar{R}_{e,\mu,\tau}$ of Fig.~\ref{fig1_2}(a1,a2,a3), we can infer that the corresponding black points in Fig.~\ref{fig1}(a2) form a strip oriented along 
the upper-right direction.
Particularly, the black points in Fig.~\ref{fig1} (a2), the value of $\bar{f}^\oplus_{\mu}$ ($\bar{f}^\oplus_{\tau}$) is greater (smaller) than that of $\bar{f}^\oplus_{0\mu}$ ($\bar{f}^\oplus_{0\tau}$), which is consistent with darker red (darker blue) region in the green square of Fig.~\ref{fig1_2} (a2) (Fig.~\ref{fig1_2} (a3)). Following the same logic, the distributions for Case-(b) and Case-(c) in Fig.~\ref{fig1} can be inferred from Fig.~\ref{fig1_2}.

For TXS 0506+056 with a higher neutrino energy, we repeated the same analysis. Since the location of neutrino production is
under investigation.
we assume two different produced ranges. 
The first range is $[10^2,10^5]R_s$, where $10^5R_s$ is the influence radius of TXS 0506+056 \cite{Cline:2022qld}.
The second range is chosen to be the same as NGC 1068, i.e. $[10,10^2]R_s$. The results of second production range are shown in Appendix \ref{appendixA}. Fig \ref{fig3} shows the correspoding flavor ratio distribution for the production range $[10^2,10^5]R_s$.
The event-averaged deviation for Fig.~\ref{fig3} is shown in Fig.~\ref{fig3_2}. The current IceCube sensitivity is able to distinguish the flavor ratio distributions from NGC 1068 and TXS 0506+056 according to their ``shapes.'' Taking Case-(d) as an example, the $m_\phi=10\,{\rm peV}$ distribution (black points) of Fig.~\ref{fig1}(d1) is a strip extended in enlarging $\bar{f}_e^\oplus$ and reducing $\bar{f}_{\mu,\tau}^\oplus$ direction, while the $m_\phi=10\,{\rm peV}$ distribution of Fig.~\ref{fig3}(d1) is like a crab claw increasing $\bar{f}_\mu^\oplus$ and decreasing $\bar{f}_\tau^\oplus$.

\begin{figure}
    \renewcommand{\thefigure}{3-1}
    \captionsetup[subfloat]{labelformat=empty}
    \centering
    \subfloat[\hspace{0.75cm}(a1)\hspace{3.1cm} (b1)\hspace{3.3cm}(c1)\hspace{3.1cm}(d1)] {\includegraphics[height=0.26\linewidth]{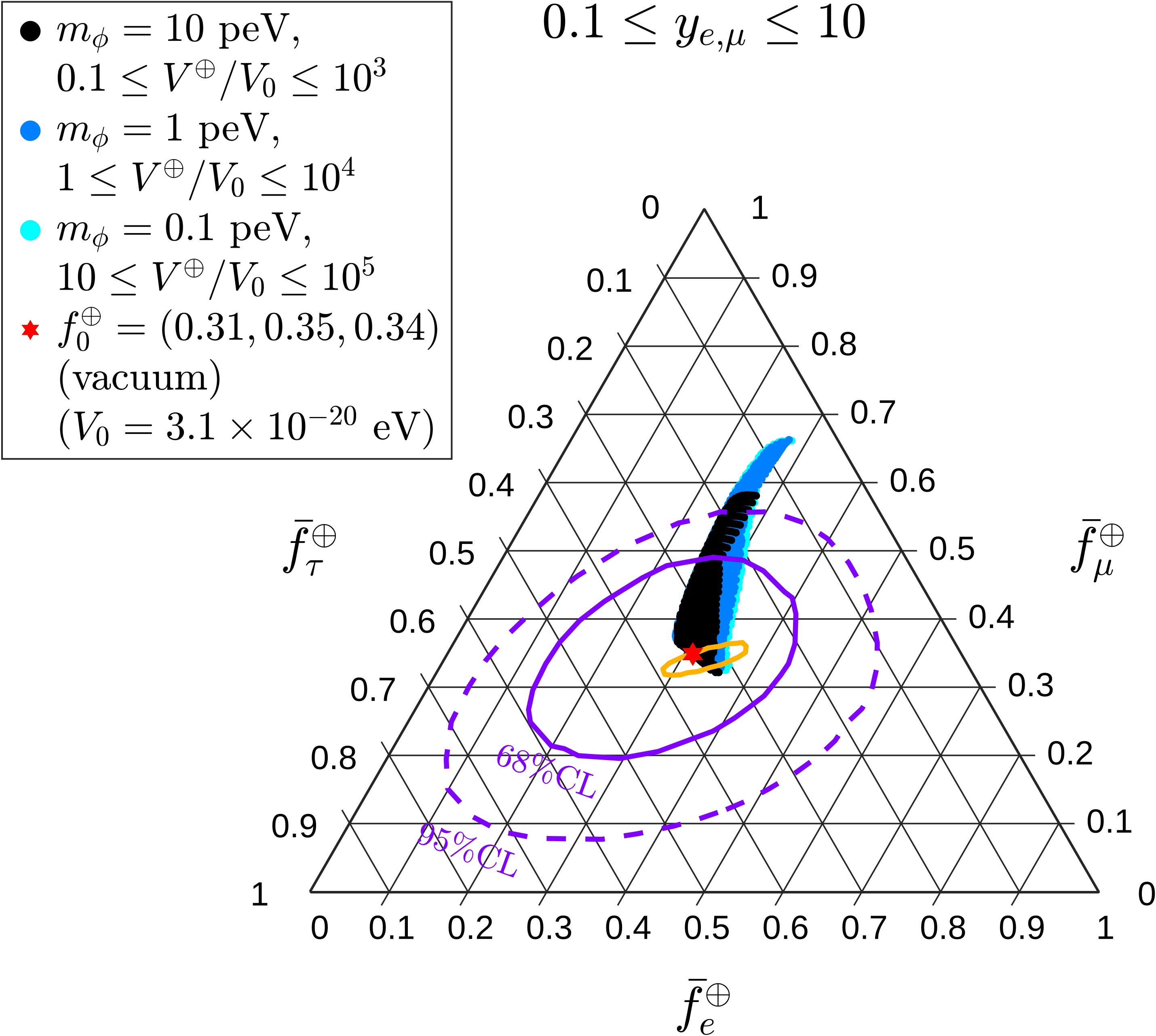}\quad
    \includegraphics[height=0.26\linewidth]{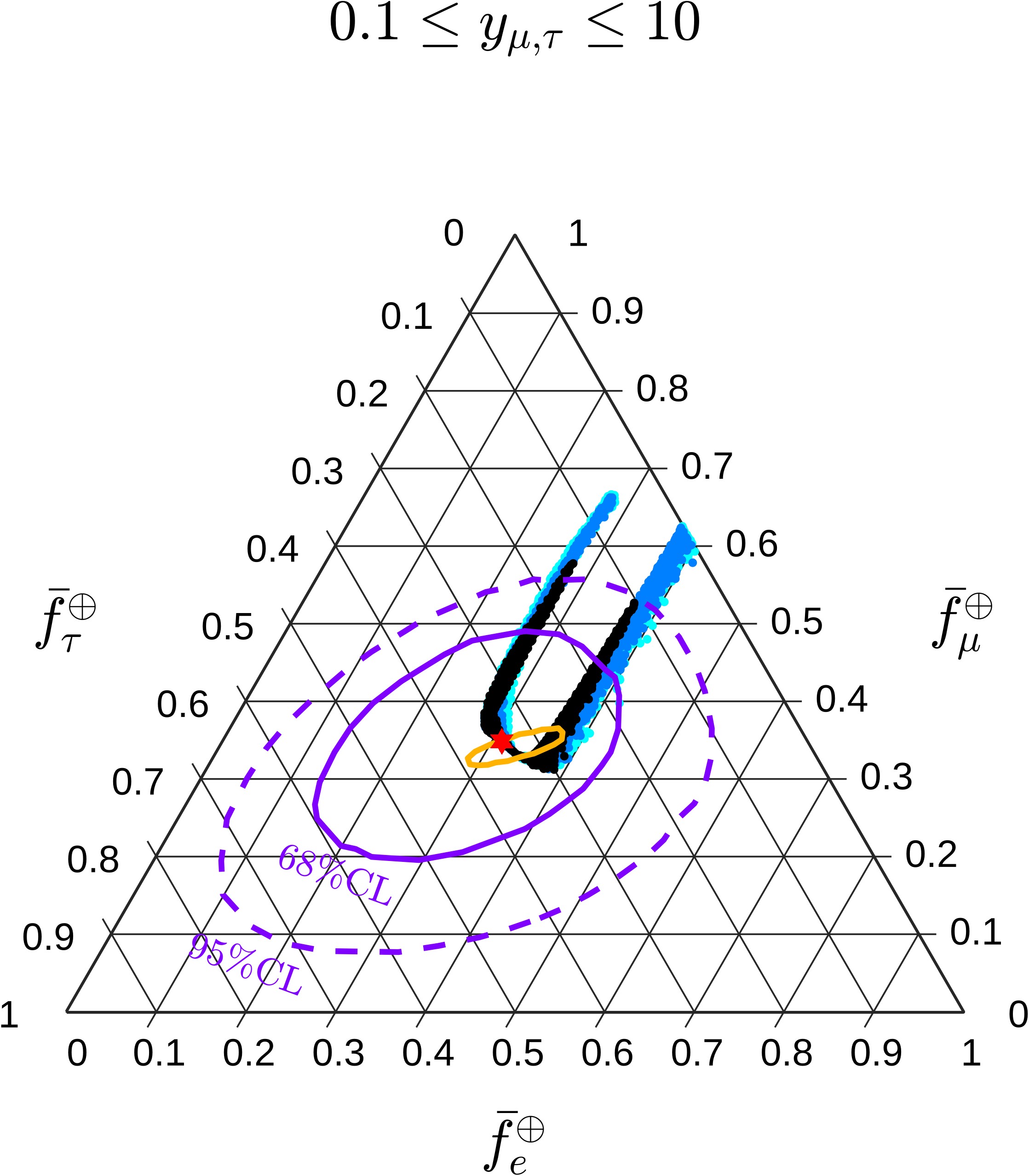}\quad
    \includegraphics[height=0.26\linewidth]{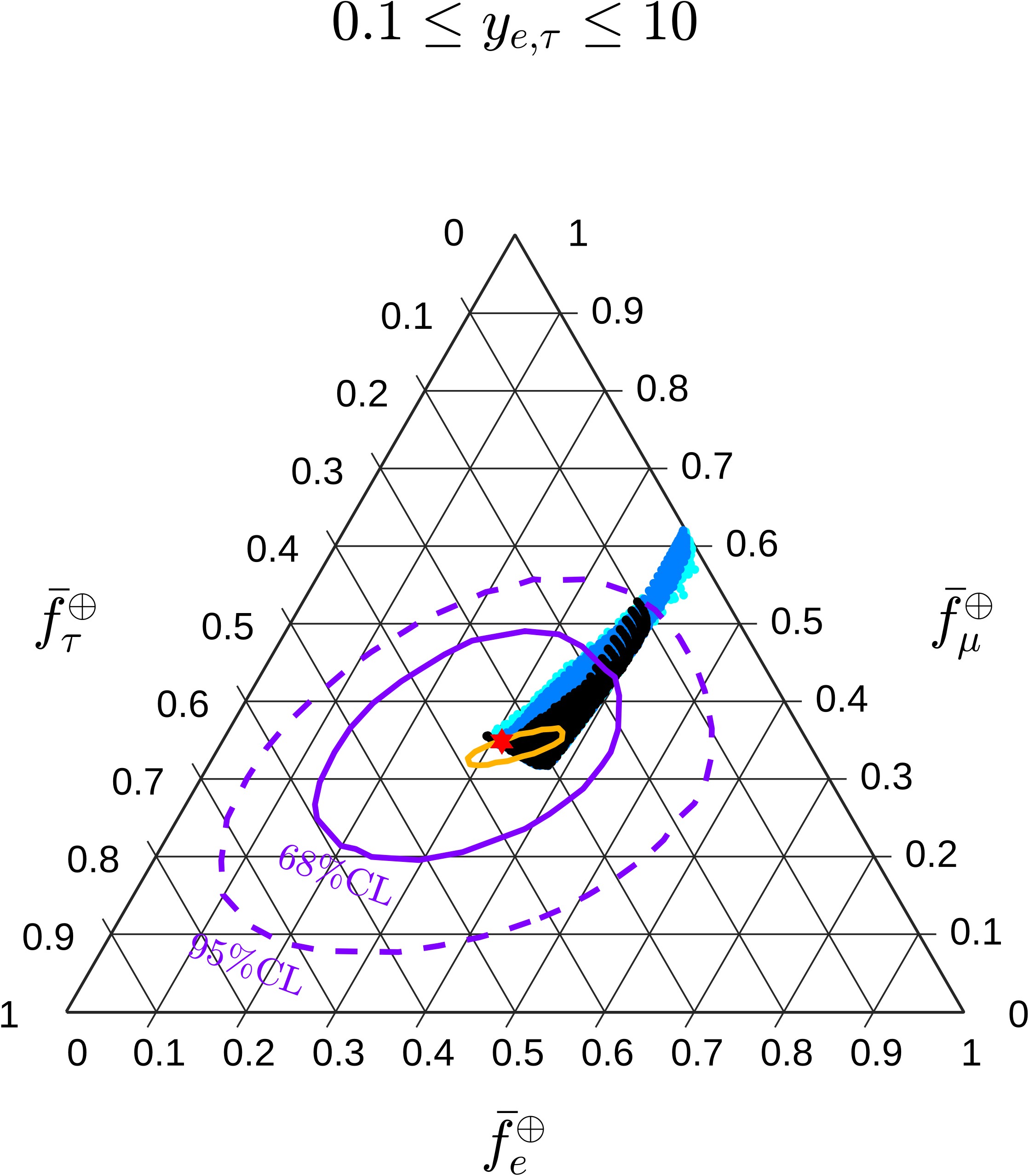}\quad
    \includegraphics[height=0.26\linewidth]{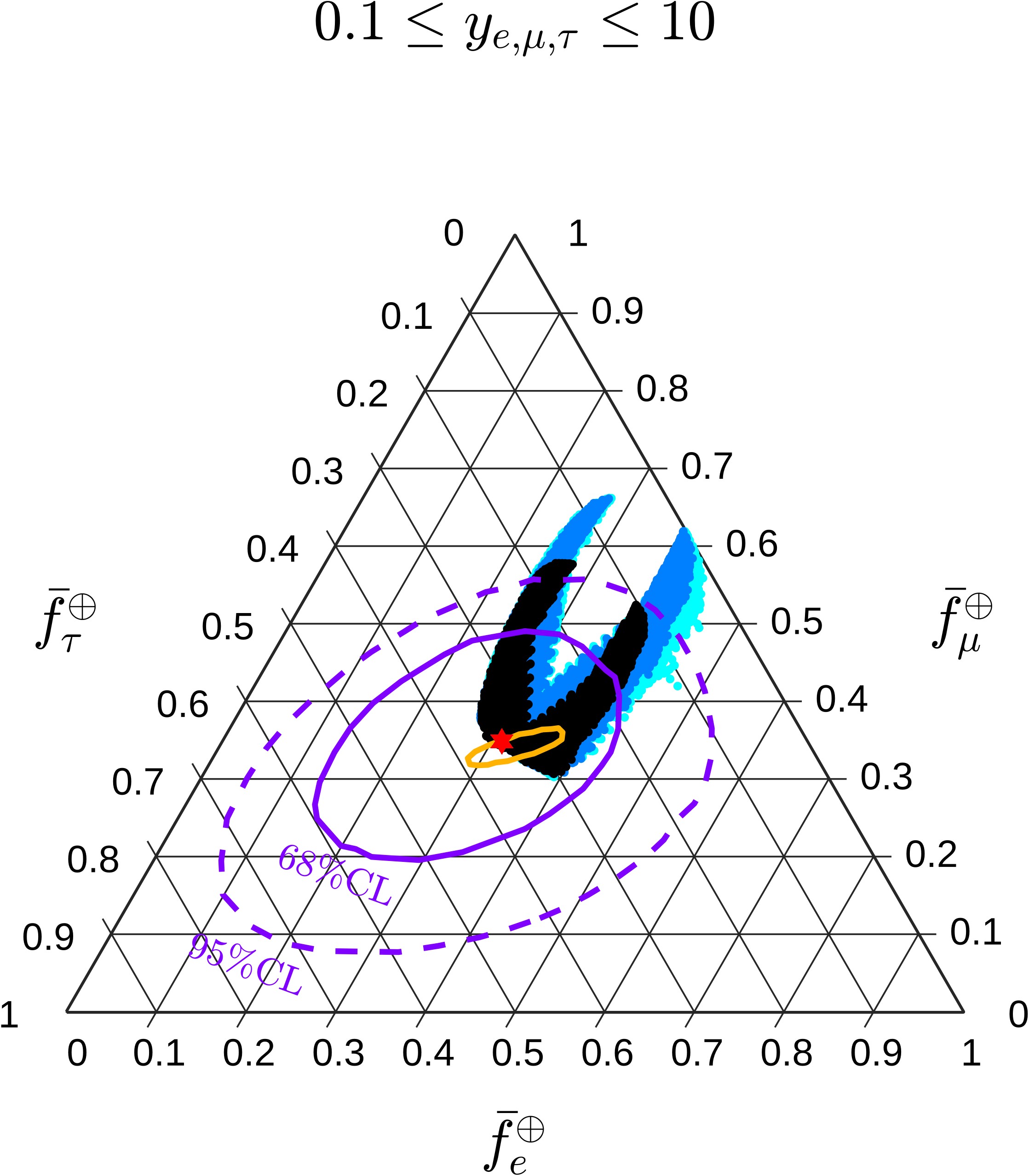}}
    
    \subfloat[\hspace{0.75cm}(a2)\hspace{3.1cm} (b2)\hspace{3.3cm}(c2)\hspace{3.1cm}(d2)]{\includegraphics[height=0.26\linewidth]{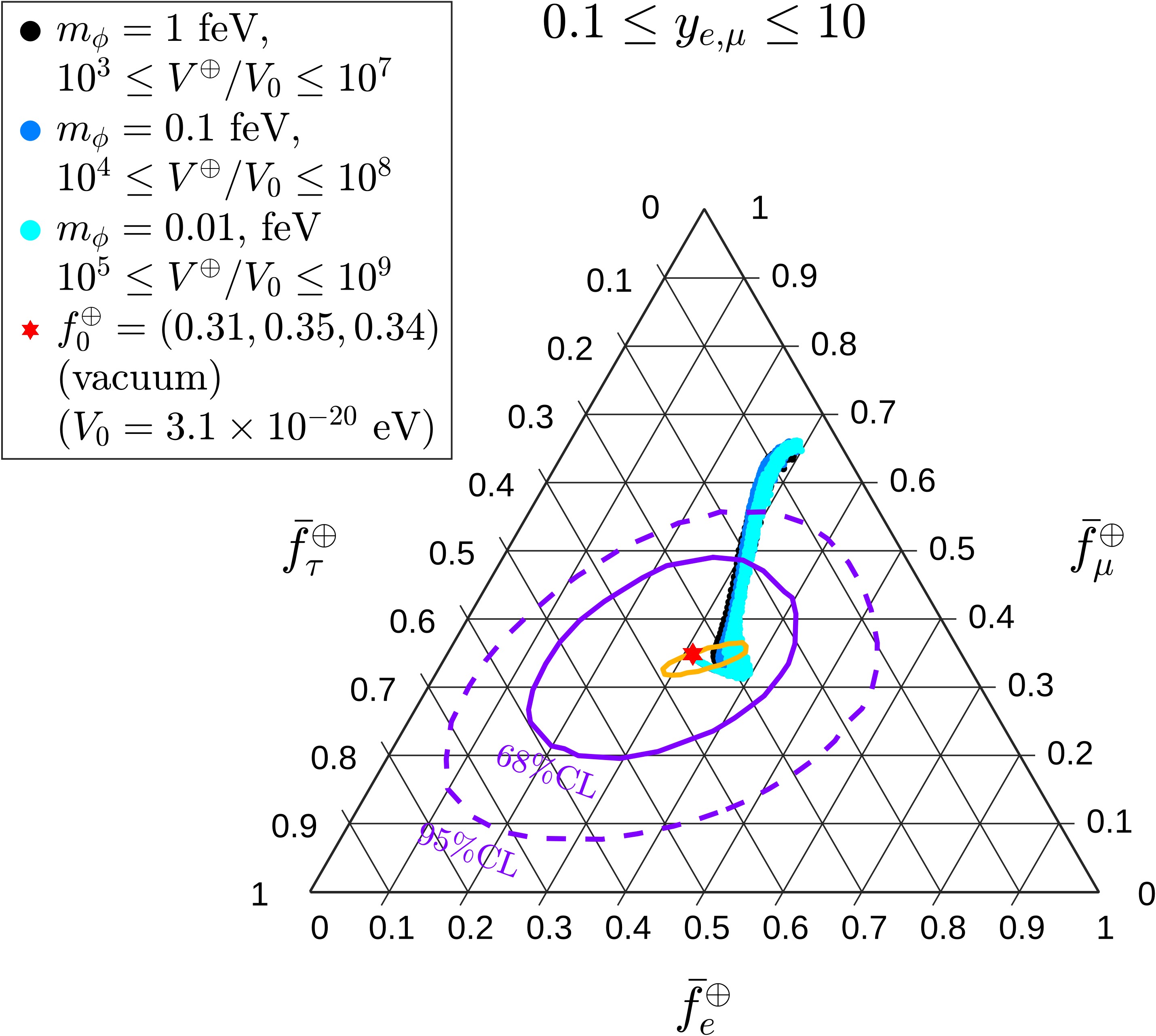}\quad
    \includegraphics[height=0.26\linewidth]{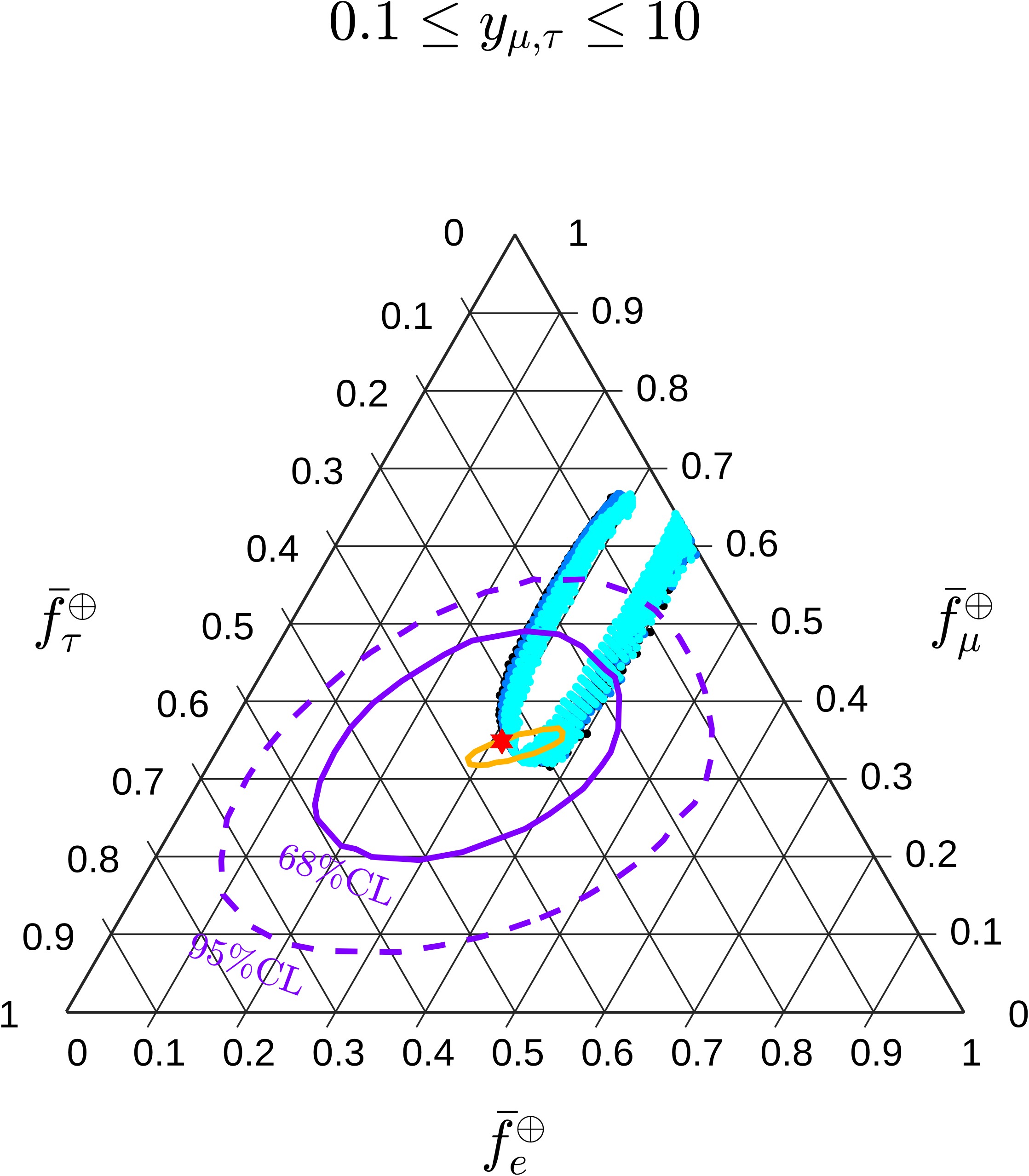}\quad
    \includegraphics[height=0.26\linewidth]{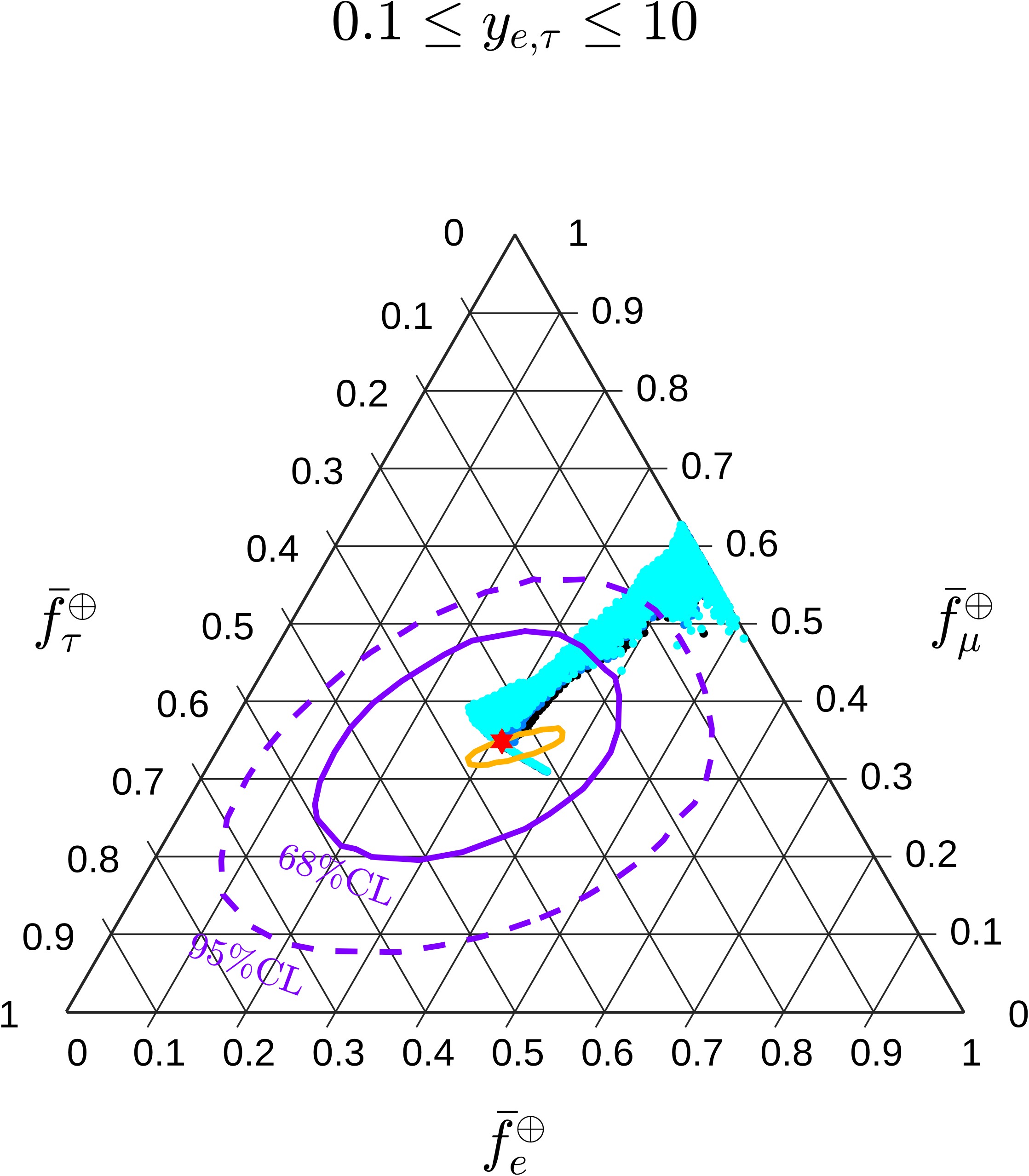}\quad
    \includegraphics[height=0.26\linewidth]{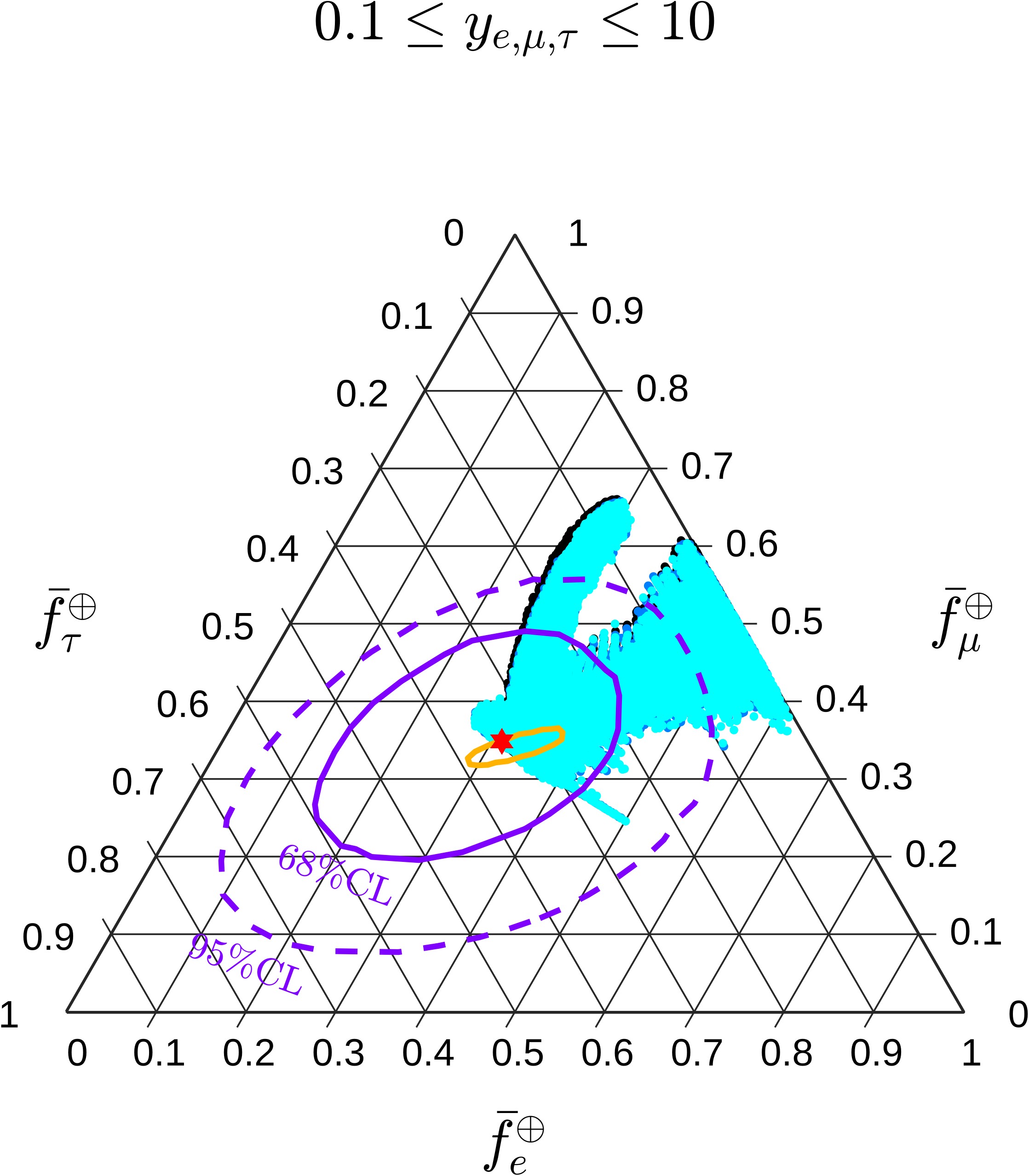}}

    \caption{The ternary plots of flavor ratio at Earth for Case-(a,b,c,d), where the neutrinos are produced from TXS 0506+056. In each plot the effective potential around TXS 0506+056 is far above $V^{31}_r$. 
    The neutrinos from TXS 0506+056 are assumed to be produced within $[10^2,10^5]R_s$.
    }
    \label{fig3}
\end{figure}


\begin{figure}
    \renewcommand{\thefigure}{3-2}
    \captionsetup[subfloat]{labelformat=empty}
    \centering
    \subfloat[(a1)\hspace{4.8cm} (a2)\hspace{4.8cm}(a3)]{\includegraphics[width=0.33\linewidth]{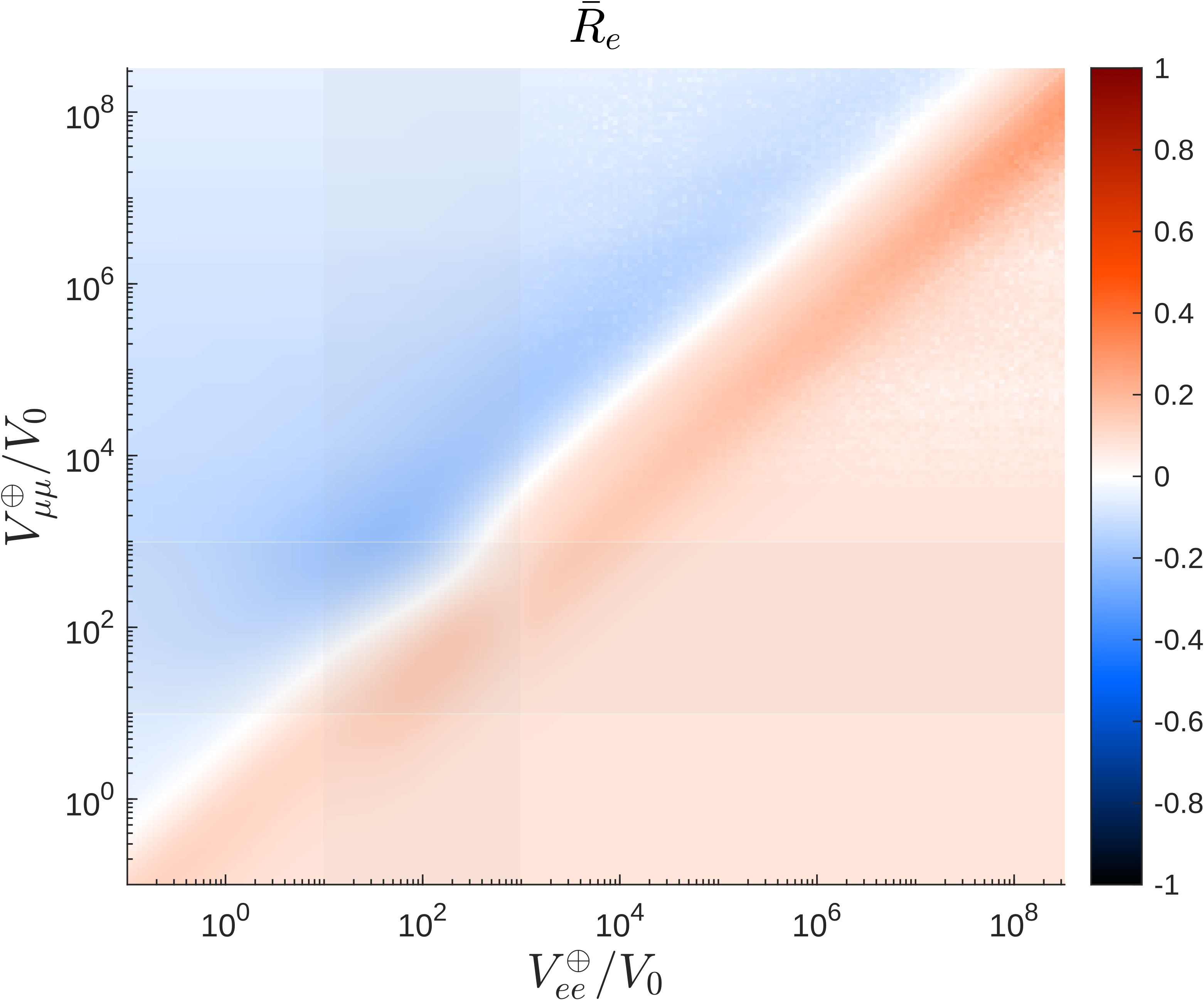}\quad
    \includegraphics[width=0.33\linewidth]{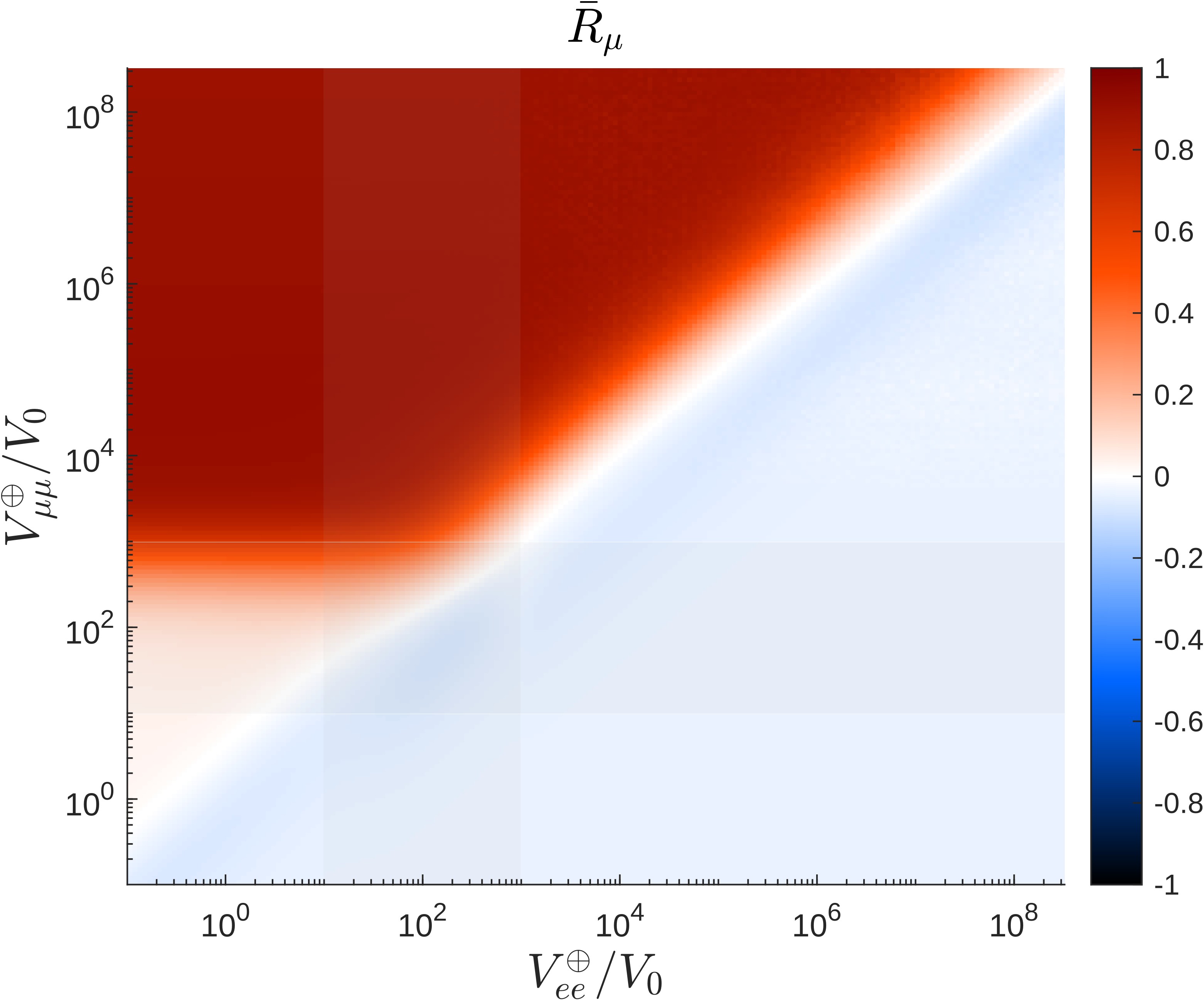}\quad
    \includegraphics[width=0.33\linewidth]{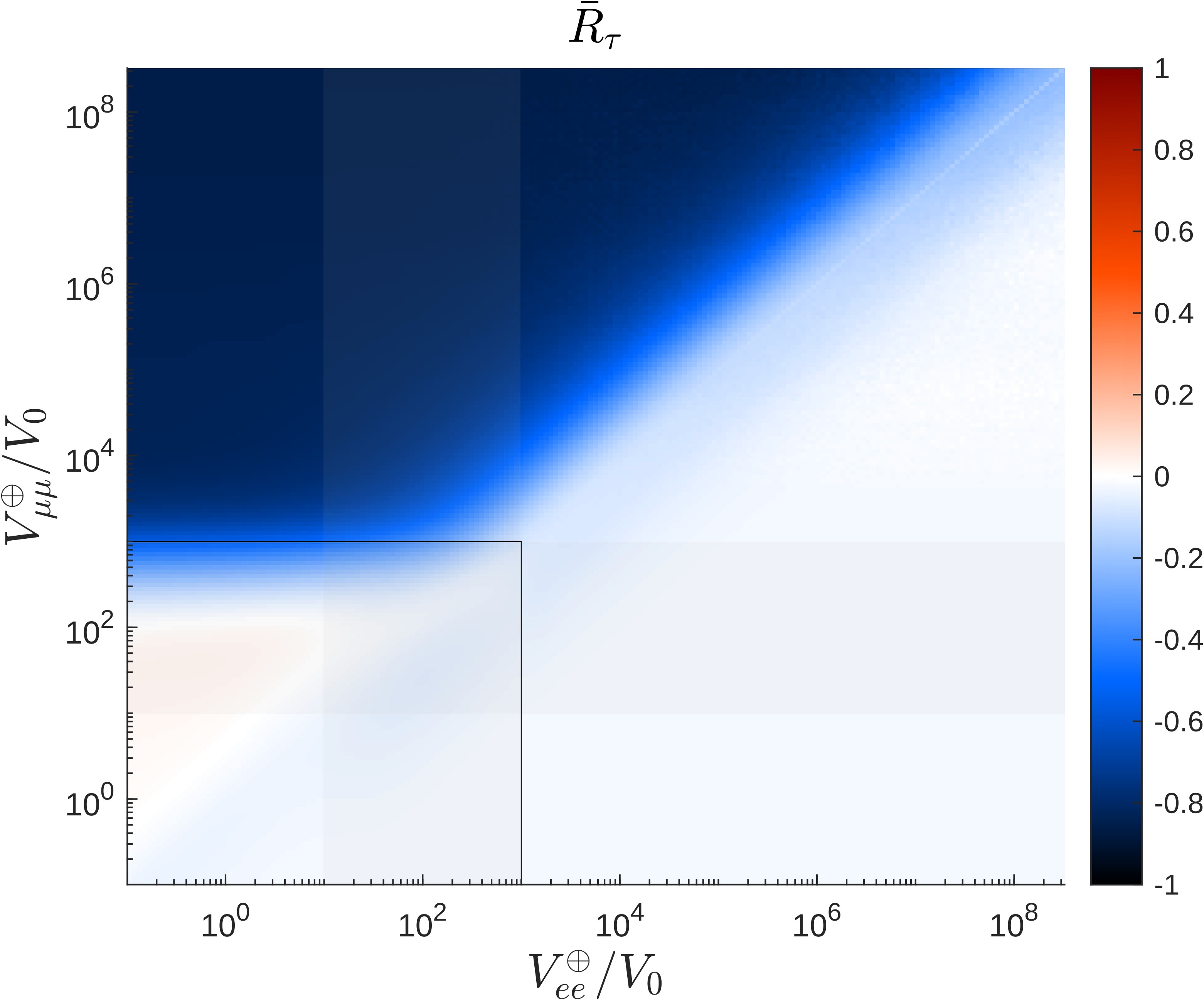}}
 
    \subfloat[(b1)\hspace{4.8cm} (b2)\hspace{4.8cm}(b3)]{\includegraphics[width=0.33\linewidth]{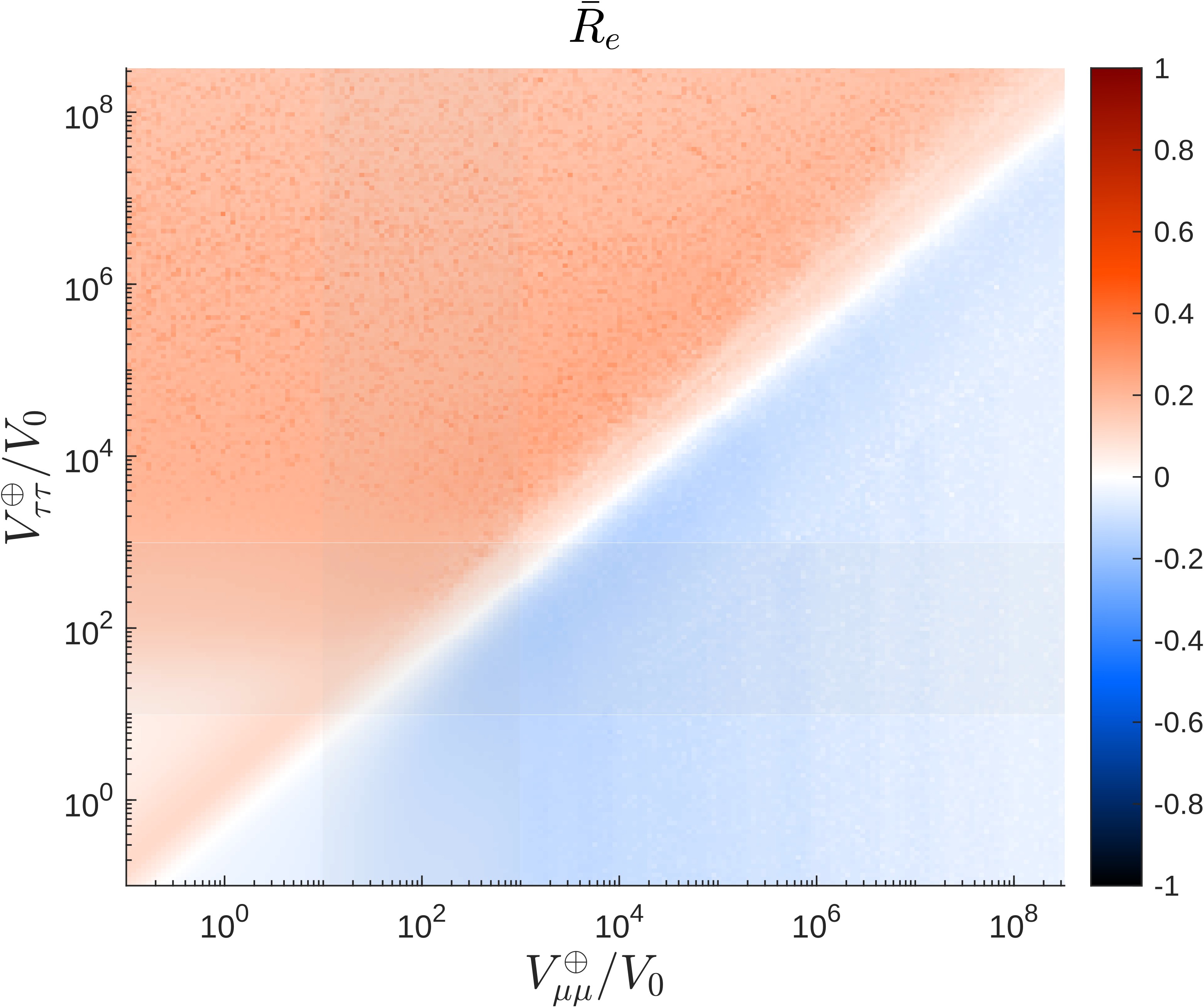}\quad
    \includegraphics[width=0.33\linewidth]{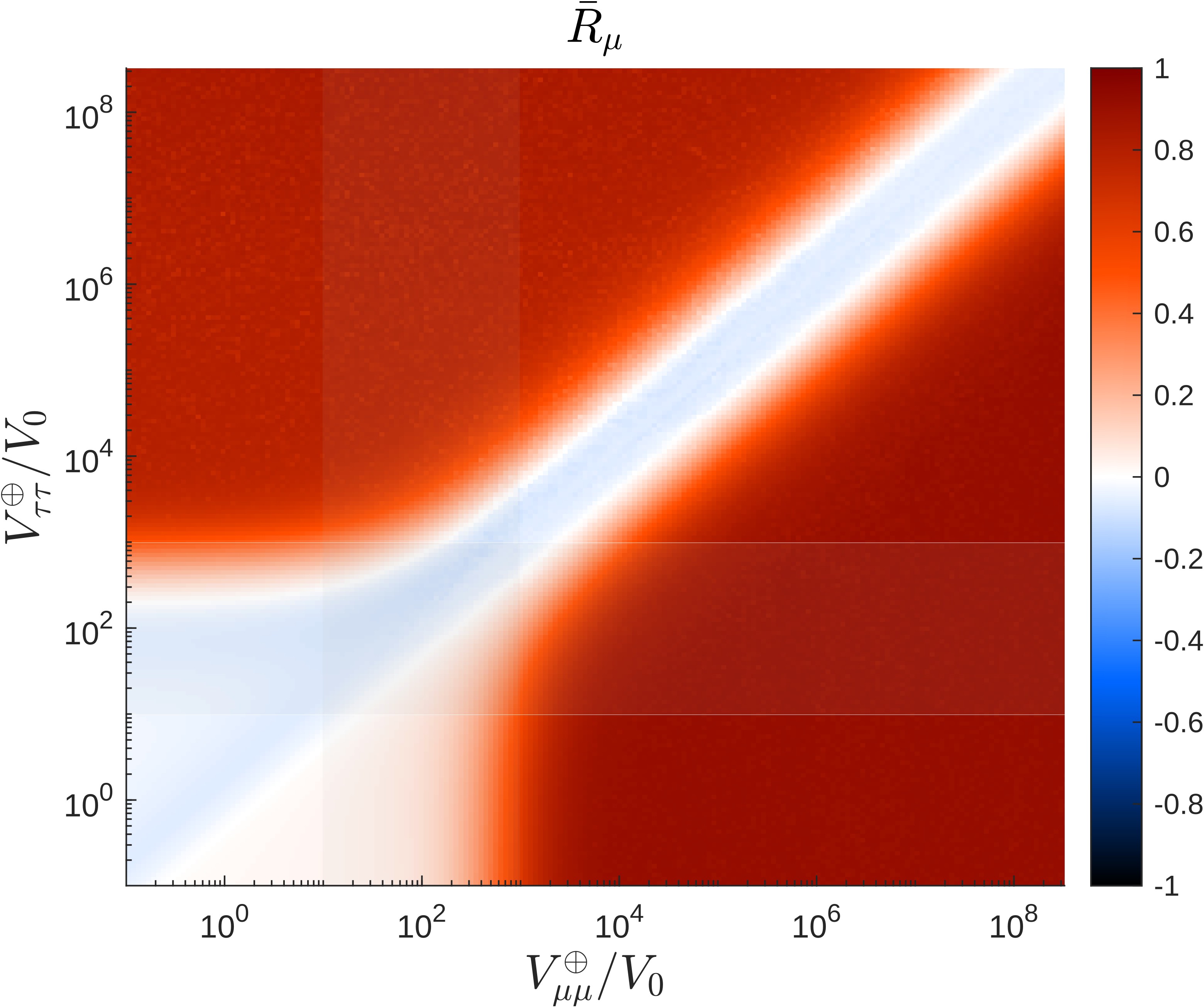}\quad
    \includegraphics[width=0.33\linewidth]{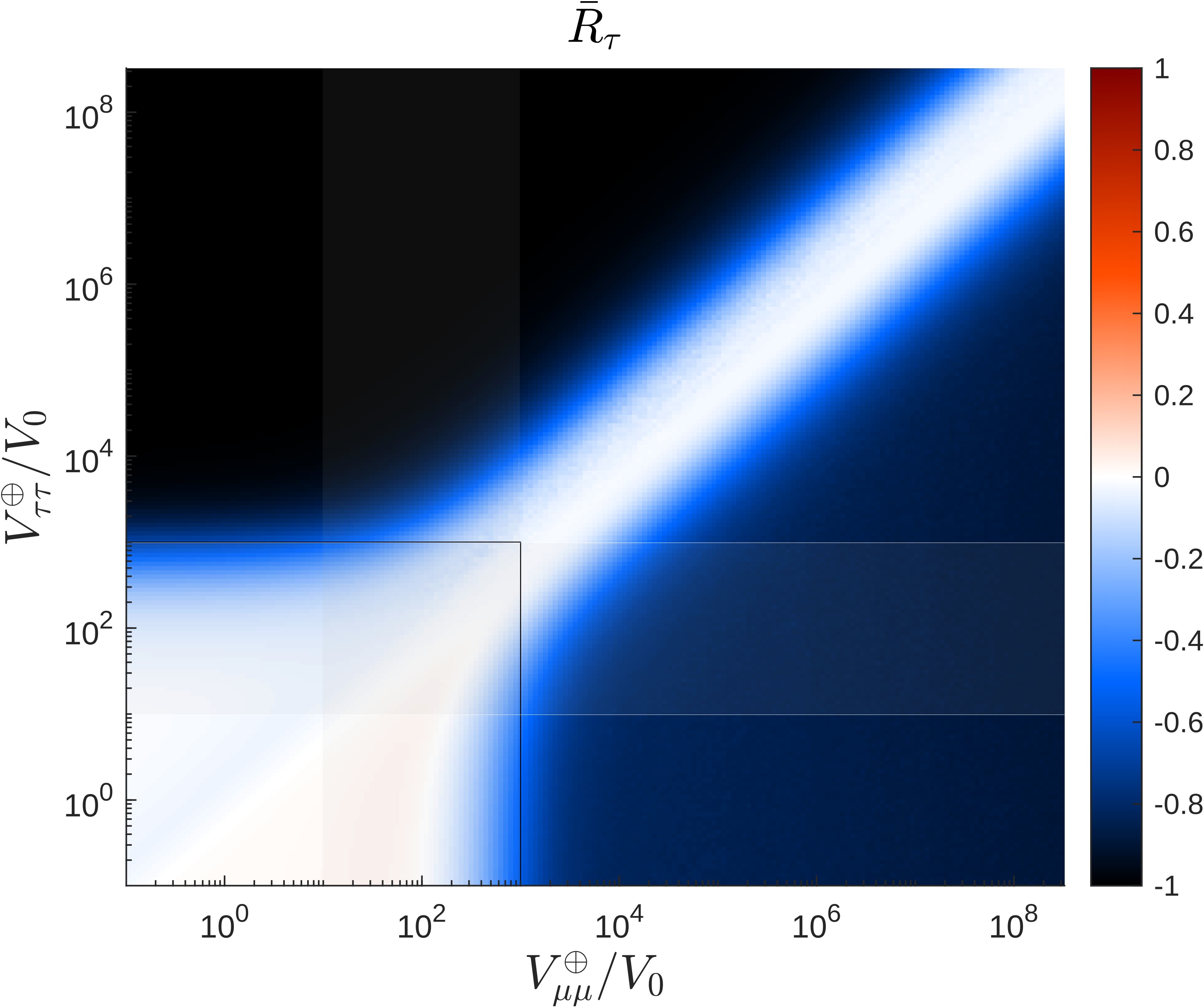}}

    \subfloat[(c1)\hspace{4.8cm} (c2)\hspace{4.8cm}(c3)]{\includegraphics[width=0.33\linewidth]{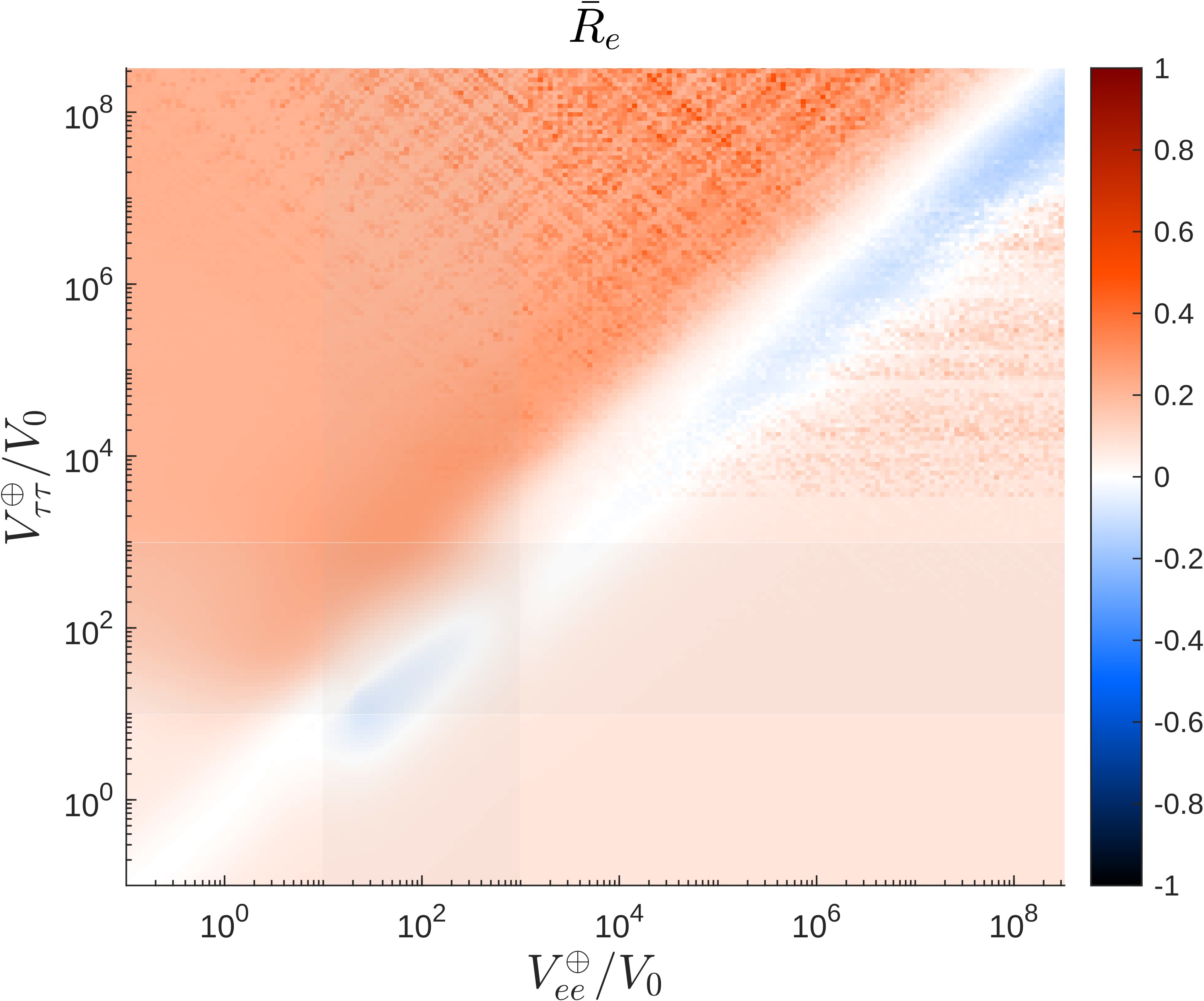}\quad
    \includegraphics[width=0.33\linewidth]{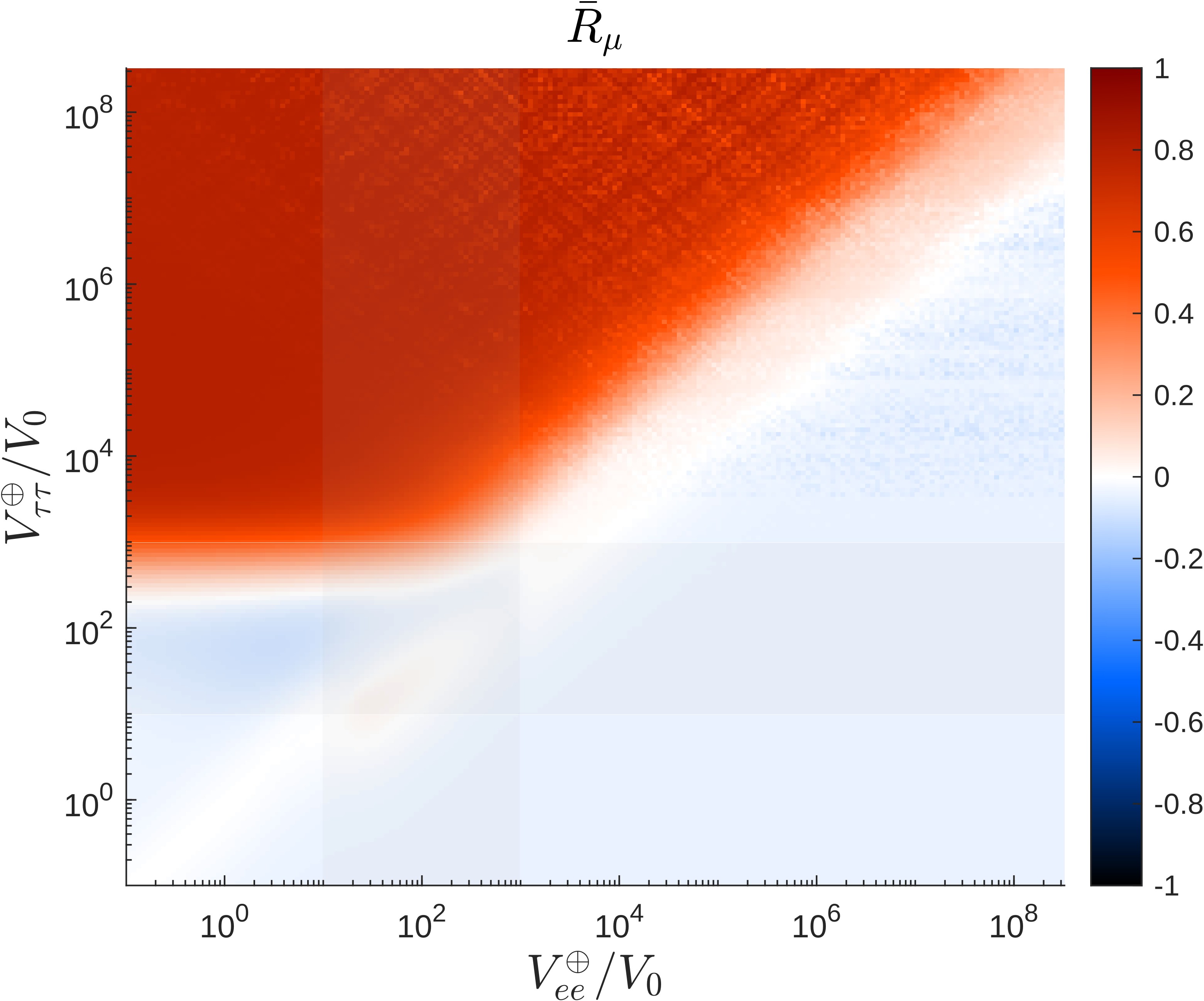}\quad
    \includegraphics[width=0.33\linewidth]{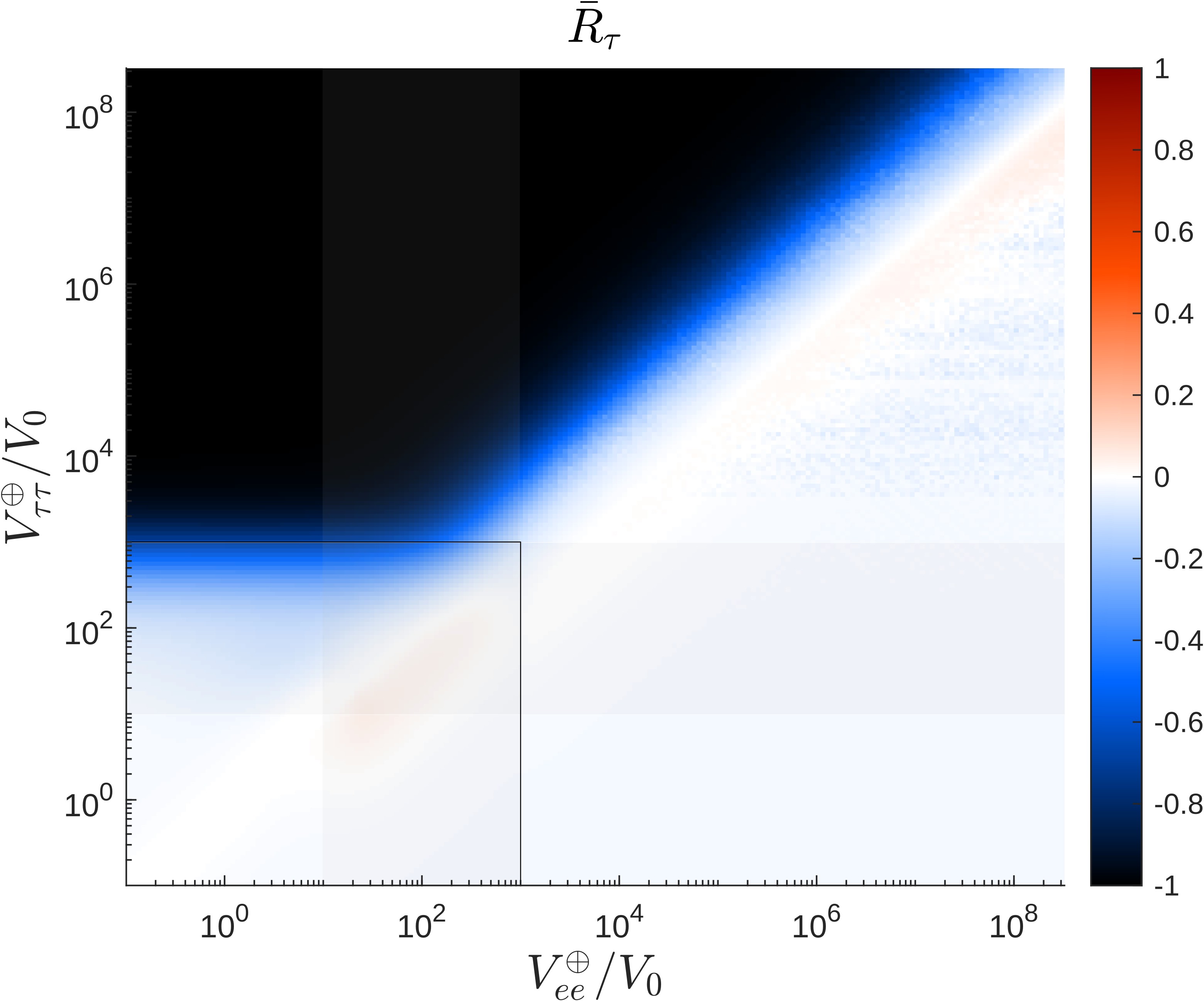}}
    \caption{The distributions of event-averaged deviation $\bar{R}_\beta$ of Fig.~\ref{fig3} for Case-(a,b,c). The square boxes in panel (a3,b3,c3) represents the potential range of $m_\phi=10$ peV. }
    \label{fig3_2}
\end{figure}

\begin{figure}
    \renewcommand{\thefigure}{4}
    \captionsetup[subfloat]{labelformat=empty}
    \centering
    \subfloat[\hspace{0.4cm}(a-NO)\hspace{4.5cm} (b-NO)\hspace{4.5cm}(c-NO)]{\includegraphics[width=0.33\linewidth]{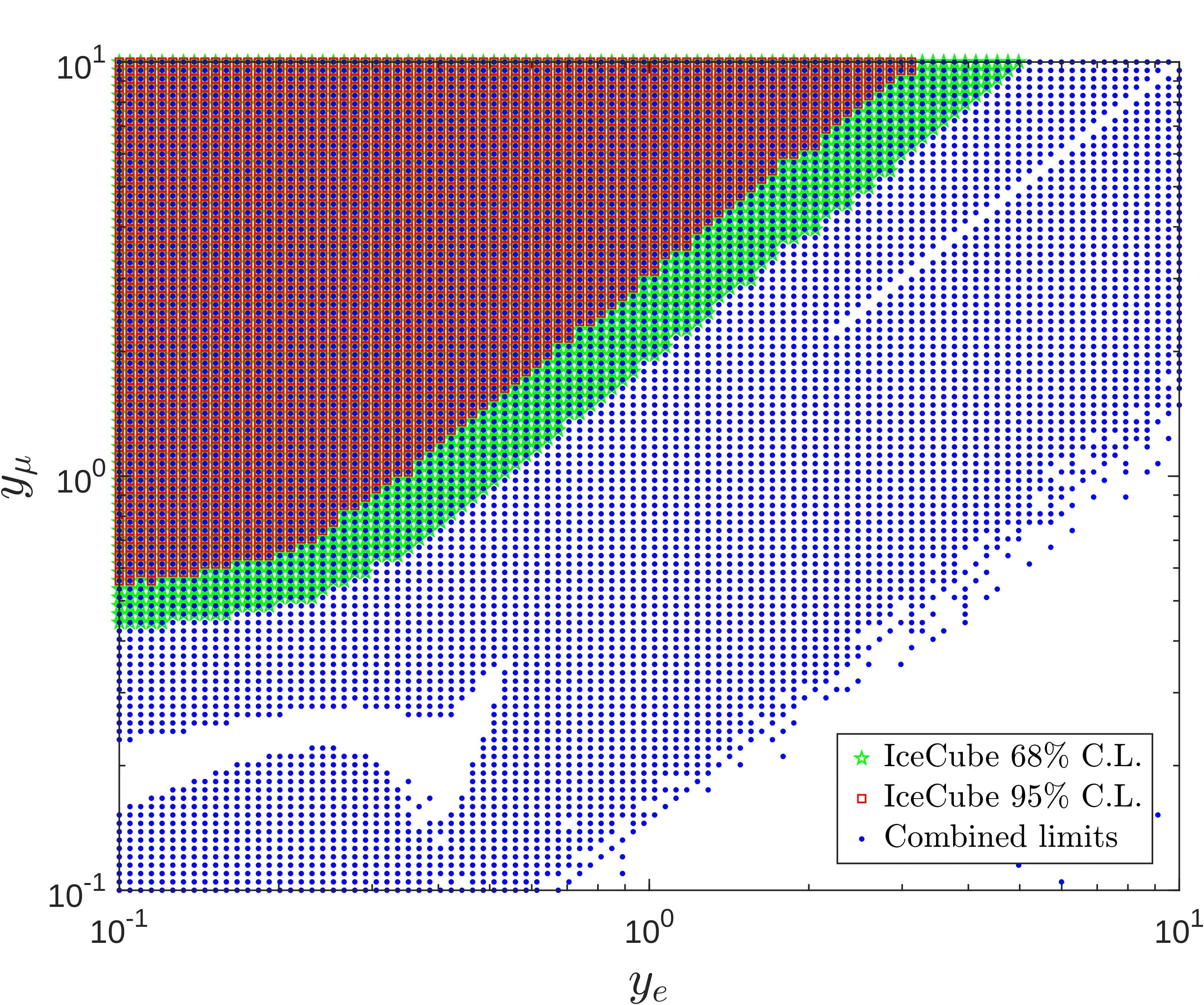}\quad
    \includegraphics[width=0.33\linewidth]{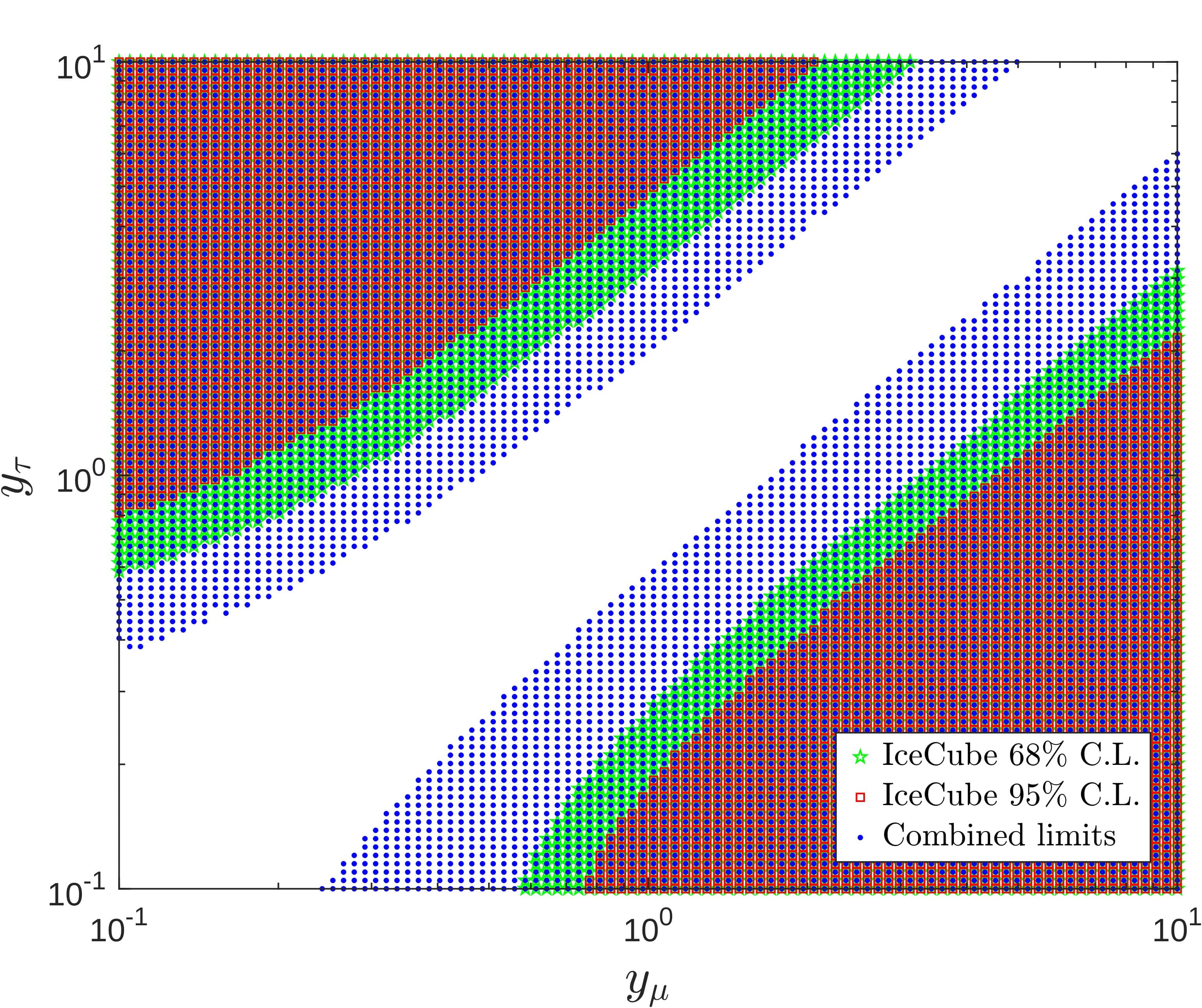}\quad
    \includegraphics[width=0.33\linewidth]{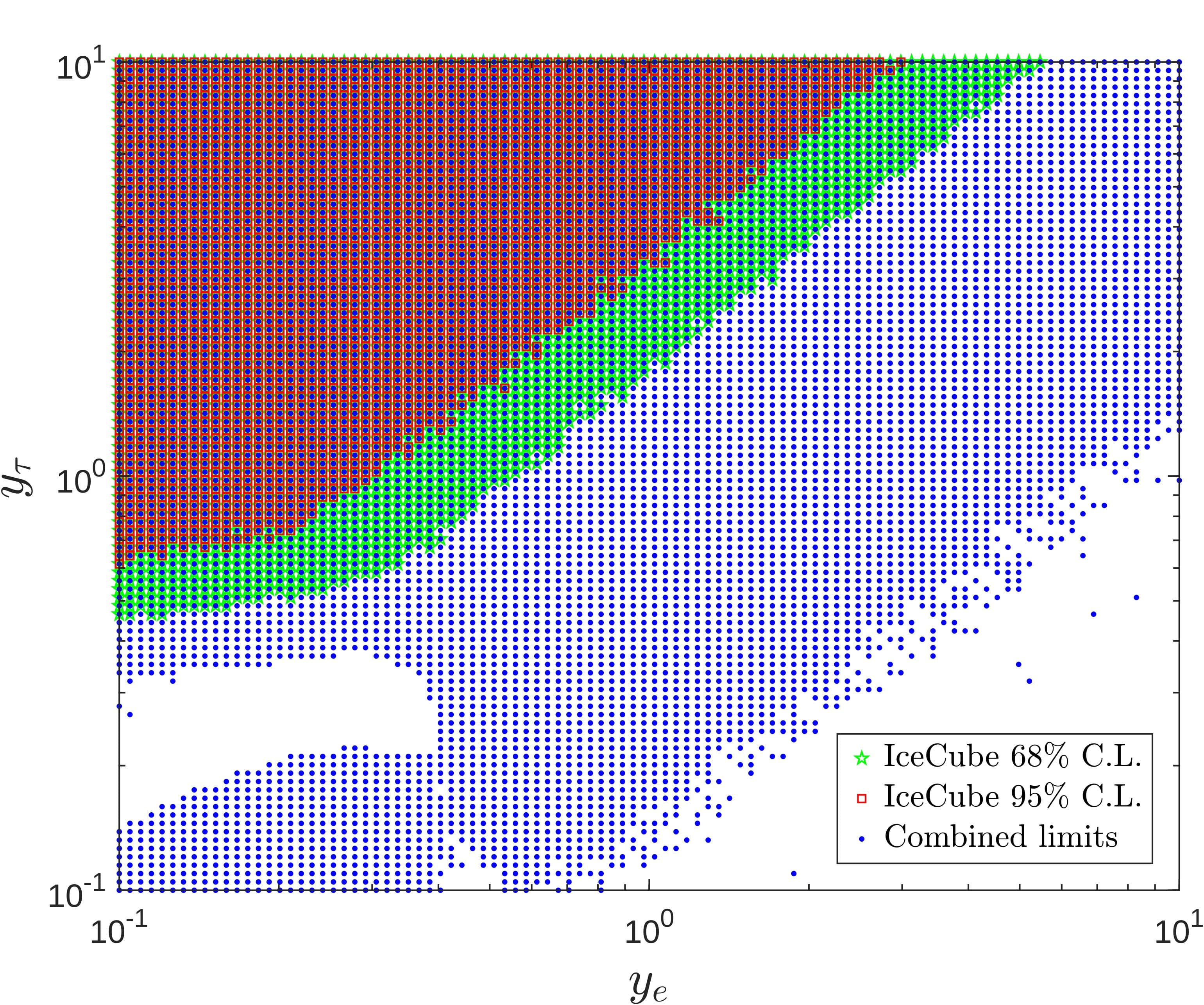}}

    \subfloat[\hspace{0.4cm}(a-IO)\hspace{4.5cm} (b-IO)\hspace{4.5cm}(c-IO)]{\includegraphics[width=0.33\linewidth]{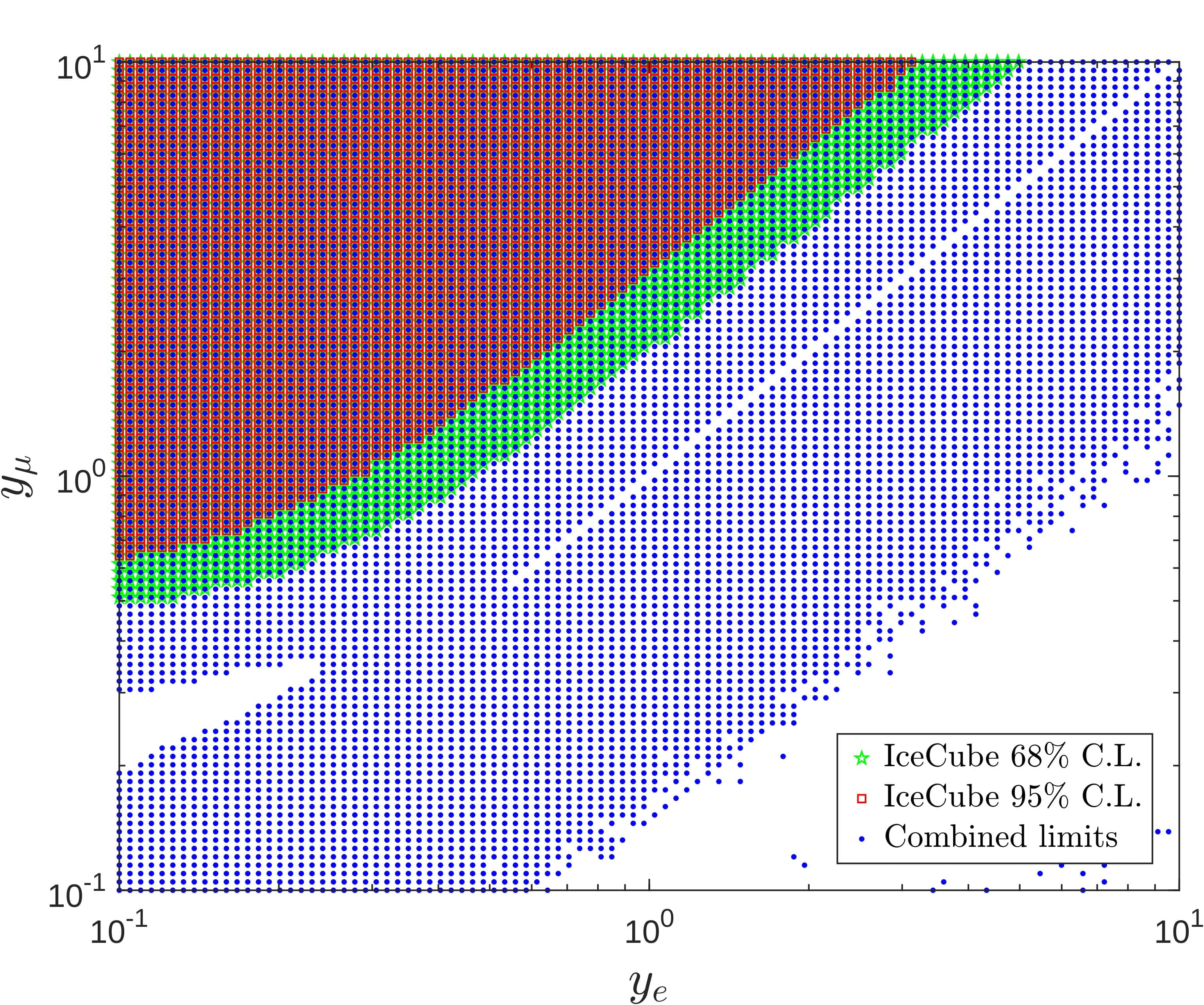}\quad
    \includegraphics[width=0.33\linewidth]{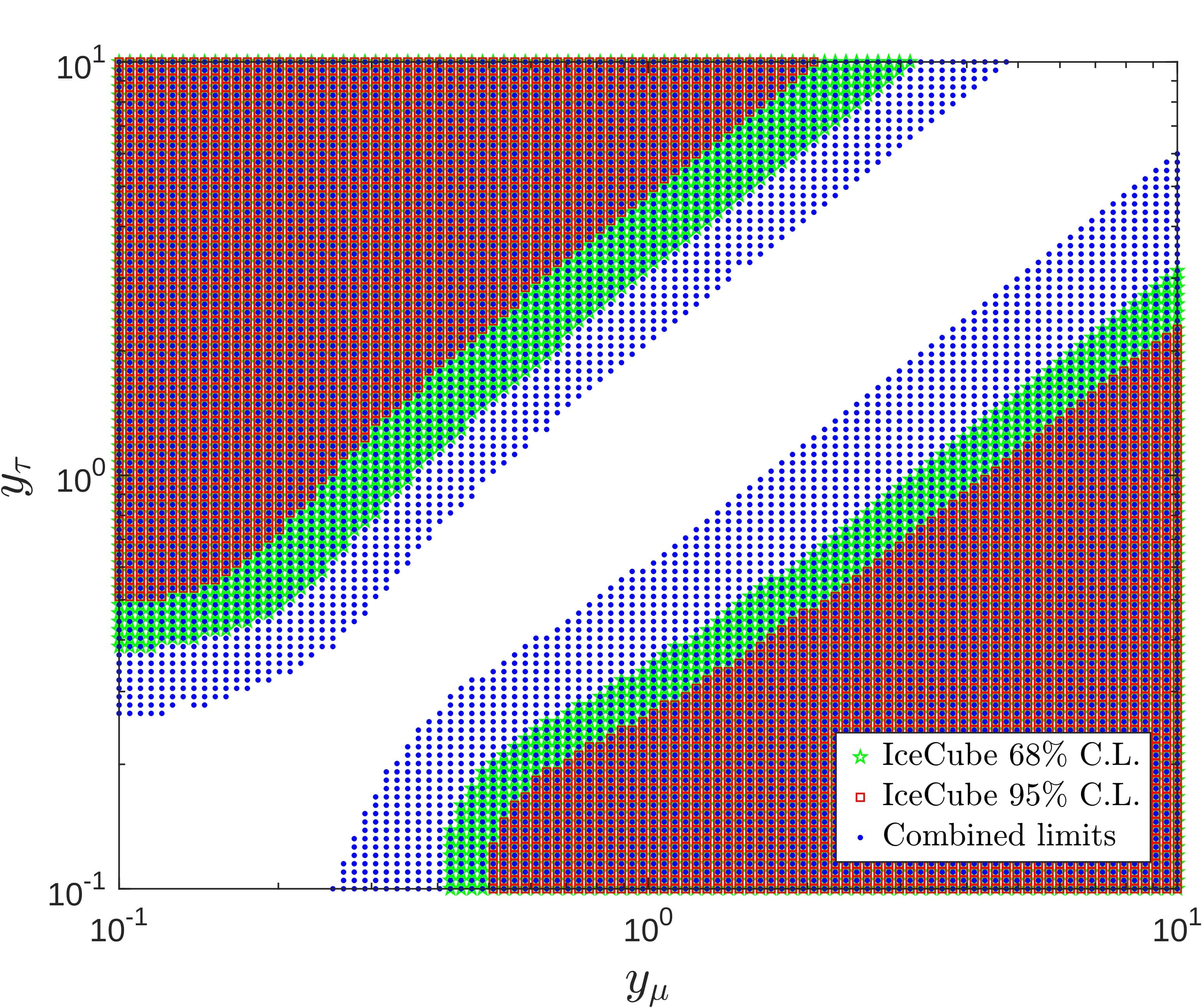}\quad
    \includegraphics[width=0.33\linewidth]{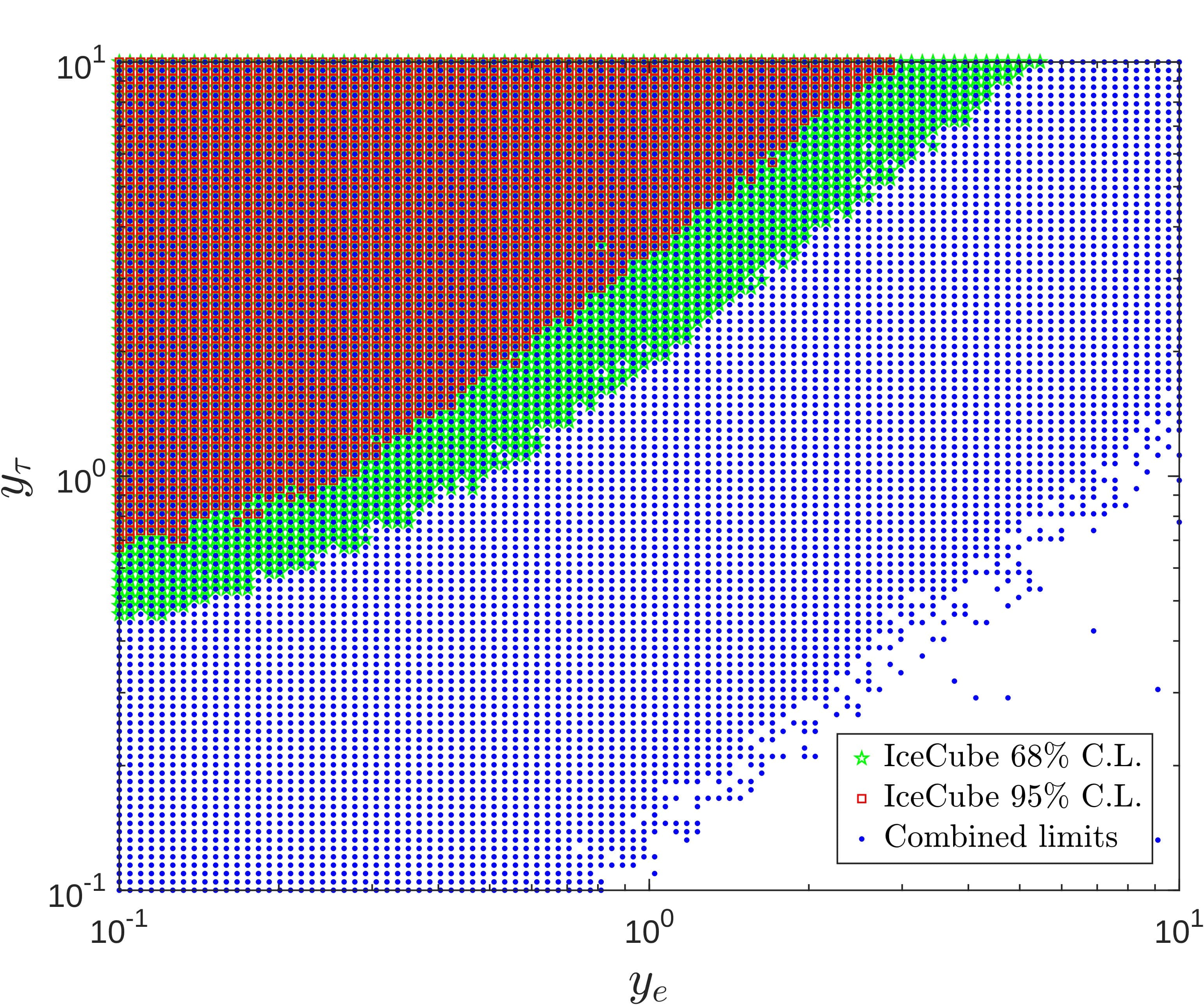}}
    \caption{The limits on coupling combinations: Case-(a) $(y_e,y_\mu,0)$, Case-(b) $(0,y_\mu,y_\tau)$, Case-(c) $(y_e,0,y_\tau)$ with $m_\phi=1\,{\rm feV}$ for NGC 1068. The first row is the limits for normal mass ordering, the second row is for inverted mass ordering.
    These color points are excluded by the IceCube MESE 68\% (red), 95\% (green) C.L. contour, and the combined $\nu$ telescopes (blue) sensitivity contour.}
    \label{fig5}
\end{figure}

\begin{figure}
    \renewcommand{\thefigure}{5}
    \captionsetup[subfloat]{labelformat=empty}
    \centering
    \subfloat[(d1-NO)\hspace{4.2cm} (d2-NO)\hspace{4.2cm}(d3-NO)]{\includegraphics[width=0.33\linewidth]{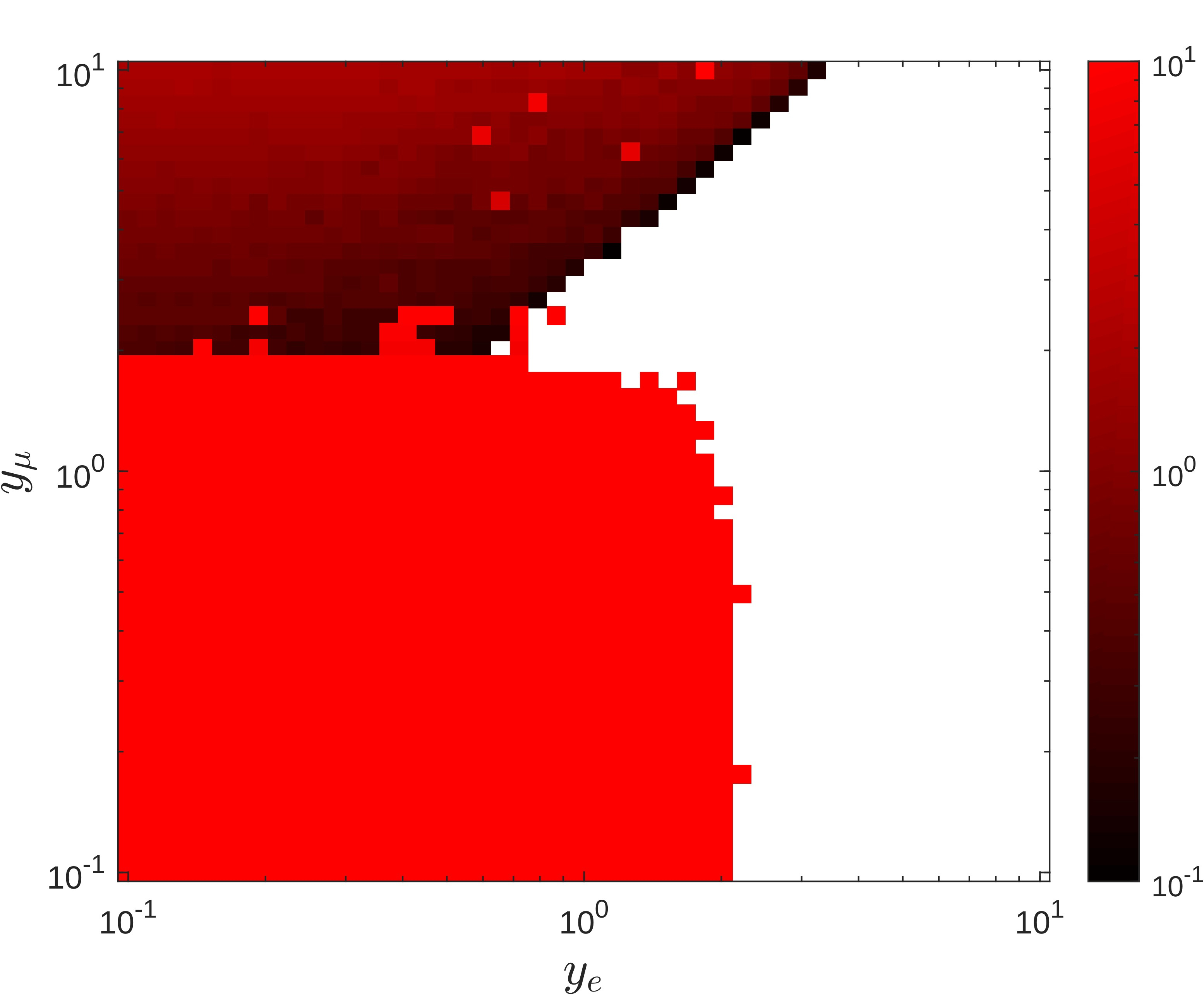}\quad
    \includegraphics[width=0.33\linewidth]{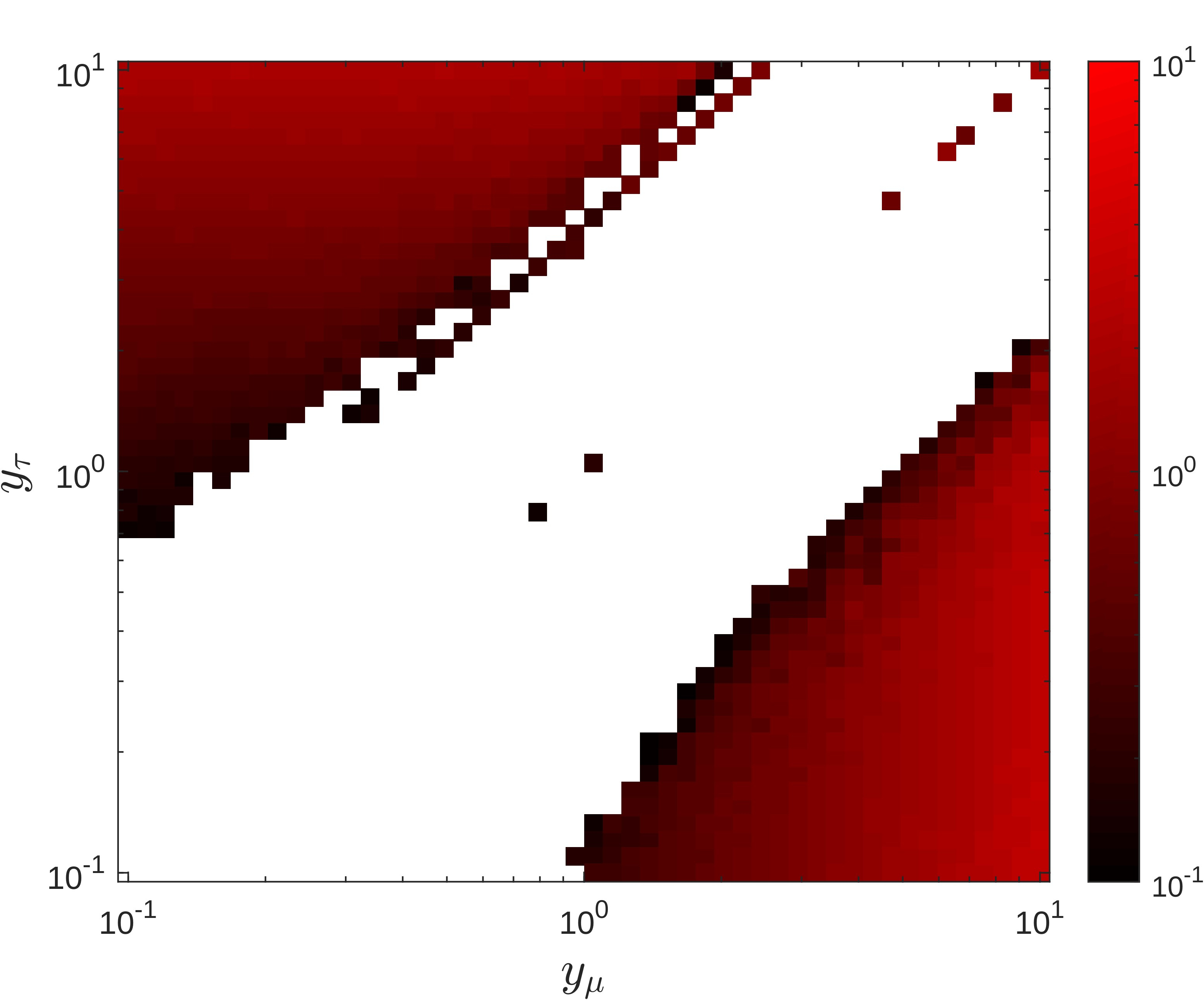}\quad
    \includegraphics[width=0.33\linewidth]{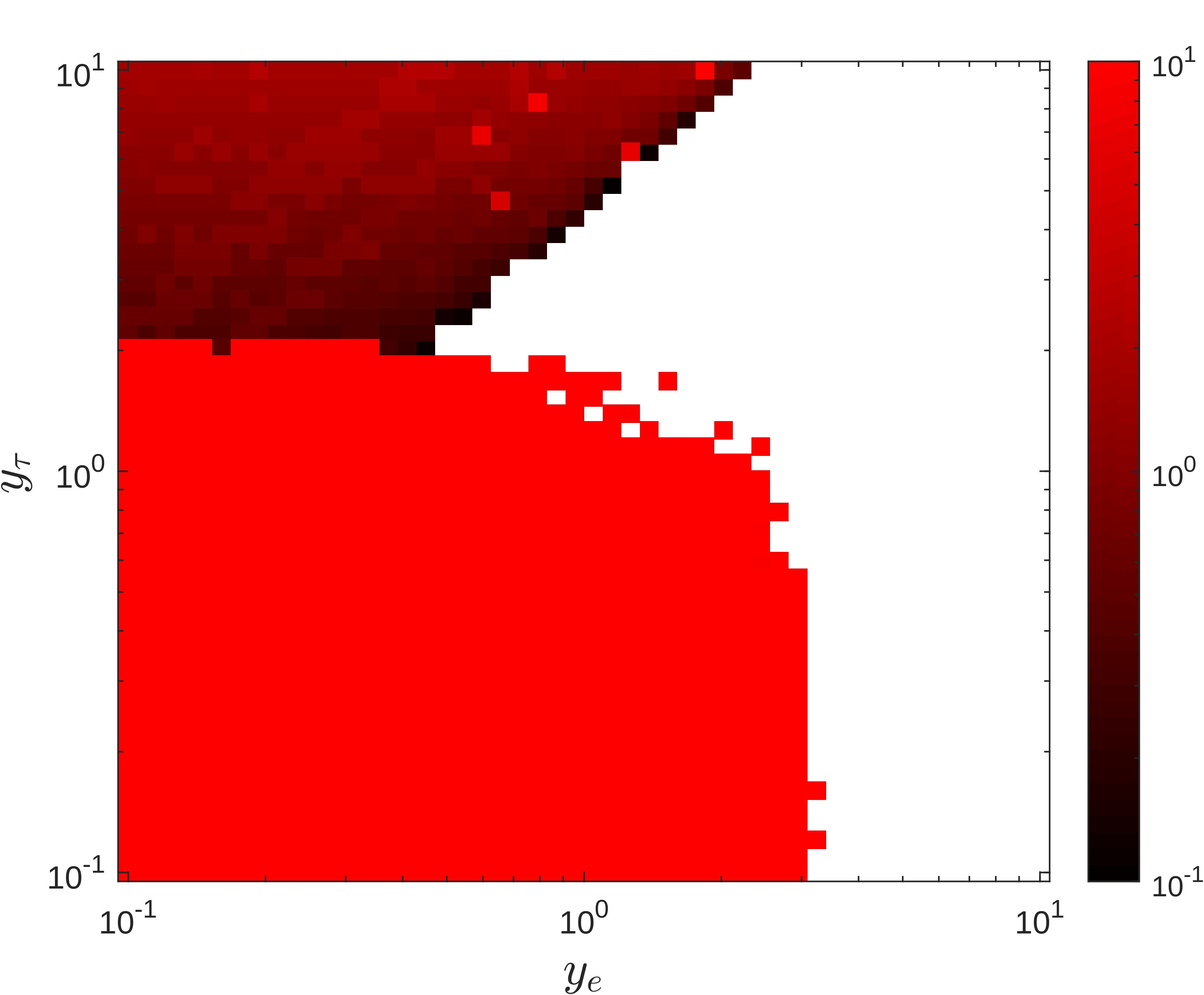}}


    \subfloat[(d1-IO)\hspace{4.2cm} (d2-IO)\hspace{4.2cm}(d3-IO)]{\includegraphics[width=0.33\linewidth]{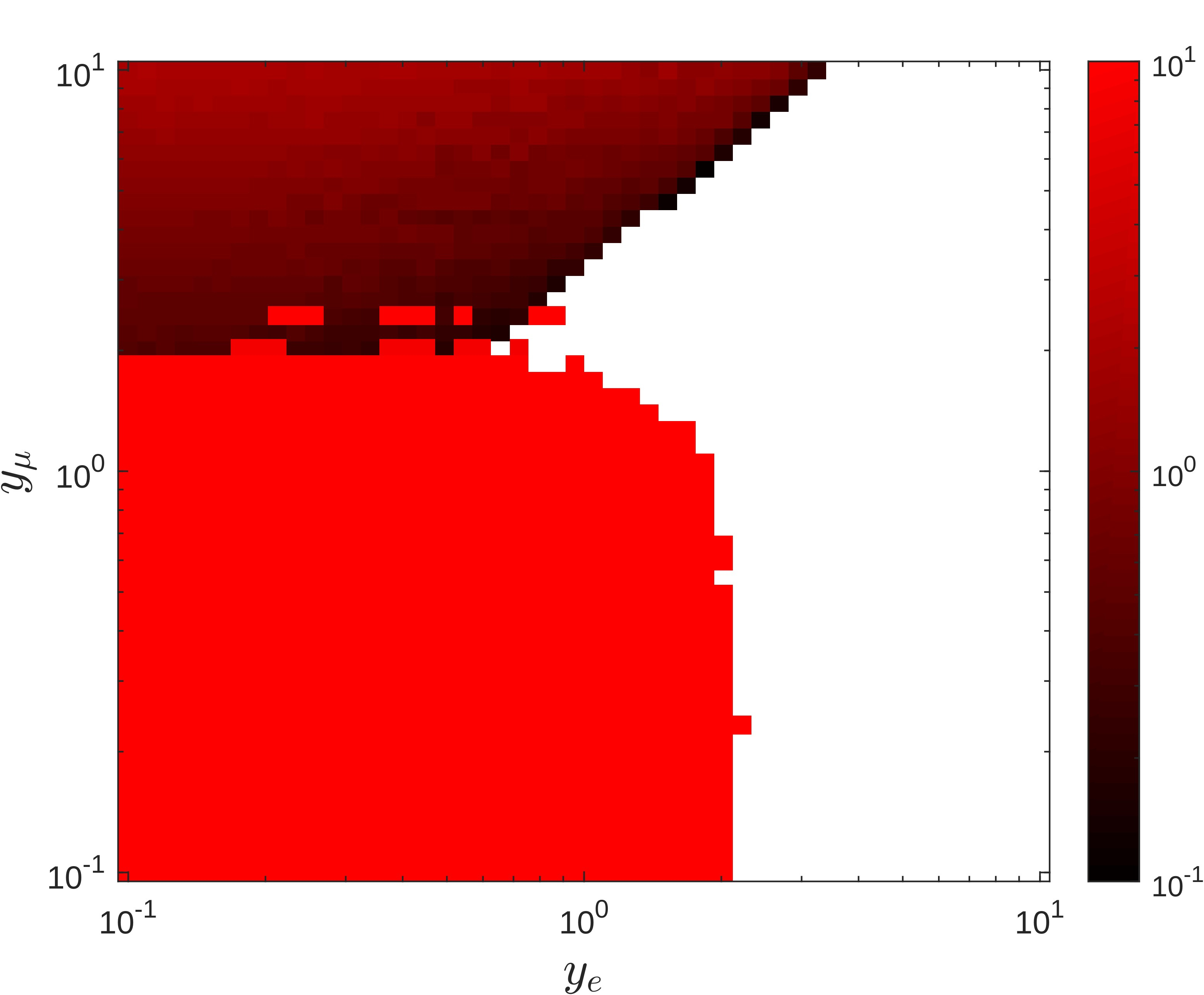}\quad
    \includegraphics[width=0.33\linewidth]{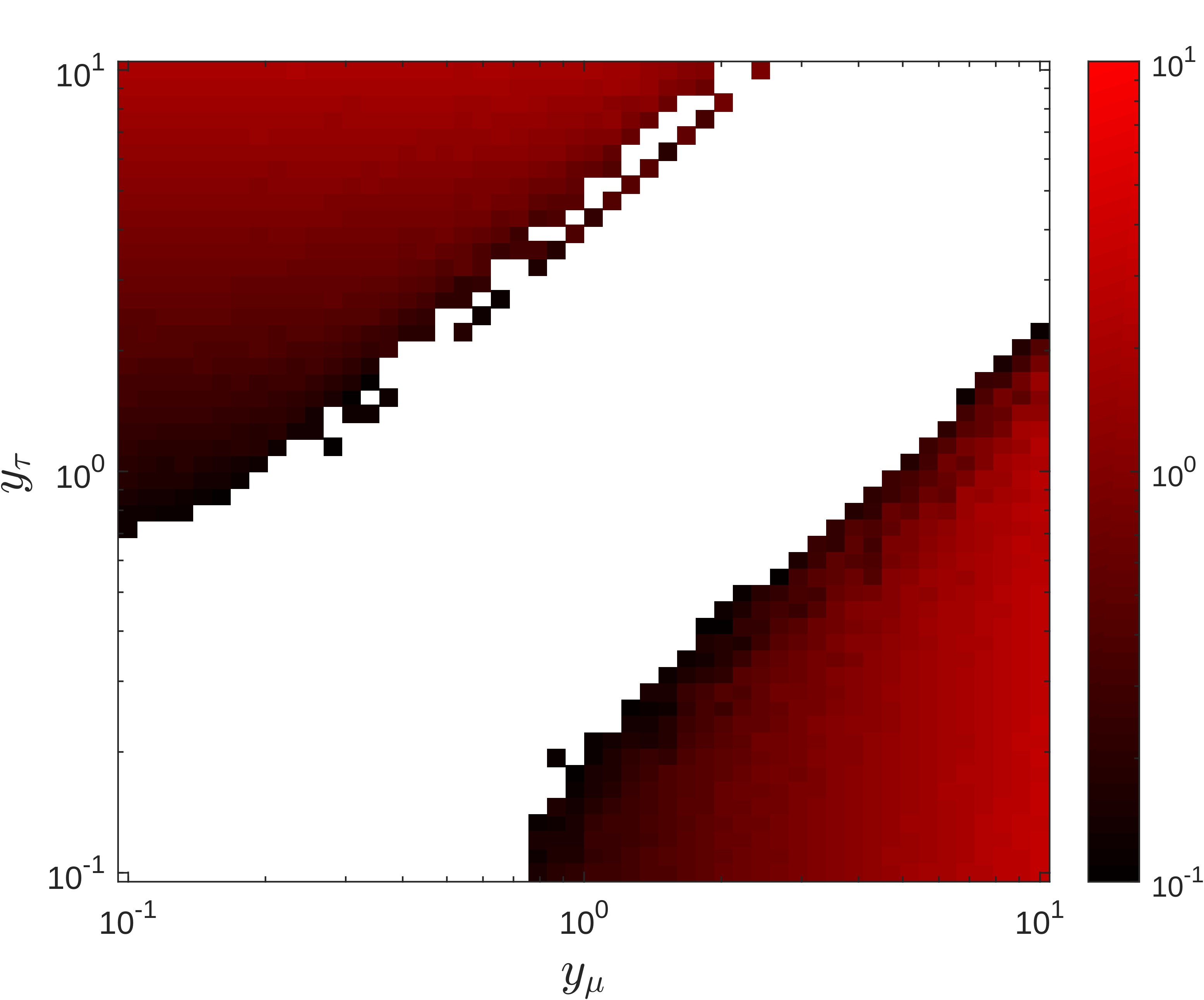}\quad
    \includegraphics[width=0.33\linewidth]{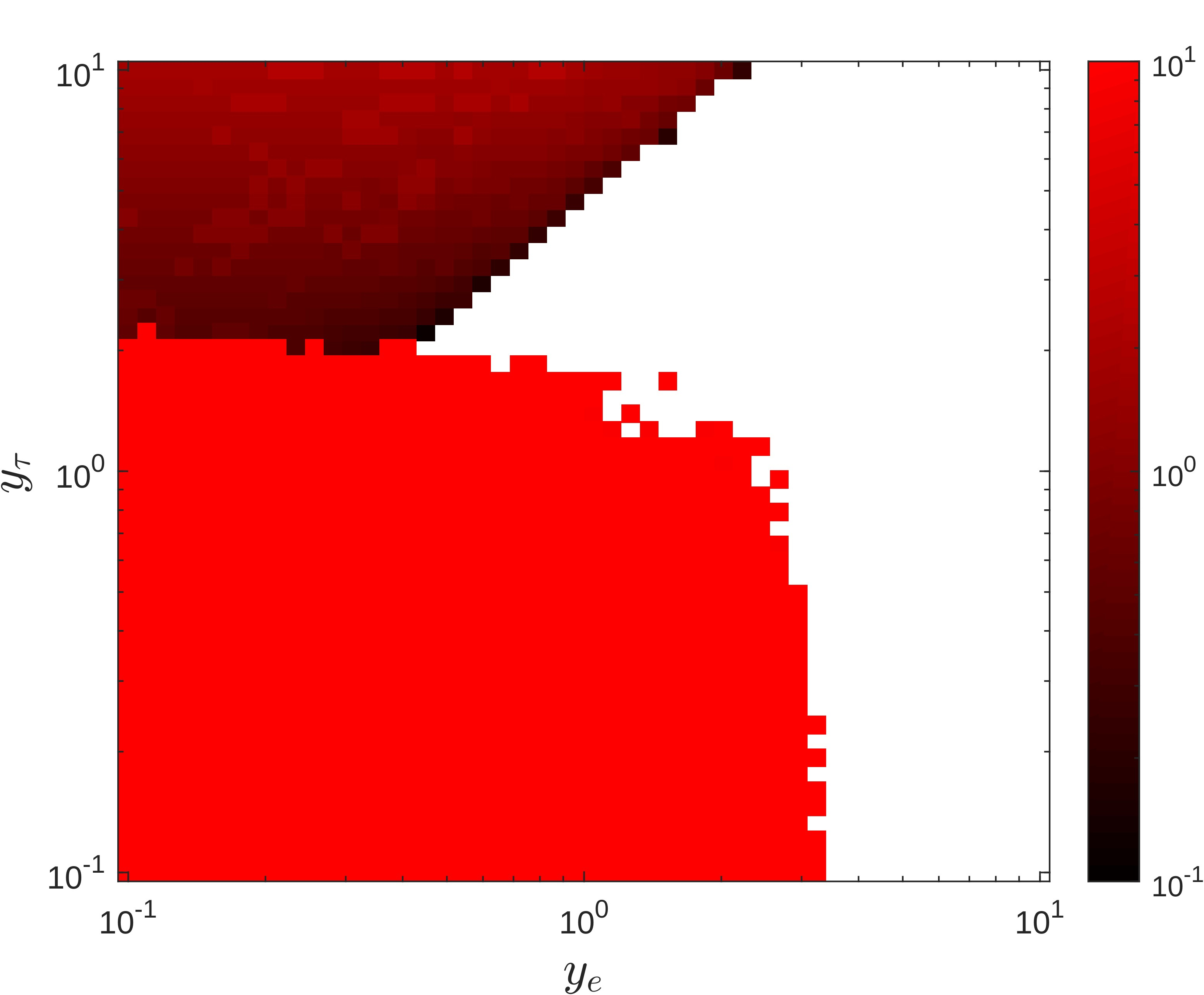}}

    \caption{The colored regions represent the couplings of Case-(d) $(y_{e}, y_{\mu}, y_{\tau})$ excluded by IceCube MESE 95\% sensitivity contour. The excluded regions with $m_\phi=1\,{\rm feV}$ by NGC 1068 are projected to two of three Yukawa couplings, and the color bar represents the maximum value of the third coupling. Upper (Lower) panels are for normal (inverted) mass ordering.
    }
    \label{fig6}
\end{figure}

Including the constraint from the combined future neutrino telescopes, IceCube-Gen2
/KM3NeT/Baikal-GVD/P-ONE/TAMBO,  
we are able to set the limits on the coupling constants. In Fig.~\ref{fig5}, we demonstrate the constraints on the coupling constants for $m_\phi=1\,{\rm feV}$.
These figures are obtained by showing the excluded regions for three coupling combinations Case-(a,b,c), selecting the points such that their $\bar{f}^\oplus_{e,\mu,\tau}$ lie outside the confidence contours, which is shown in Fig.~\ref{fig1}. 
The labels (a-NO), (b-NO), and (c-NO) correspond to the coupling combinations Case-(a,b,c) with normal-ordering assumption of neutrino mass.  
We can understand these patterns by referring to the $m_\phi=1\,{\rm feV}$ green squares in Fig.~\ref{fig1_2} (a1,a2,a3), (b1,b2,b3), and (c1,c2,c3), where the coupling constants 
are related to the effective potential via Eq.(\ref{effV}).
The constraint boundaries of the first row of Fig.~\ref{fig5} are roughly consistent with the color boundaries in Fig.~\ref{fig1_2} because darker regions indicate points that significantly deviate from $f^\oplus_0$, and thus be excluded by the experiments. 
The dark scalar mass $m_\phi$ also affects the value of $V^\oplus_{\alpha\beta}$ which scales as $m^{-1}_\phi$ in Eq.(\ref{effV}). According to Fig.~\ref{fig5} for $m_\phi=1\,{\rm feV}$ and Fig.~\ref{fig1_2}, we can derive the boundaries for other values of $m_\phi$. For completeness, we include the corresponding limits on coupling constants considering inverted mass ordering in (a-IO), (b-IO), and (c-IO) of Fig.~\ref{fig5}. Comparing the first and second rows in Fig.~\ref{fig5}, the future combined $\nu$ telescope will have the sensitivity to differentiate normal ordering from inverted ordering.
The parameters $(y_{e}, y_{\mu} , y_{\tau})$ of Case-(d) excluded by IceCube MESE 95\% C.L. contour are shown in Fig.~\ref{fig6}. 
Similar to Fig.~\ref{fig5}, the constraint boundaries of the first row of Fig. \ref{fig6} are roughly consistent with the color boundaries in the $m_\phi=1\,{\rm feV}$ green squares of Fig.~\ref{fig1_2}, except for the light red patch at the lower left corner of Fig.~\ref{fig6}(d1-NO) and (d3-NO). This patch arises from the nonzero value of the third coupling constant. The coupling constants excluded in this patch contributes to the region $\left(0.45\leq \bar{f}^\oplus_e\leq 0.85,\ 0.15\leq \bar{f}^\oplus_\mu\leq0.45,\ 0\leq \bar{f}^\oplus_\tau\leq0.1 \right)$ of $m_\phi=1\,{\rm feV}$ flavor ratio distribution in Fig.~\ref{fig1}(d2).
The same calculation is repeated using the combined $\nu$-telescope contour, and the results are shown in Appendix~\ref{appendixB}.

\subsection{DSNB Neutrino Flavor Ratio in Hyper-K, DUNE, and JUNO}
\begin{figure}[H]
    \renewcommand{\thefigure}{6-1}
    \captionsetup[subfloat]{labelformat=empty}
    \centering
    \subfloat[\hspace{0.75cm}(a)\hspace{3.4cm} (b)\hspace{3.4cm}(c)\hspace{3.4cm}(d)]{
    \includegraphics[height=0.26\linewidth]{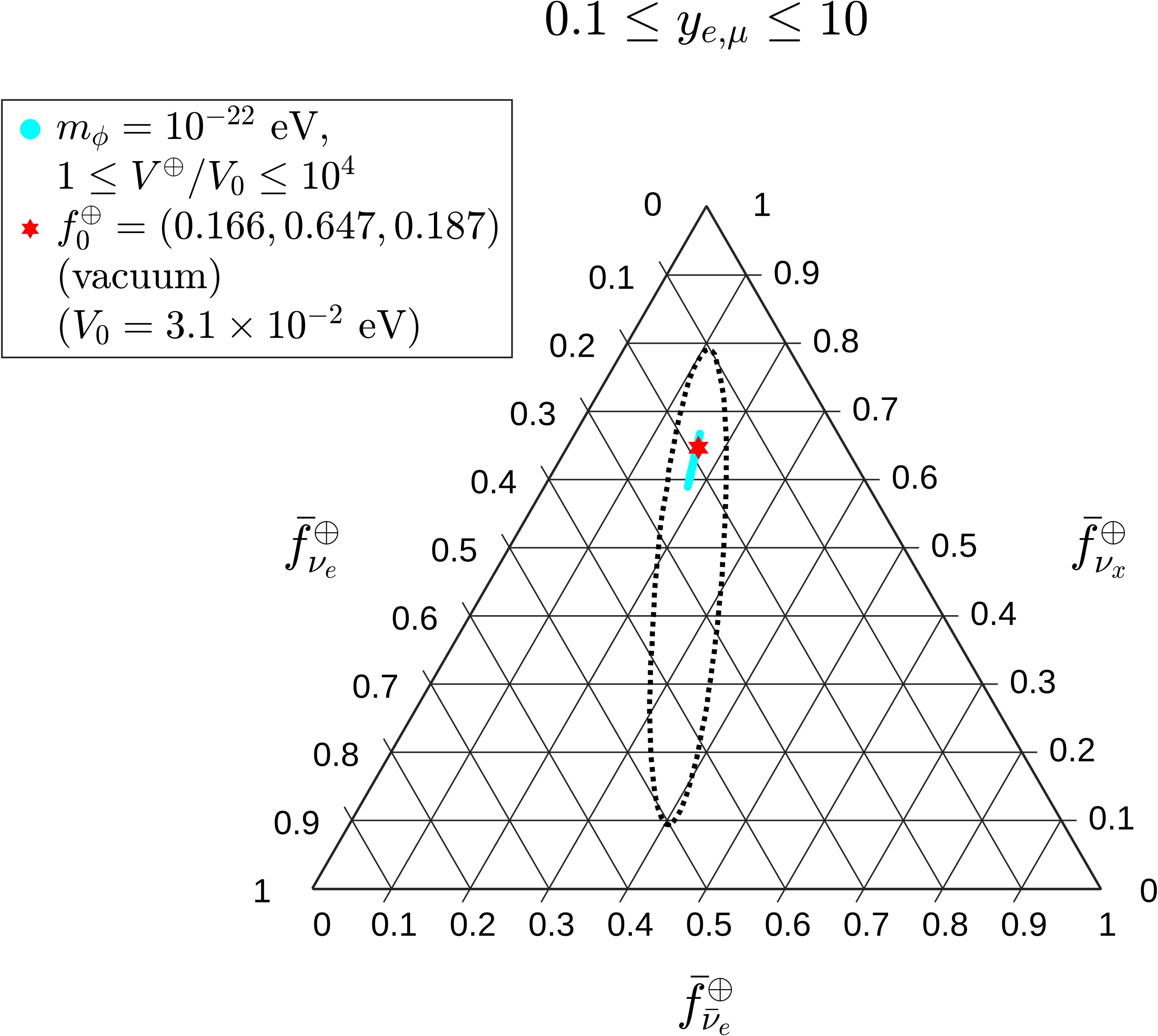}\quad

    \includegraphics[height=0.26\linewidth]{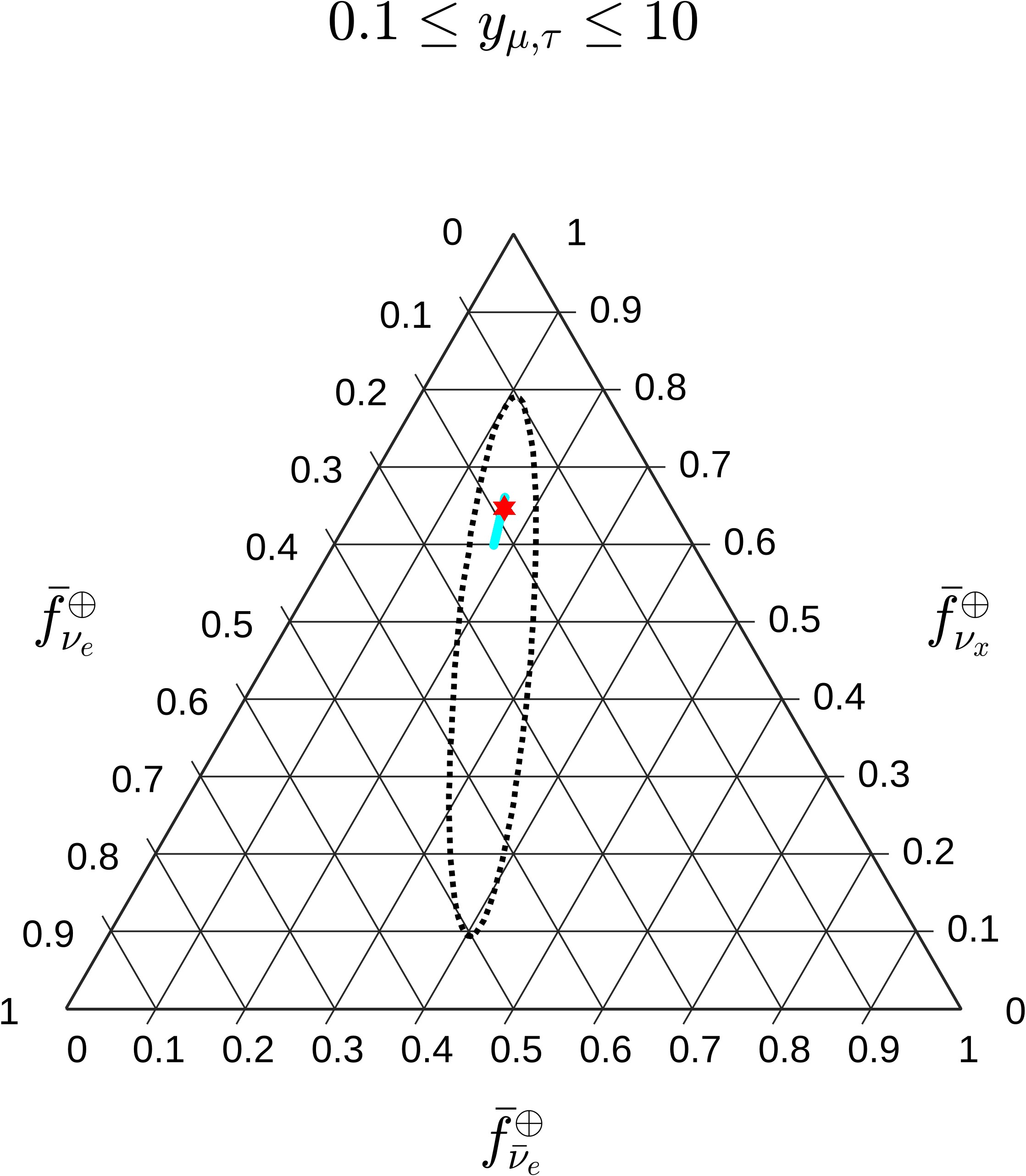}\quad

    \includegraphics[height=0.26\linewidth]{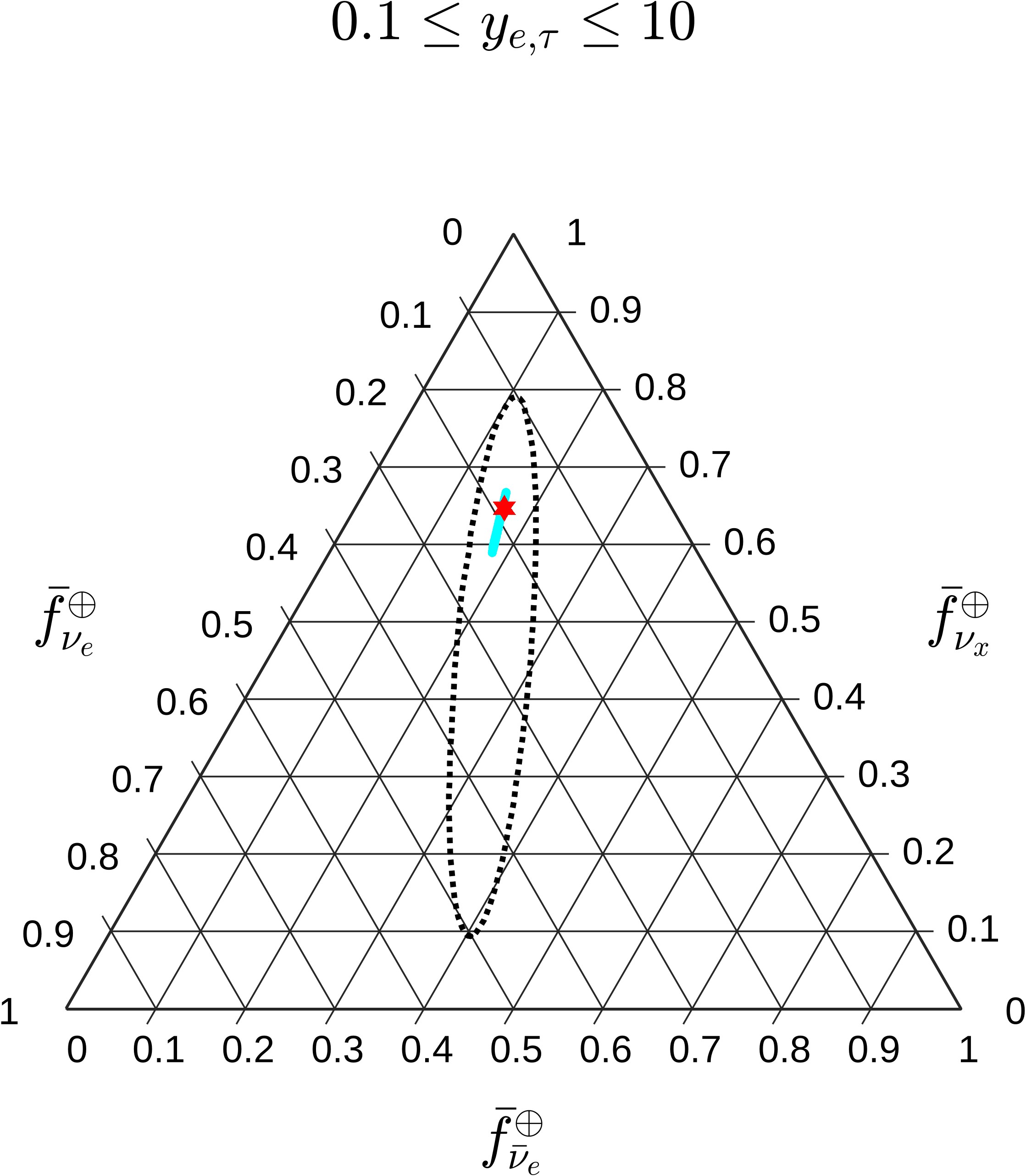}\quad

    \includegraphics[height=0.26\linewidth]{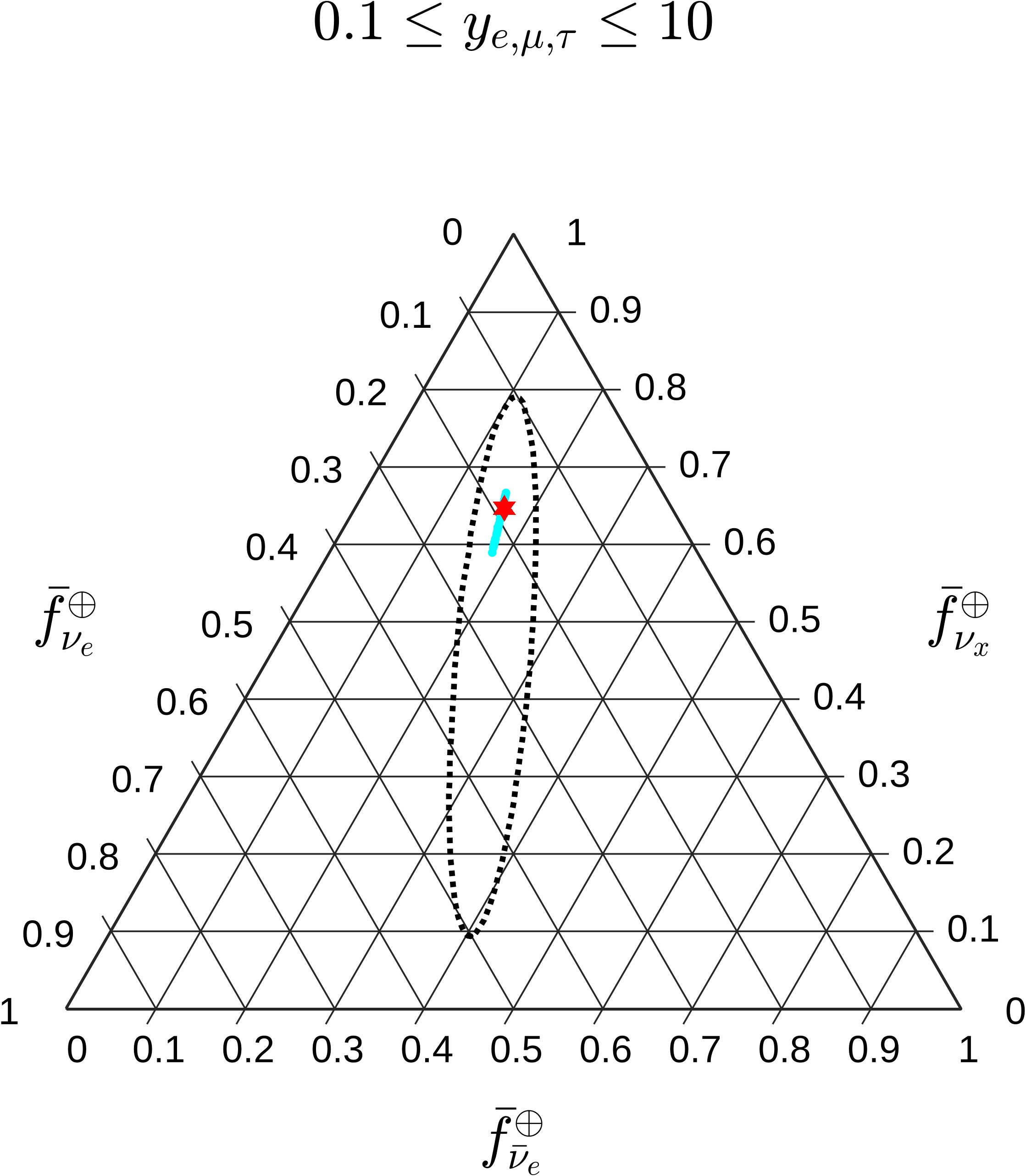}
    }

    \caption{The ternary plots of DSNB for Case-(a,b,c,d) with $m_\phi= 10^{-22}$ eV and $m_F=0.1$ GeV. The dotted contour is the combined limit from HK, DUNE, and JUNO.}
    \label{fig7}
\end{figure}

\begin{figure}[H]
    \renewcommand{\thefigure}{6-2}
    \captionsetup[subfloat]{labelformat=empty}
    \centering
    \subfloat[(a1)\hspace{4.8cm} (a2)\hspace{4.8cm}(a3)]{\includegraphics[width=0.33\linewidth]{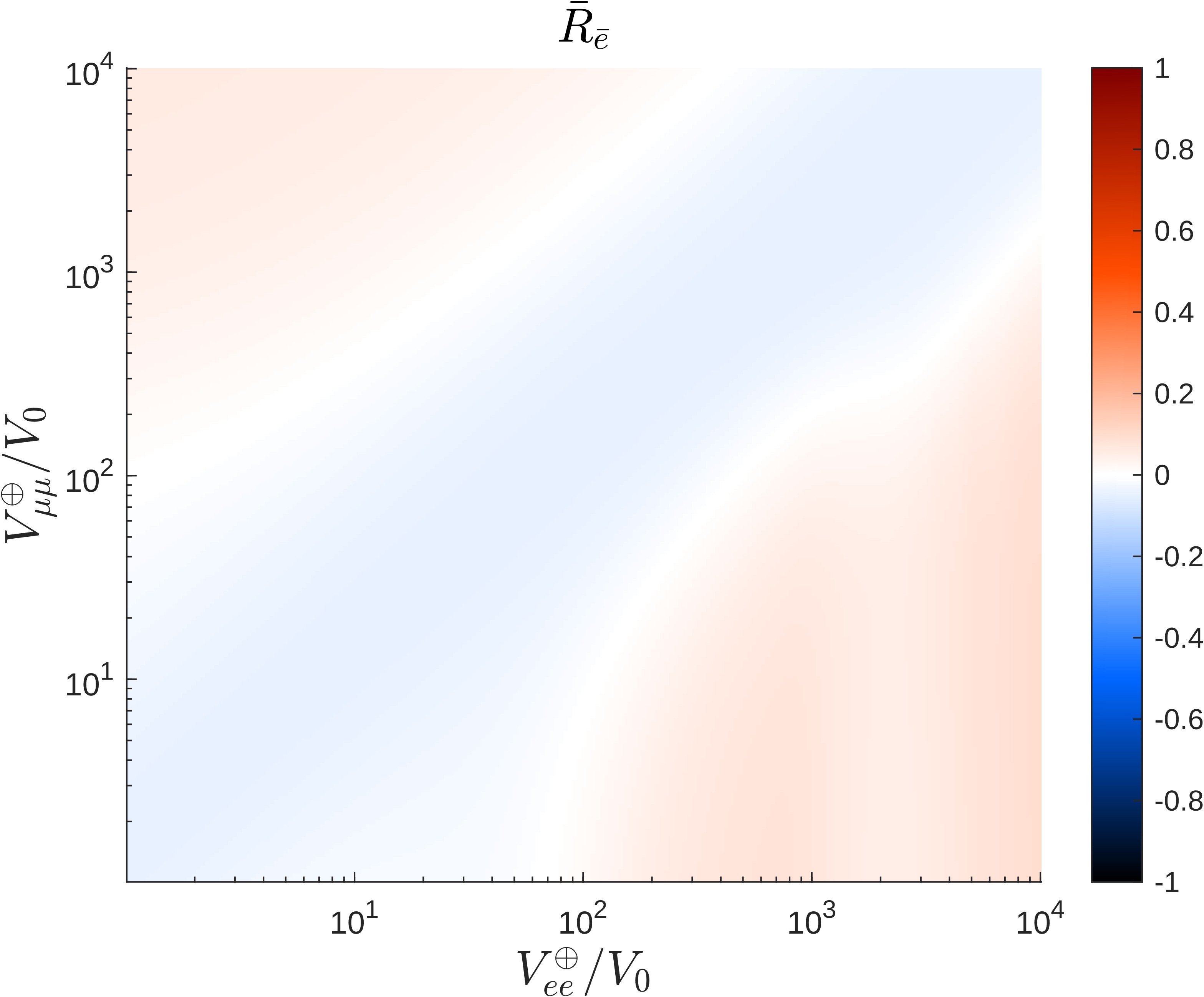}\quad
    \includegraphics[width=0.33\linewidth]{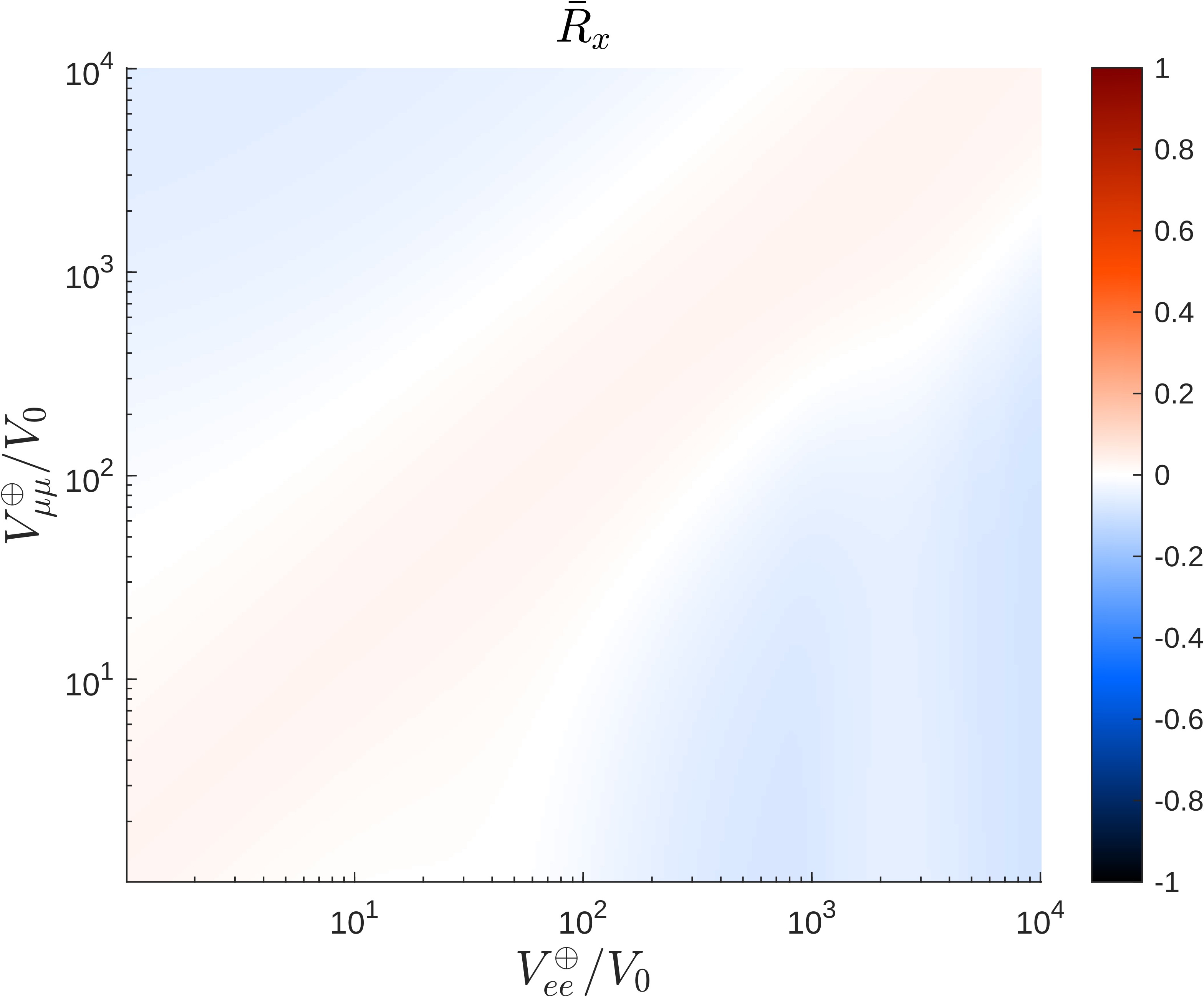}\quad
    \includegraphics[width=0.33\linewidth]{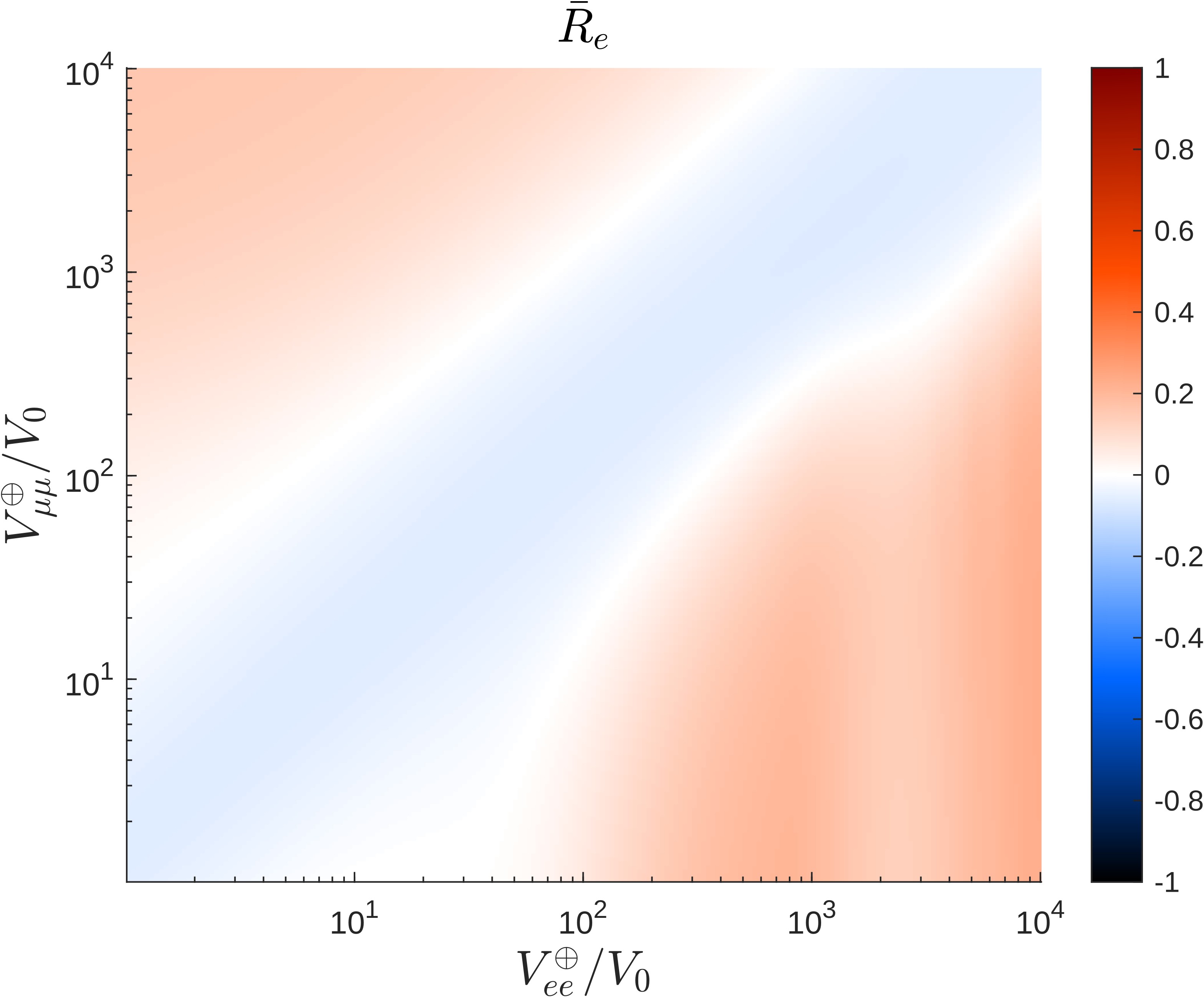}}
 
    \subfloat[(b1)\hspace{4.8cm} (b2)\hspace{4.8cm}(b3)]{\includegraphics[width=0.33\linewidth]{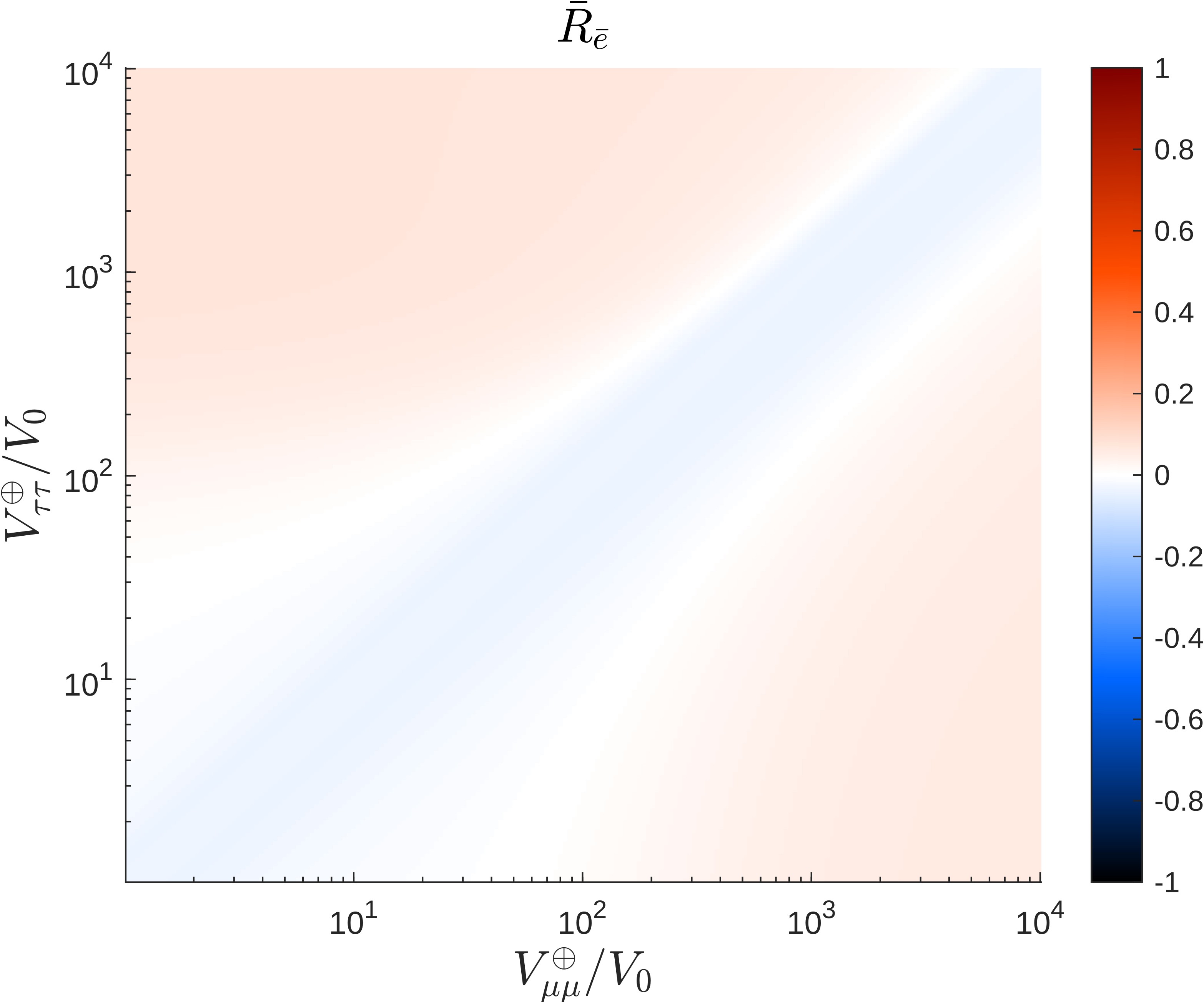}\quad
    \includegraphics[width=0.33\linewidth]{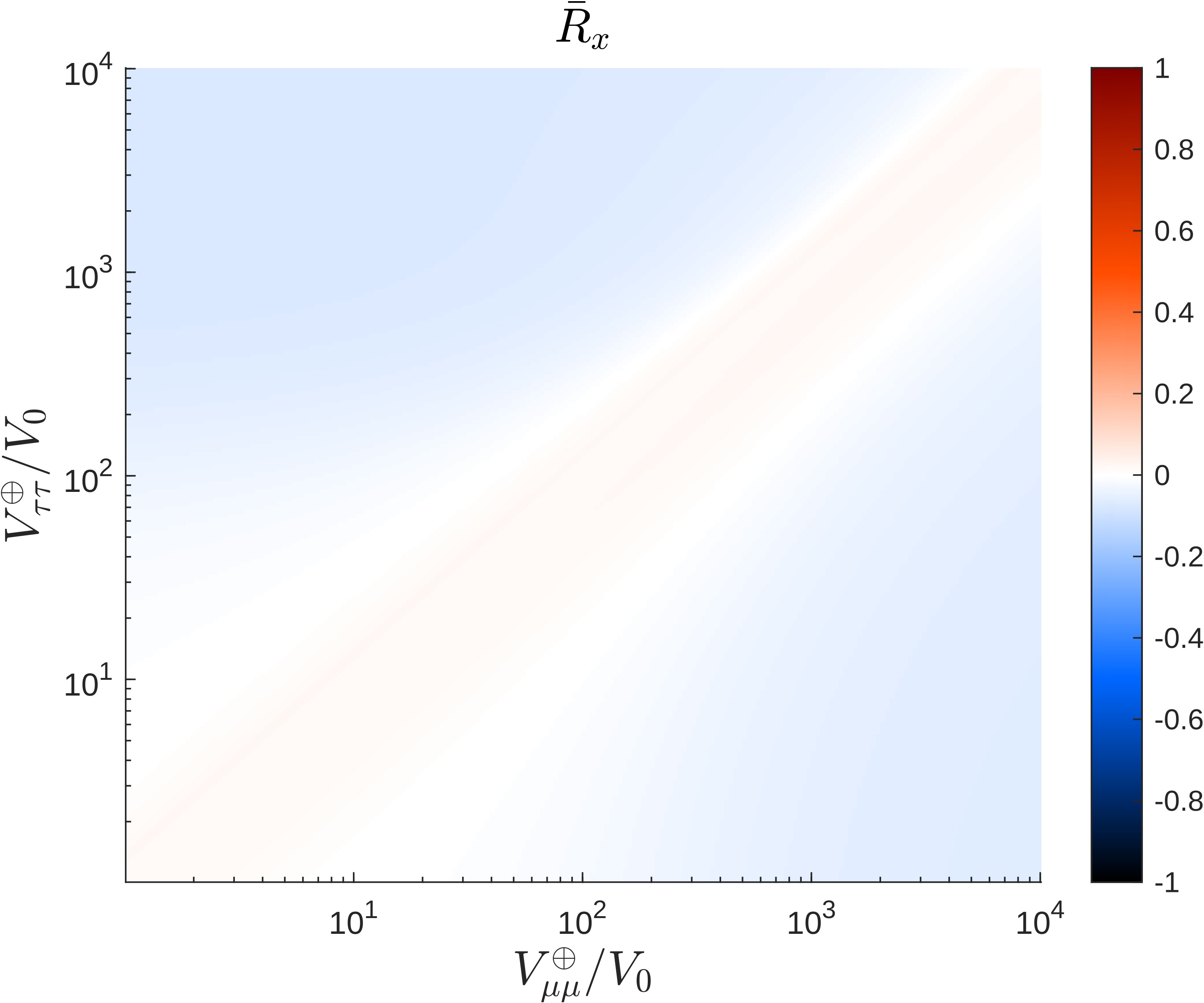}\quad
    \includegraphics[width=0.33\linewidth]{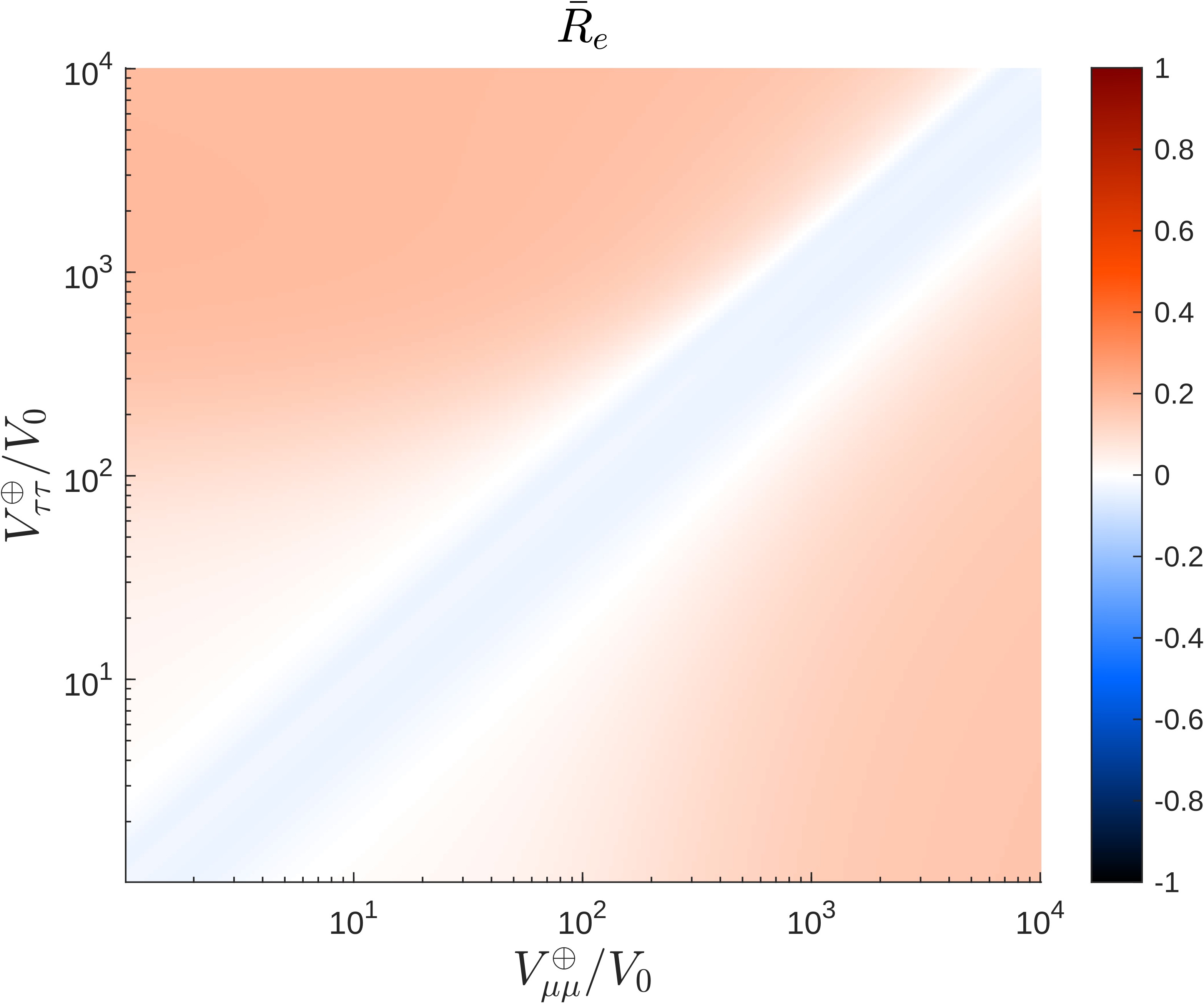}}

    \subfloat[(c1)\hspace{4.8cm} (c2)\hspace{4.8cm}(c3)]{\includegraphics[width=0.33\linewidth]{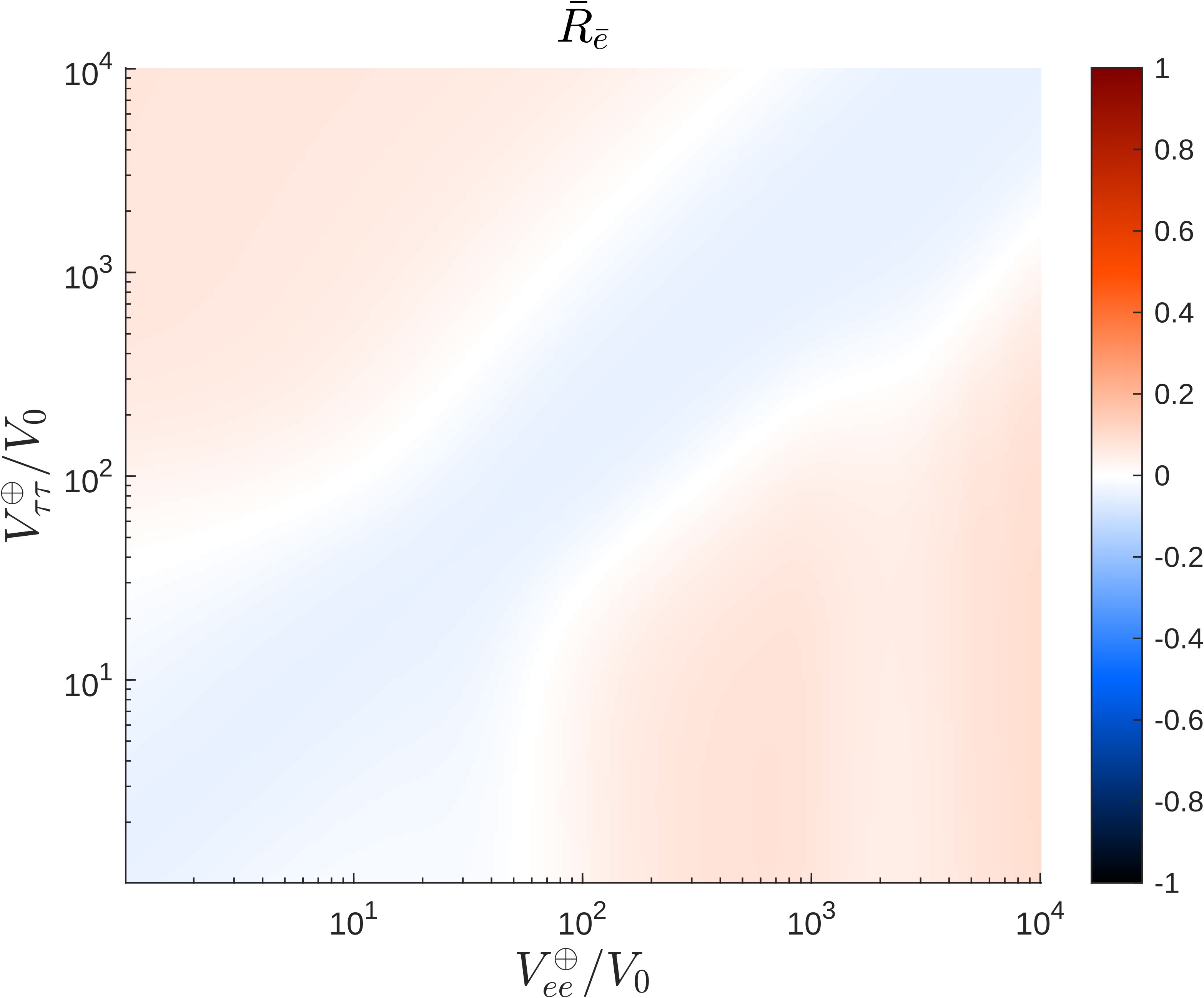}\quad
    \includegraphics[width=0.33\linewidth]{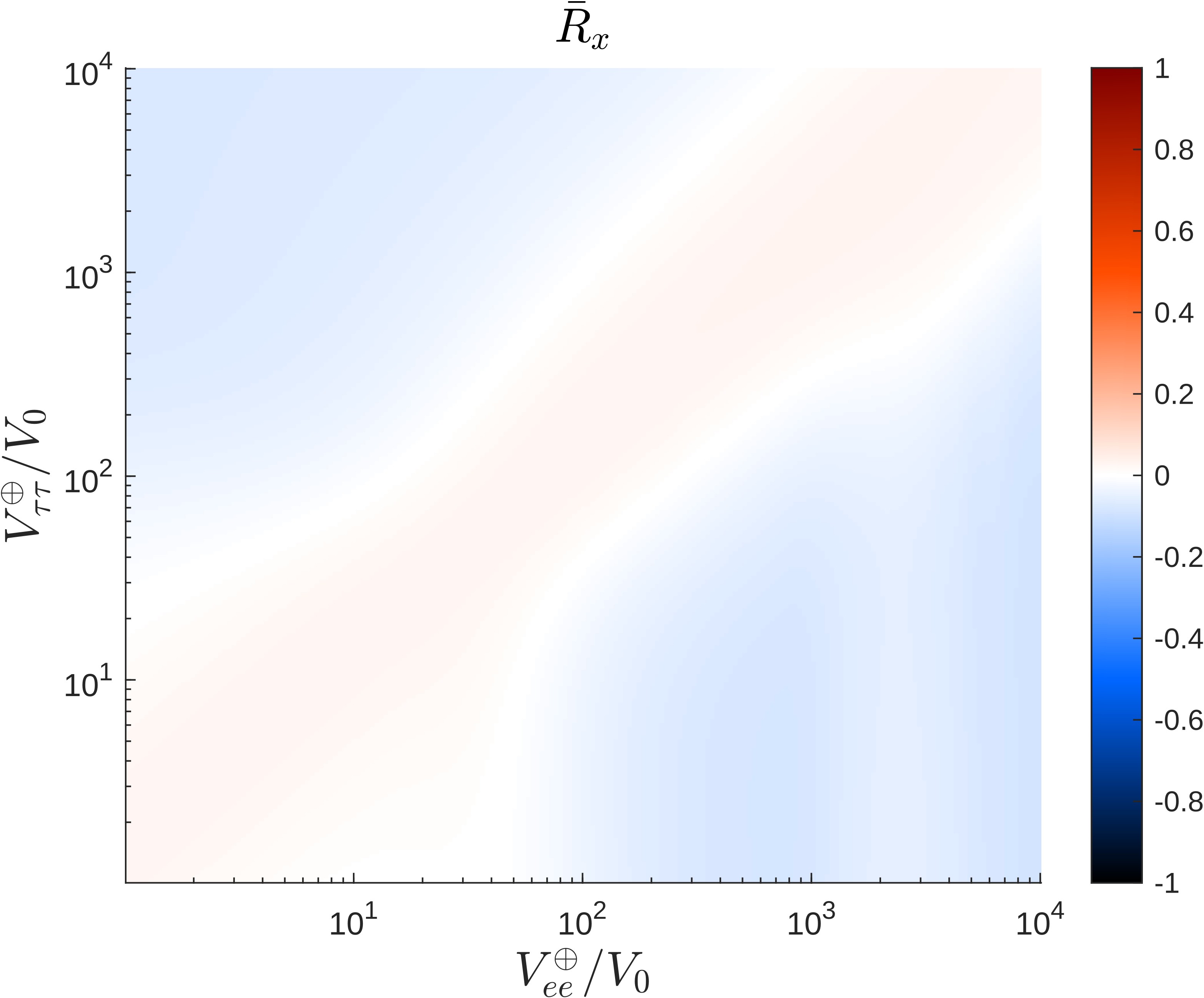}\quad
    \includegraphics[width=0.33\linewidth]{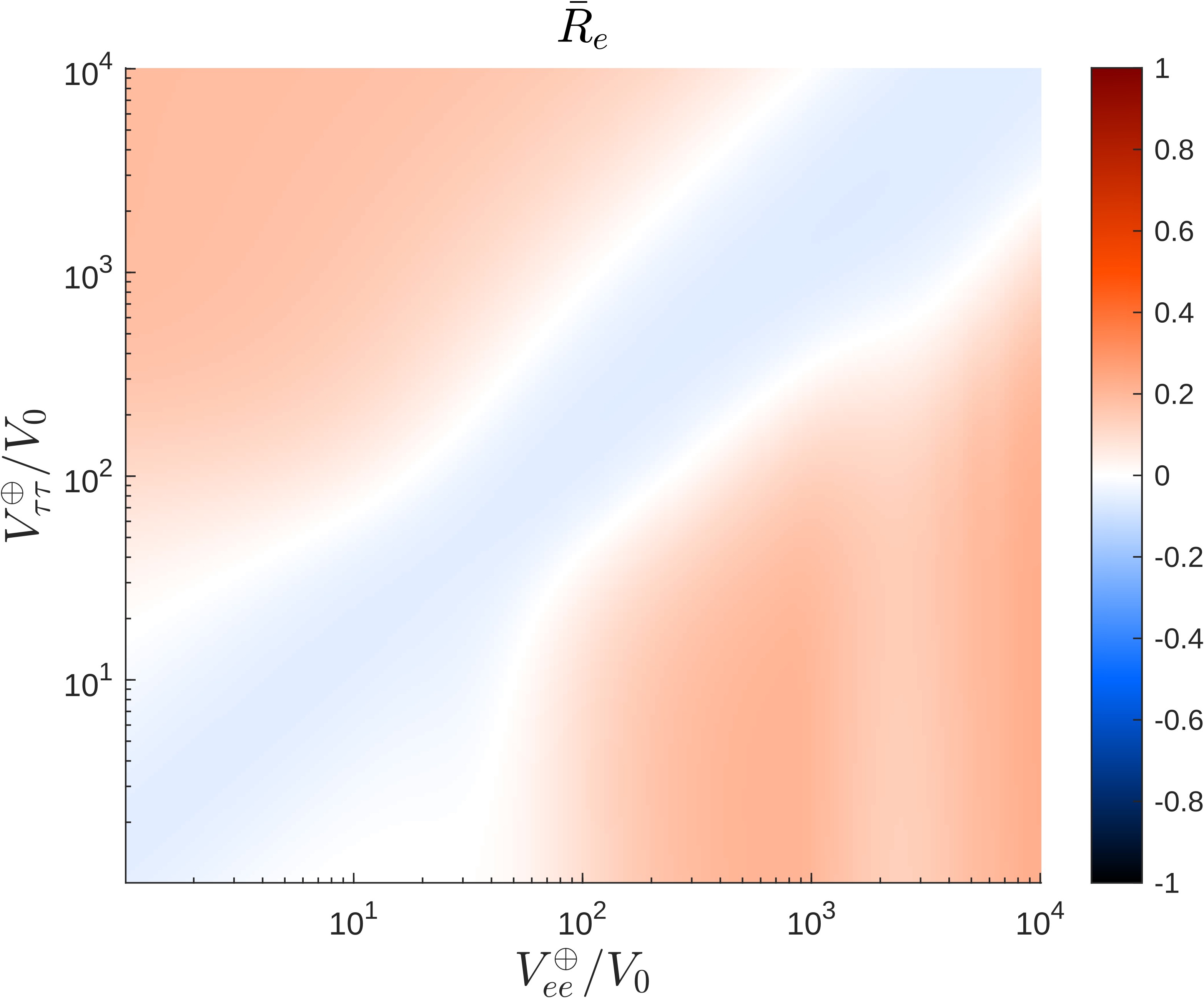}}
    \caption{The distributions of event-averaged deviation $\bar{R}_\beta$ of Fig.~\ref{fig7} for Case-(a,b,c).}
    \label{fig7_2}
\end{figure}

The similar calculations for DSNB neutrinos are shown in Fig.~\ref{fig7} and \ref{fig7_2}. Since the experiments that may detect the DSNB flux is more sensitive to $\nu_e$ and $\bar{\nu}_e$, we adopt the three axes of ternary plot to be $(\bar{\nu}_e,\nu_x,\nu_{e})$. The initial flavor ratio $f_0=(f_{0\bar{e}},f_{0x},f_{0e})=(0.179,0.605,0.216)$ is obtained from Eq.(\ref{ini_fDSNB}a) with the temperature $T_{\nu_e}=5\,{\rm MeV}$, $T_{\bar{\nu}_e}=6\,{\rm MeV}$, and $T_{\nu_x}=7\,{\rm MeV}$. After the propagation in vacuum, the final flavor ratio at Earth is $f^\oplus_0=(0.166,0.647,0.187)$. 
The final flavor ratio of DSNB neutrinos, shown in Fig.~\ref{fig7}, is obtained by weighted averaging the initial flavor ratio of each energy bin Eq.(\ref{ini_fDSNB}b) and energy distribution of the event number in Eq.(\ref{DeltaN_N}). 
The combined limit from HK, DUNE, and JUNO \cite{Tabrizi:2020vmo} is included in Fig.~\ref{fig7}. The flavor ratio distributions are almost identical for the four coupling combinations Case-(a,b,c,d),
and the distribution points lie within the combined experimental contour. 
Only Case-(b) exhibits a mild difference, and this can be seen from the color map of Fig.~\ref{fig7_2} (d1,d2,d3) in which the diagonal regions are narrower than that of the corresponding panels for Case-(a,b).
As a result, the distribution in Fig. \ref{fig7}(b) is ``shorter'' than others.
Overall, the neutrino dark matter interaction tends to decrease the fractions of $\mu$ and $\tau$ flavor, and the fraction of electron neutrino is increased. The anti-electron neutrino fraction is mildly shifted away from the vacuum case (red point). Since the neutrino energy from diffuse supernovae is several order smaller than AGN neutrino, we consider $m_\phi=10^{-22}\,{\rm eV}$ and $m_F=0.1\,{\rm GeV}$ to enhance the effective potential. 
Even though, the $m_\phi=10^{-22}\,{\rm eV}$ already reaches the mass lower bound of ultra-light DM, it is still not enough to make the flavor distribution distinguishable from $f^\oplus_0$.

\bigskip


\section{Summary}\label{sum}
We consider high-energy neutrinos from astrophysical sources oscillating under the presence of neutrino-scalar DM interactions, including four coupling combinations: Case-(a) $(y_e,y_\mu,0)$, Case-(b) $(0,y_\mu,y_\tau)$, Case-(c) $(y_e,0,y_\tau)$, and Case-(d) $(y_e,y_\mu,y_\tau)$. The effective potential $V_{\alpha\beta}$ depends on the coupling constant $y_{e,\mu,\tau}$, scalar DM mass $m_\phi$, and the mediator mass $m_F$. We take the AGNs and the DSNB as our sources that produce neutrinos at different energy scales. 
The $\mathcal{O}({\rm TeV})$ neutrinos are assumed to be produced from NGC 1068 and TXS 0506+056 via pion decay with initial flavor ratio $(\nu_e:\nu_\mu:\nu_\tau)=(1:2:0)$. The neutrino production regions for NGC 1068 and TXS 0506+056 are assumed to be $[10,10^2]R_s$ (corona region) and $[10^2,10^5]R_s$, respectively.
The distances between Earth and AGNs are much longer than the oscillation length, and the adiabatic parameter is extremely small, so the neutrino propagates adiabatically and the final averaged flavor ratio at Earth is obtained from Eq.(\ref{ratio_at_earth}) with $n=1$. The coupling constant is restricted by the mean free path of neutrino, which depends on the spike density of AGN. The upper limit for $y_{e,\mu,\tau}$ is 10, and we fix the lower limit at 0.1.

Considering two sets of dark matter mass, $m_\phi=10, 1, 0.1\,{\rm peV}$ and $m_\phi=1, 0.1, 0.01\,{\rm feV}$, with the mediator mass fixed at $m_F=1\,{\rm TeV}$, and assuming the neutrinos are produced randomly inside corona region, we generate the ternary plots of $\bar{f}^\oplus_{e,\mu,\tau}$ with four cases, $(y_e, y_\mu, 0)$, $(0, y_\mu, y_\tau)$, $(y_e, 0, y_\tau)$, and all $y_{e,\mu,\tau}$ nonzero.
From all of the ternary plots we see that the final tau fraction $\bar{f}^\oplus_\tau$ under the influence of $\nu\phi$ interaction tends to be decreased from $f^\oplus_{0\tau}$, 
and $\bar{f}^\oplus_\mu$ are mostly be increased from $f^\oplus_{0\mu}$.
The limits from IceCube MESE, IceCube-Gen2, and future neutrino telescope are included, and all of the distributions are partially overlap with the experimental sensitivities. We set the constraint on the coupling constants by excluding the event average points outside the experiment sensitivity contours, and the constraint for $m_\phi=1\,{\rm feV}$ is shown in Fig.~\ref{fig5} and \ref{fig6}. In these figures, the results of normal and inverted mass ordering are both included, but from the shape of the constraint we find the effect of the mass ordering is minor.
We also define the deviation parameters $\bar{R}_{e,\mu,\tau}$ and generate their color maps for the distributions in the ternary plots. The color maps indicate the difference between $\bar{f}^\oplus_{e,\mu,\tau}$ and $f^\oplus_{0e,0\mu,0\tau}$ with different values of the effective potential around the Earth. Accordingly, given a set of $(V^\oplus_{\alpha\alpha},V^\oplus_{\beta\beta})$, we can roughly infer the position of $\bar{f}^\oplus_{e,\mu,\tau}$ relative to $f^\oplus_{0e,0\mu,0\tau}$ in a ternary plot. 

The neutrino energy from the DSNB is much smaller than that from the AGNs, so the vacuum term of the Hamiltonian Eq.(\ref{flavor_totH}) becomes larger. Furthermore, the neutrinos from CCSN propagate in the extragalactic medium, where the DM density is extremely low.  The effective potential should then be enhanced to make significant change in flavor ratio. We adopt the lower limit of $m_\phi=10^{-22}\,{\rm eV}$ and fix the mediator mass $m_F$ to $0.1\,{\rm GeV}$, then repeat the same calculation on the flavor ratio. Given the temperature of $\nu_e$, $\bar{\nu}_e$, and $\nu_x$, we may compute the DSNB flux and the initial flavor ratio. 
The flavor distributions in Fig.~\ref{fig7} for Case-(a,b,c,d) extend along the common direction and tend to have smaller values of $\bar{f}_{\nu_x}^\oplus$ and larger values of $\bar{f}_{\nu_e}^\oplus$.
However, shapes of the distributions are almost identical for Case-(a,c,d), only Case-(b) $(0,y_\mu,y_\tau)$ exhibits a mild difference. 
This can also be inferred from Fig.~\ref{fig7_2}, where the color map corresponding to the coupling constant $(0,y_\mu,y_\tau)$ differs slightly from the other two cases. 
We include the combined constraint from HK/DUNE/JUNO and find that all the points stay inside the sensitivity contour.

\bigskip

\section*{Acknowledgment}

We acknowledge the kind support of the National Science and Technology Council of Taiwan, with grant number NSTC 115-2112-M-007-010-.
P.Y.T. acknowledges support from the Physics Division of the National Center for Theoretical Sciences of Taiwan with grant NSTC 114-2124-M-002-003.
Y.M.Y. is supported in part by NSTC Grant No. 115B0061I4.

\bigskip
\newpage

\bigskip
\newpage

\appendix
\section{TXS 0506+056 Analysis with Neutrino Production Region $[10,10^2]R_s$}\label{appendixA}
\begin{figure}[H]
    \renewcommand{\thefigure}{7-1}
    \captionsetup[subfloat]{labelformat=empty}
    \centering
    \subfloat[\hspace{0.75cm}(a1)\hspace{3.1cm} (b1)\hspace{3.3cm}(c1)\hspace{3.1cm}(d1)]{\includegraphics[height=0.26\linewidth]{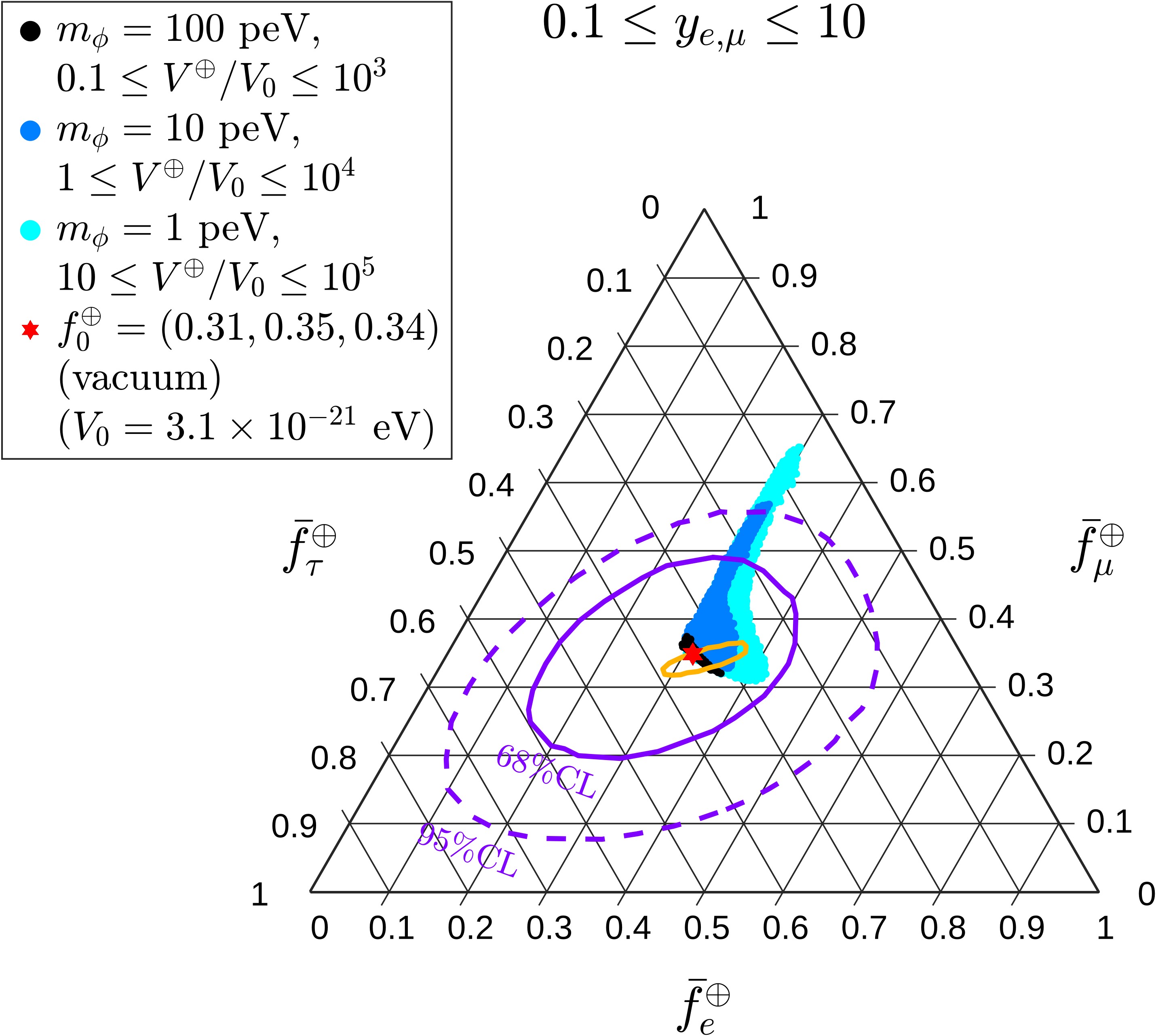}\quad
    \includegraphics[height=0.26\linewidth]{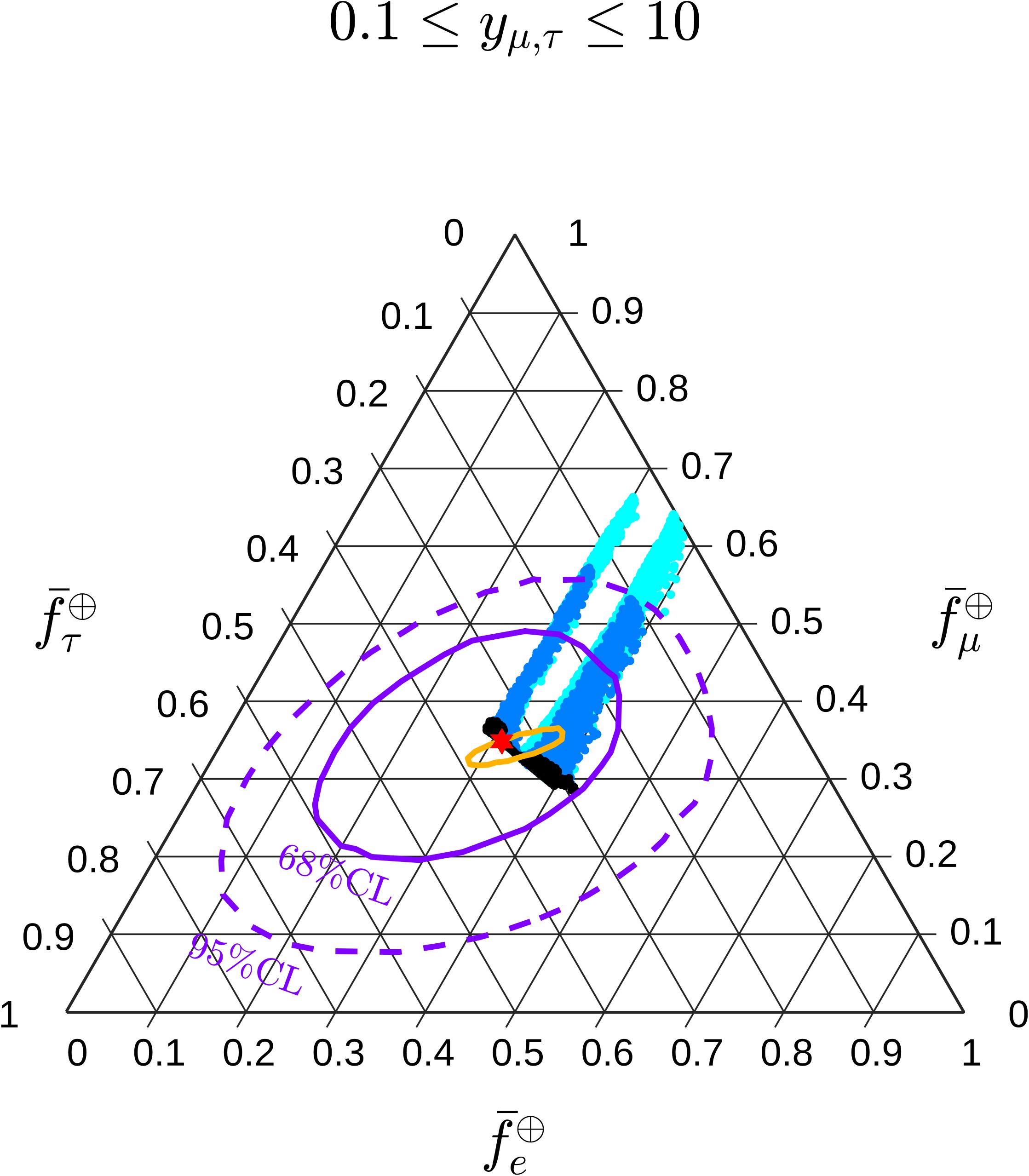}\quad
    \includegraphics[height=0.26\linewidth]{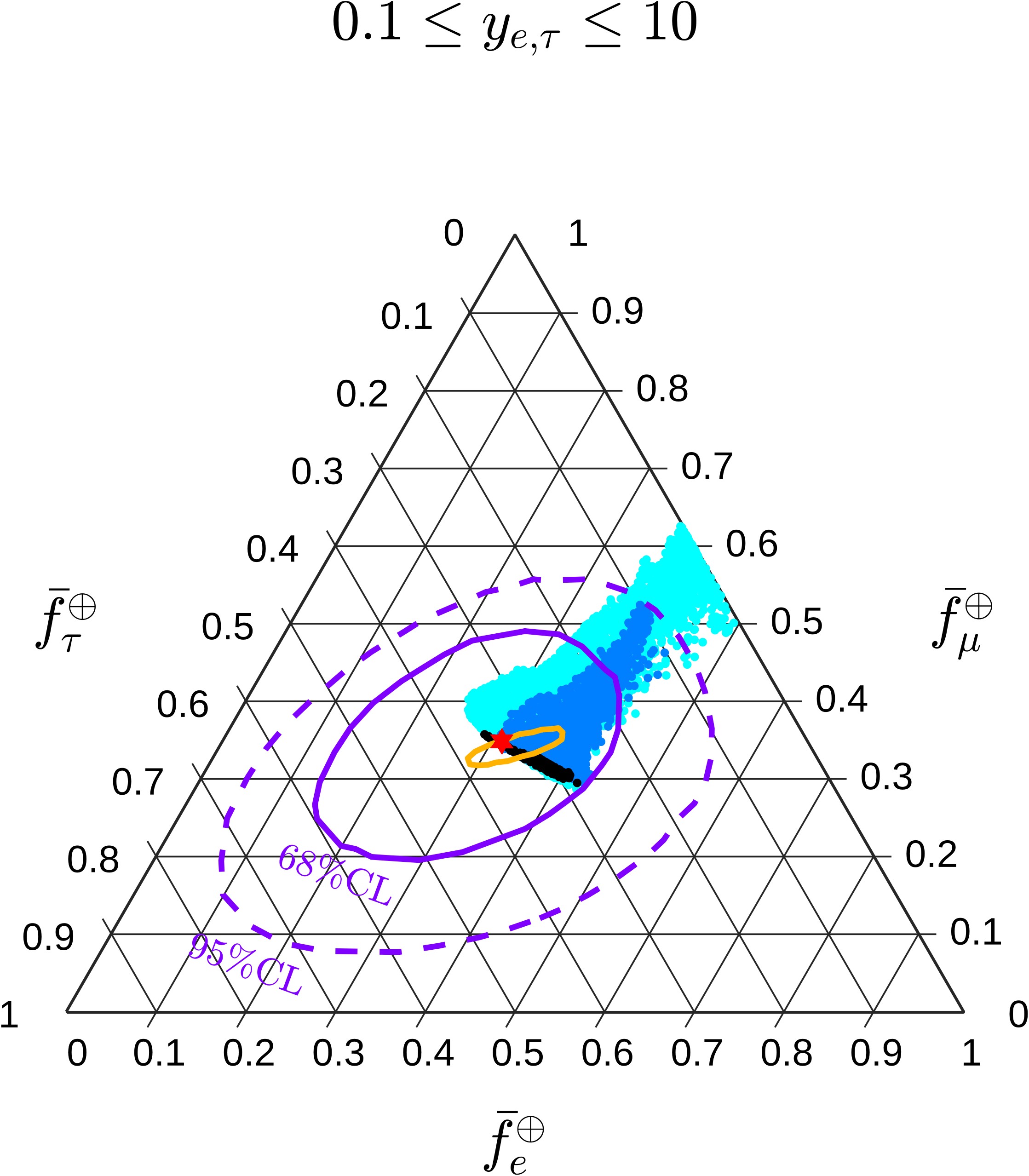}\quad
    \includegraphics[height=0.26\linewidth]{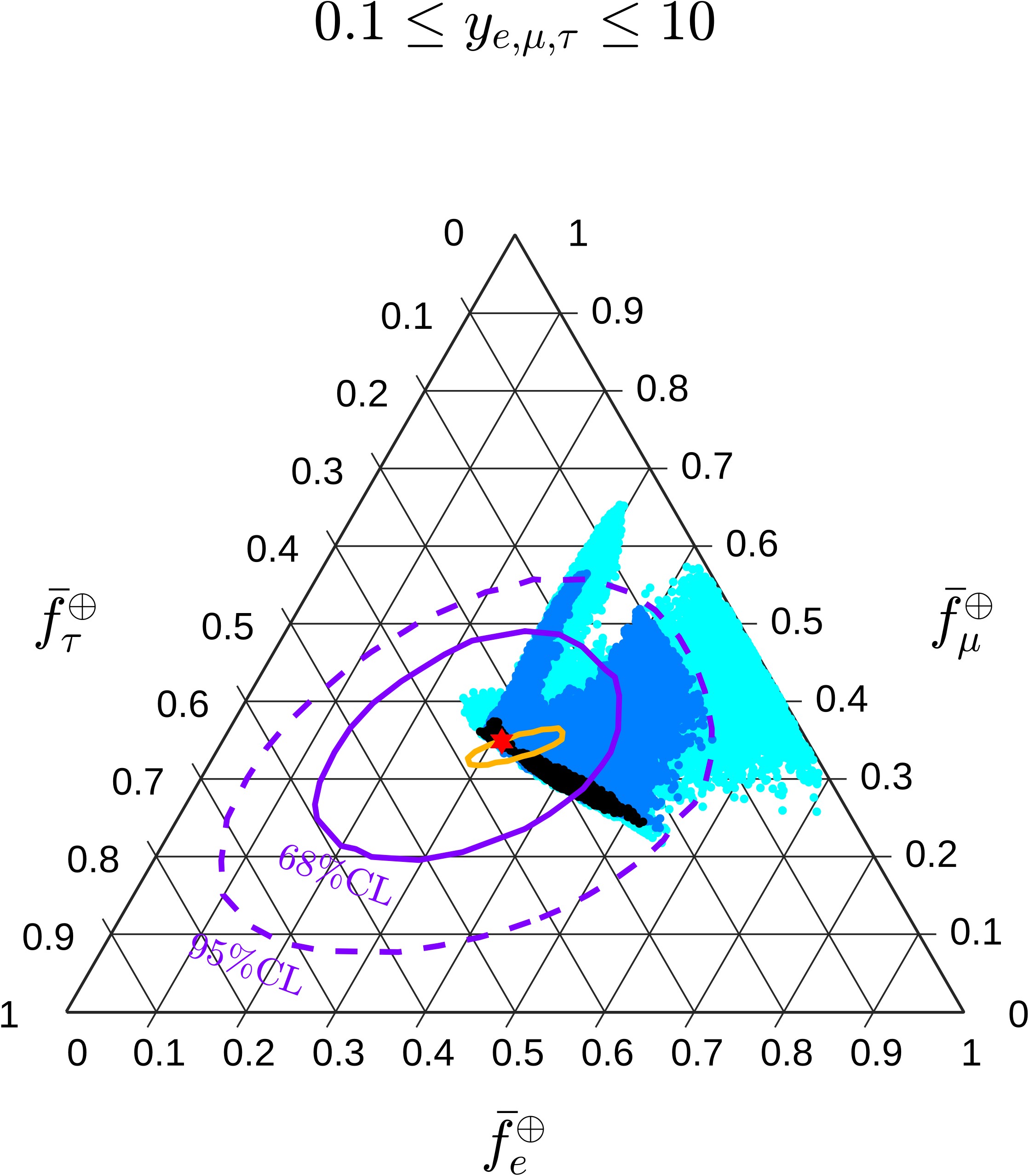}}
    
    \subfloat[\hspace{0.75cm}(a2)\hspace{3.1cm} (b2)\hspace{3.3cm}(c2)\hspace{3.1cm}(d2)]{\includegraphics[height=0.26\linewidth]{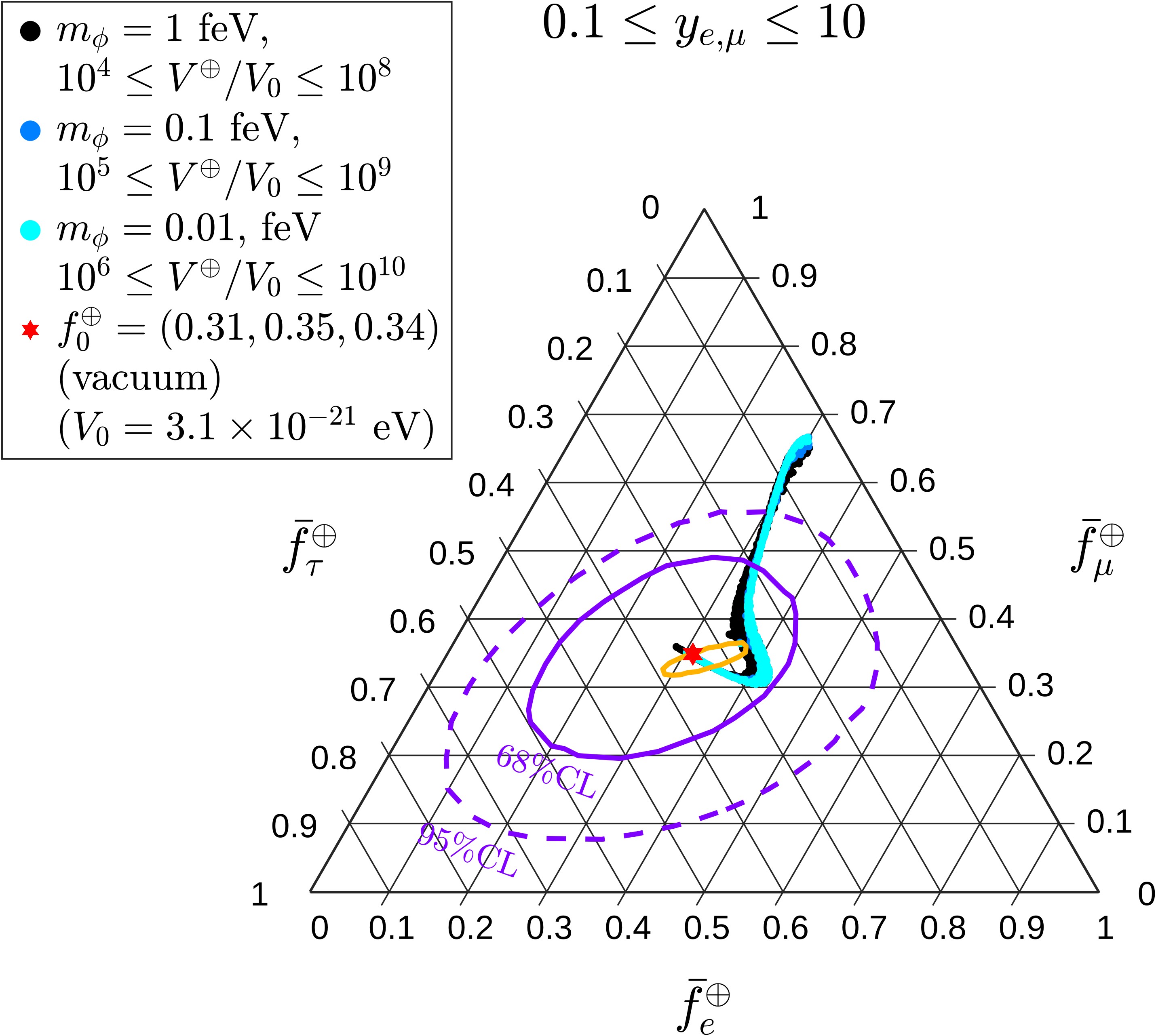}\quad
    \includegraphics[height=0.26\linewidth]{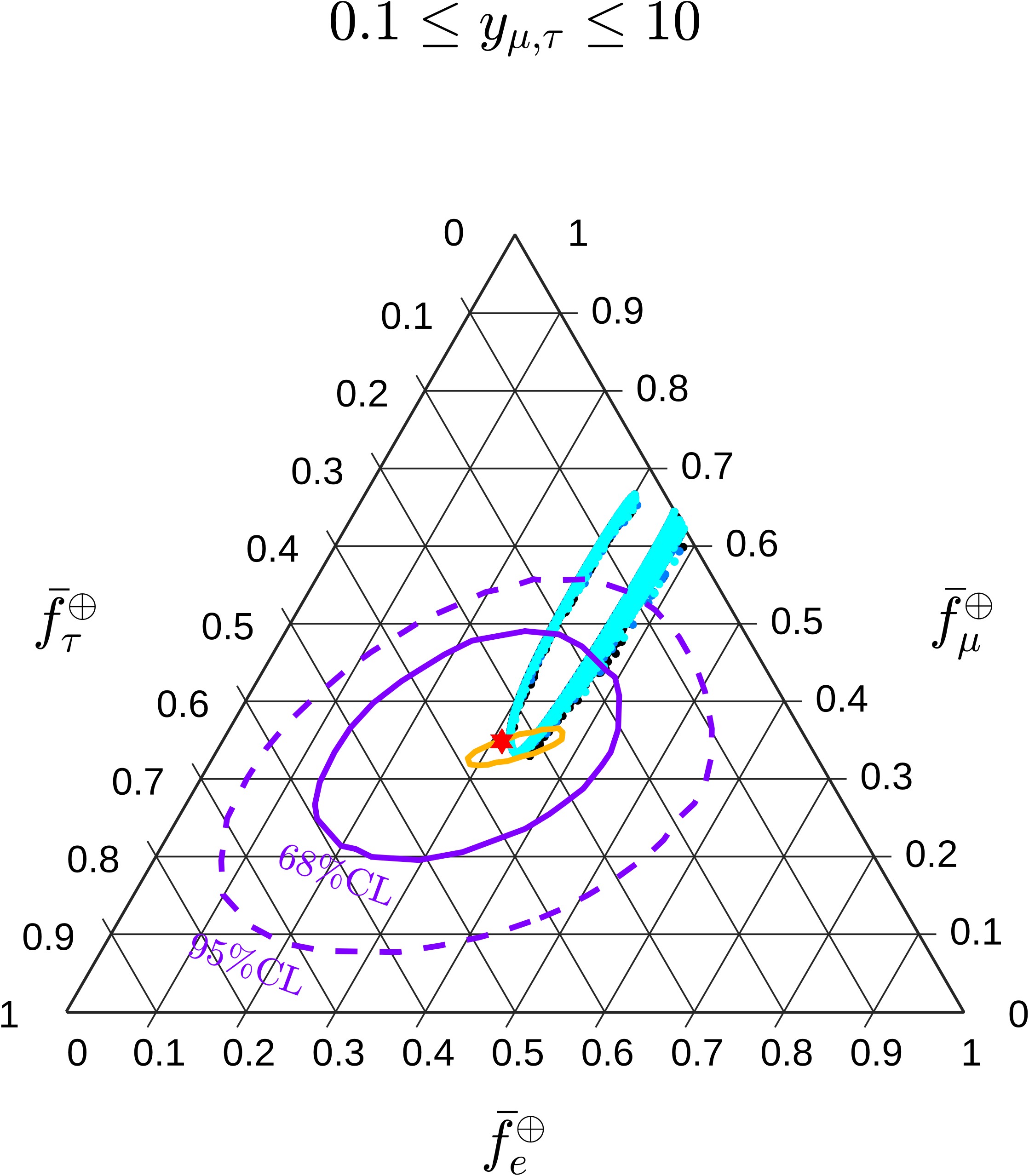}\quad
    \includegraphics[height=0.26\linewidth]{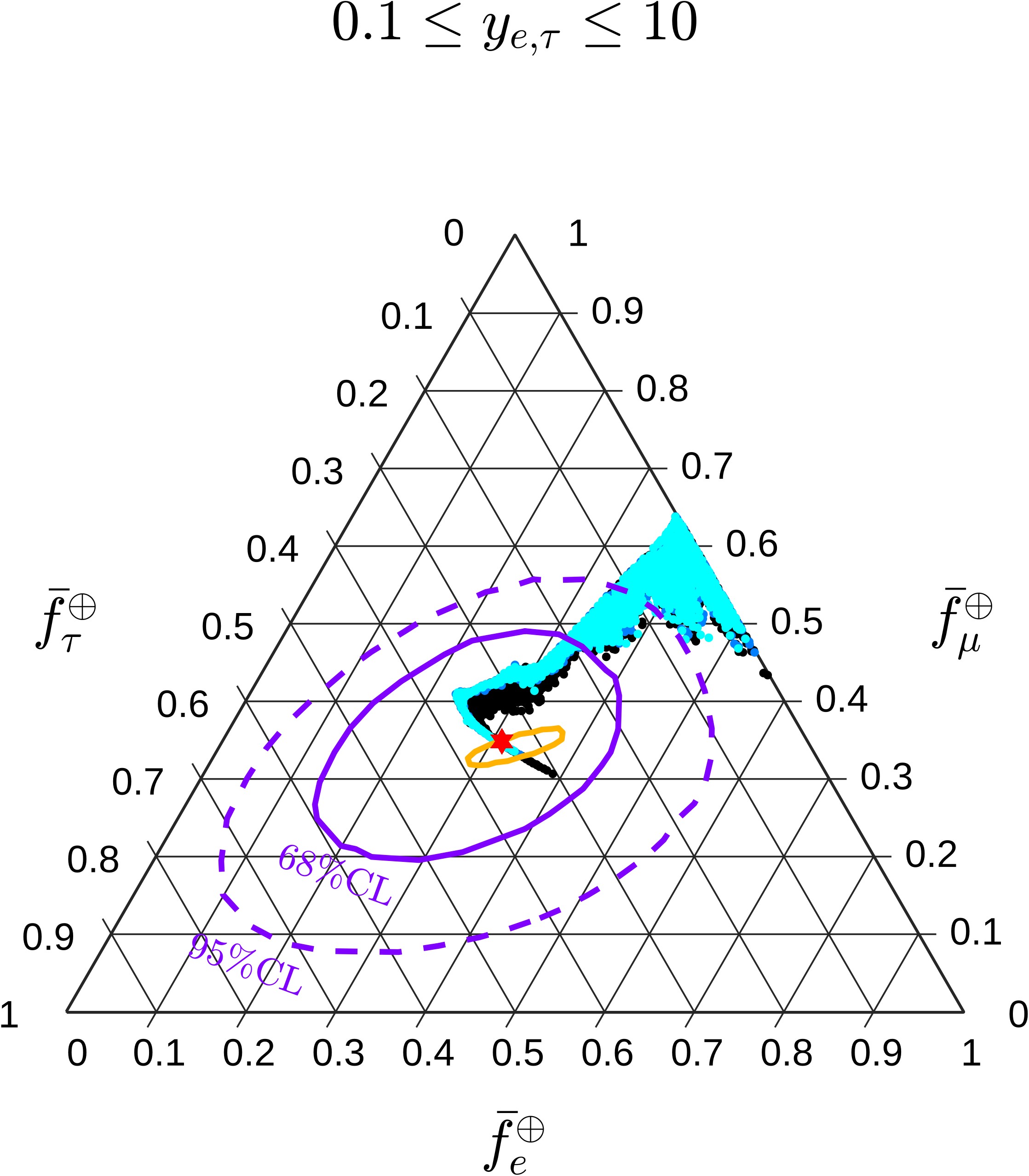}\quad
    \includegraphics[height=0.26\linewidth]{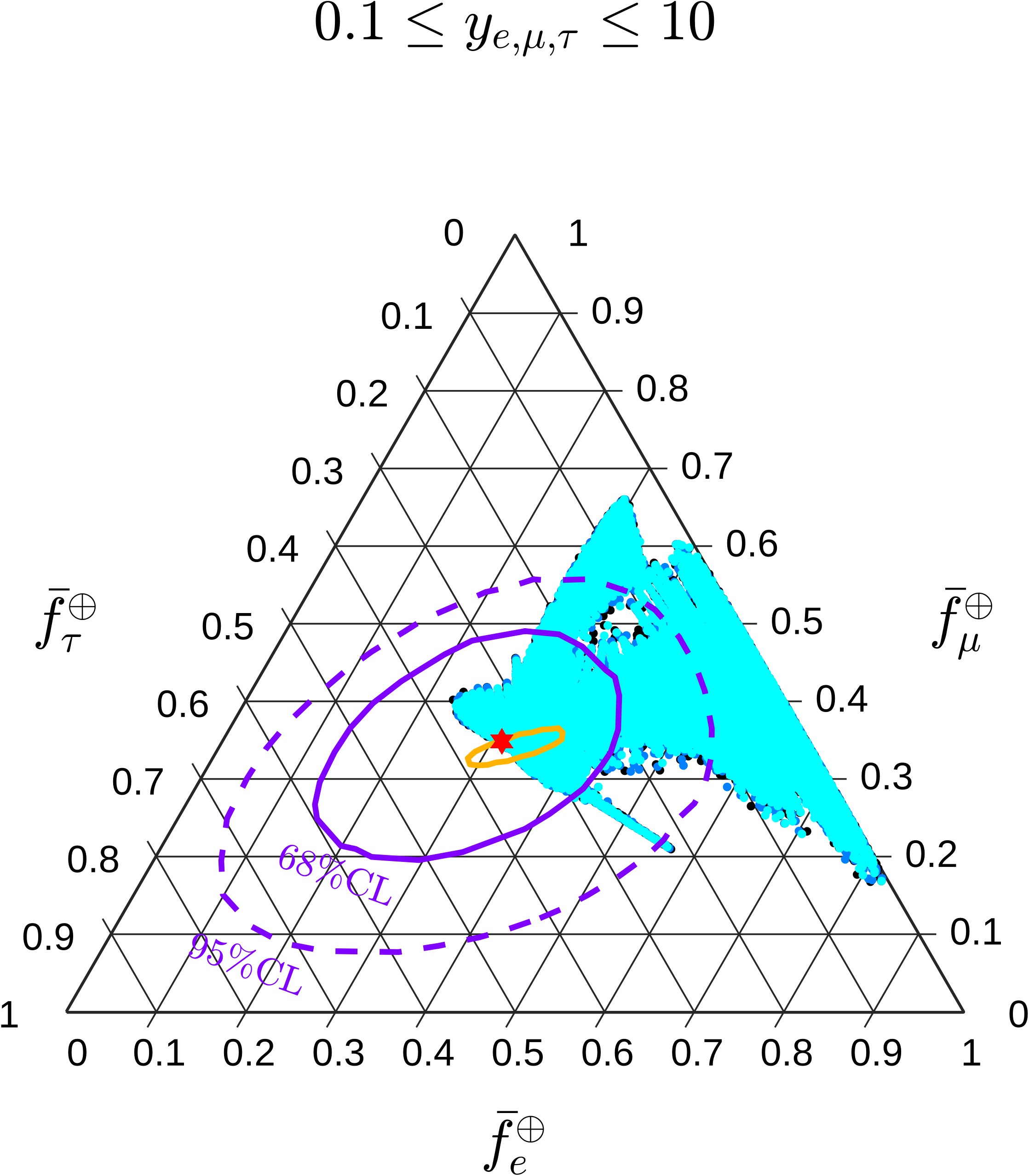}}
    
    \caption{Same setup as in Fig.~\ref{fig3} but the neutrinos from TXS 0506+056 are assumed to be produced within $[10,10^2]R_s$.}
    \label{fig2}
\end{figure}

In this section we consider the neutrino production location of TXS 0506+056 to be corona region, i.e. $[10,10^2]R_s$, and repeate the analysis in Section \ref{analy}. Unfortunately, the $m_\phi=0.1\,{\rm peV}$ and $m_\phi=1\,{\rm peV}$ distributions nearly completely overlap. Hence we replace the set of DM mass from $m_\phi=10, 1, 0.1\,{\rm peV}$  by $m_\phi=100, 10, 1\,{\rm peV}$ to obtain three distinguishable regions in Fig. \ref{fig2}(a1,b1,c1,d1).
Comparing the patterns of Fig.~\ref{fig3} and Fig.~\ref{fig2} from the same $m_\phi$, we see that the flavor ratio distributions are barely influenced by the production location and thus the DM density around AGN. 

\begin{figure}[H]
    \renewcommand{\thefigure}{7-2}
    \captionsetup[subfloat]{labelformat=empty}
    \centering
    \subfloat[(a1)\hspace{4.8cm} (a2)\hspace{4.8cm}(a3)]{\includegraphics[width=0.33\linewidth]{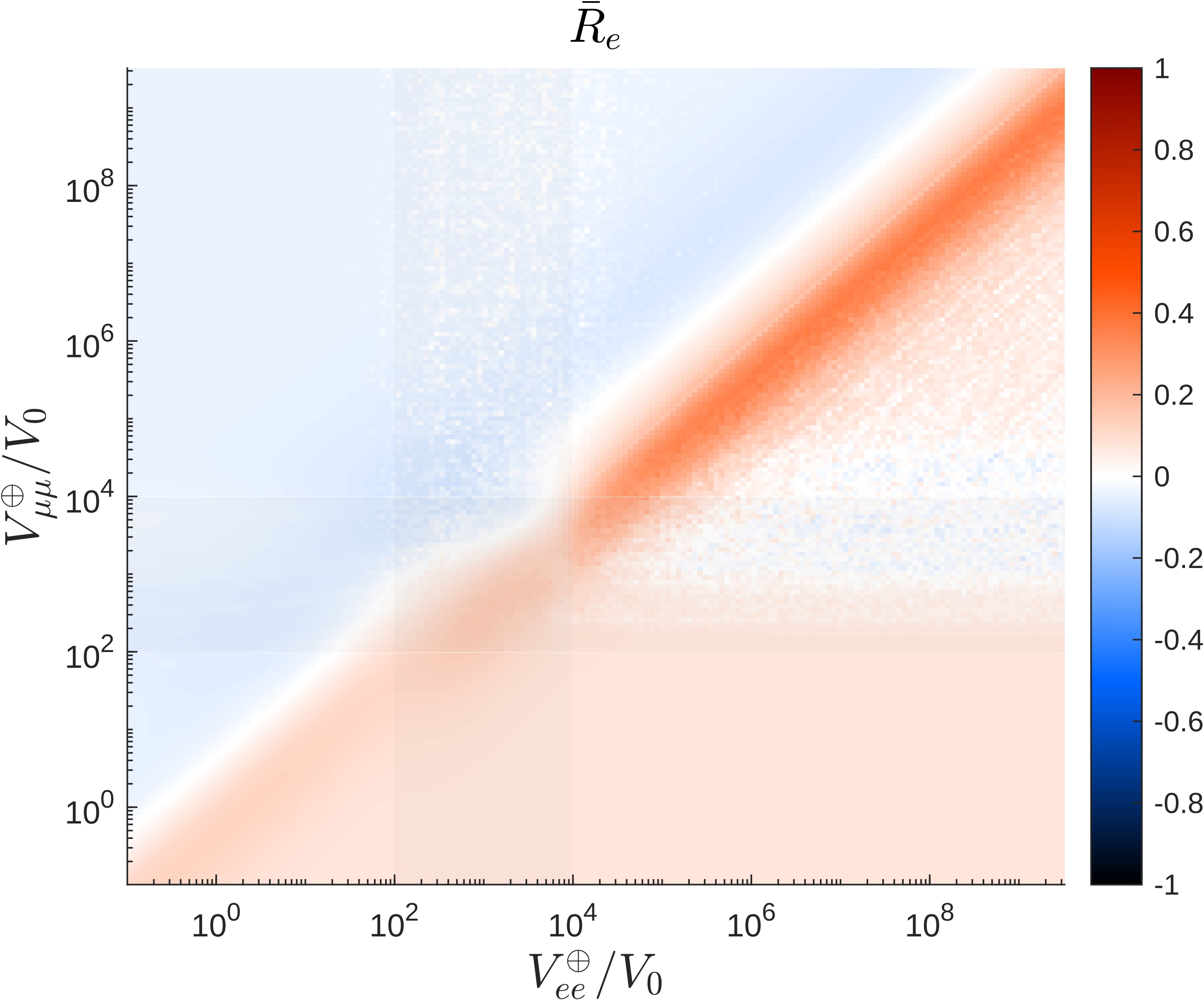}\quad
    \includegraphics[width=0.33\linewidth]{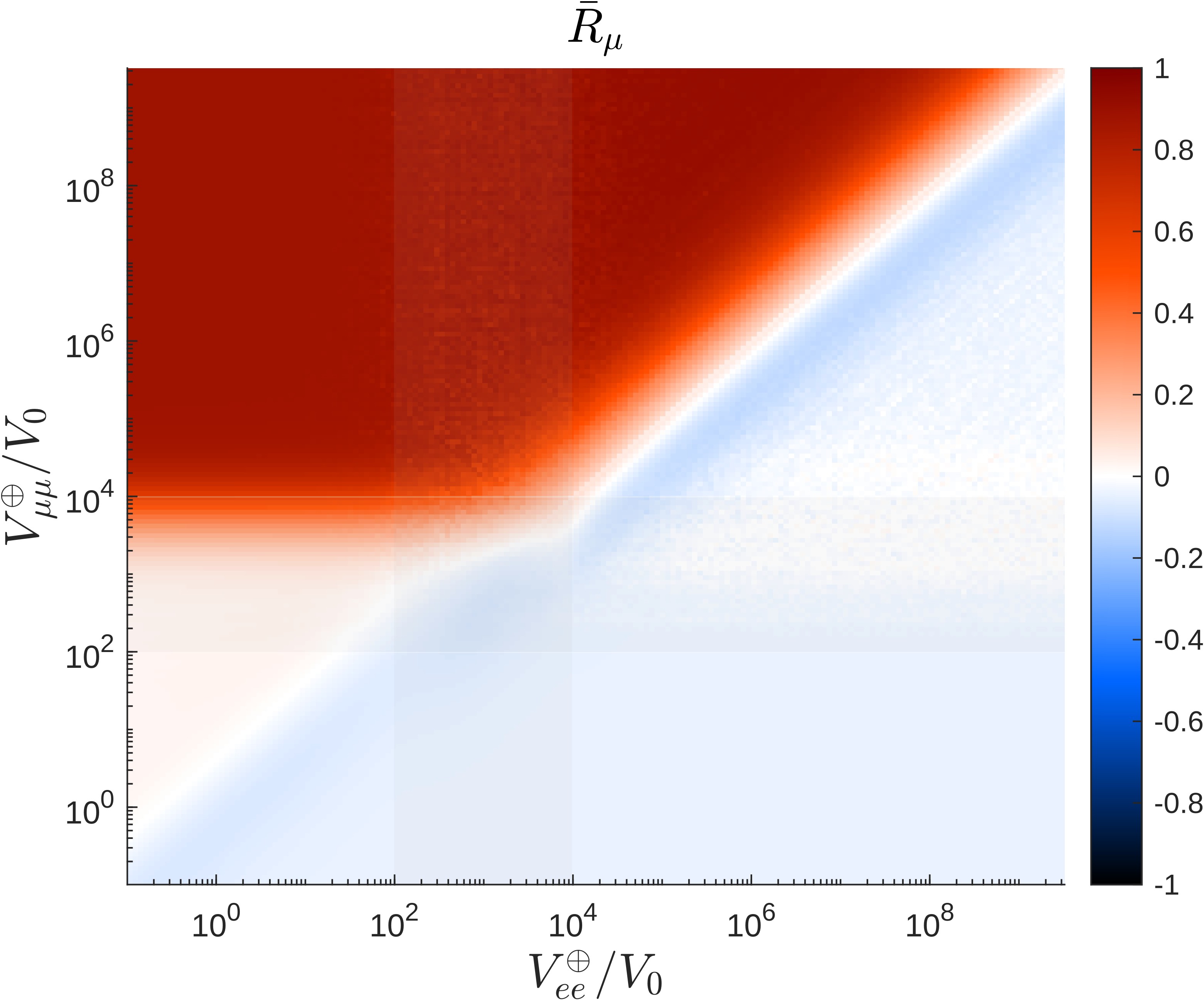}\quad
    \includegraphics[width=0.33\linewidth]{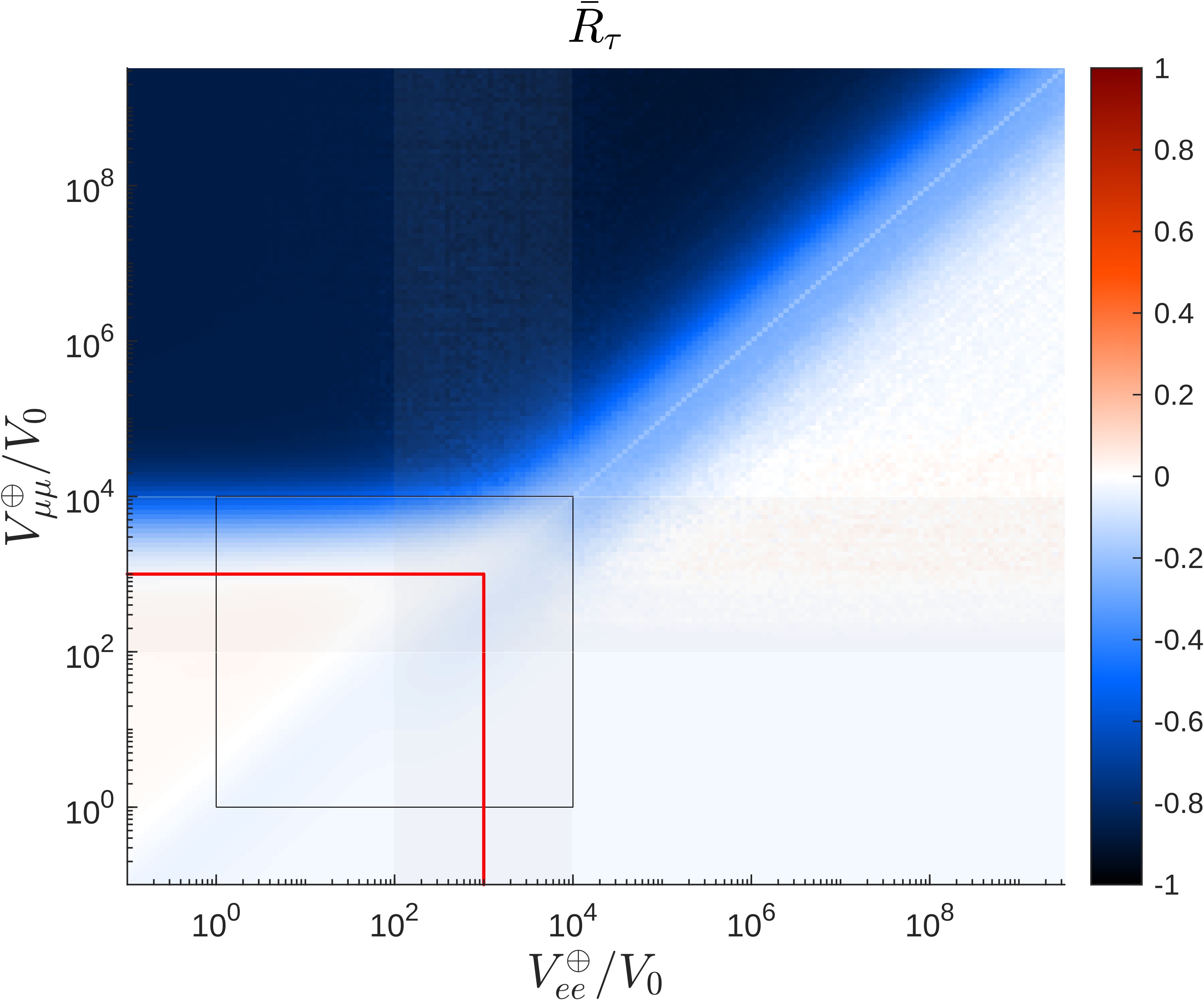}}

    \subfloat[(b1)\hspace{4.8cm} (b2)\hspace{4.8cm}(b3)]{\includegraphics[width=0.33\linewidth]{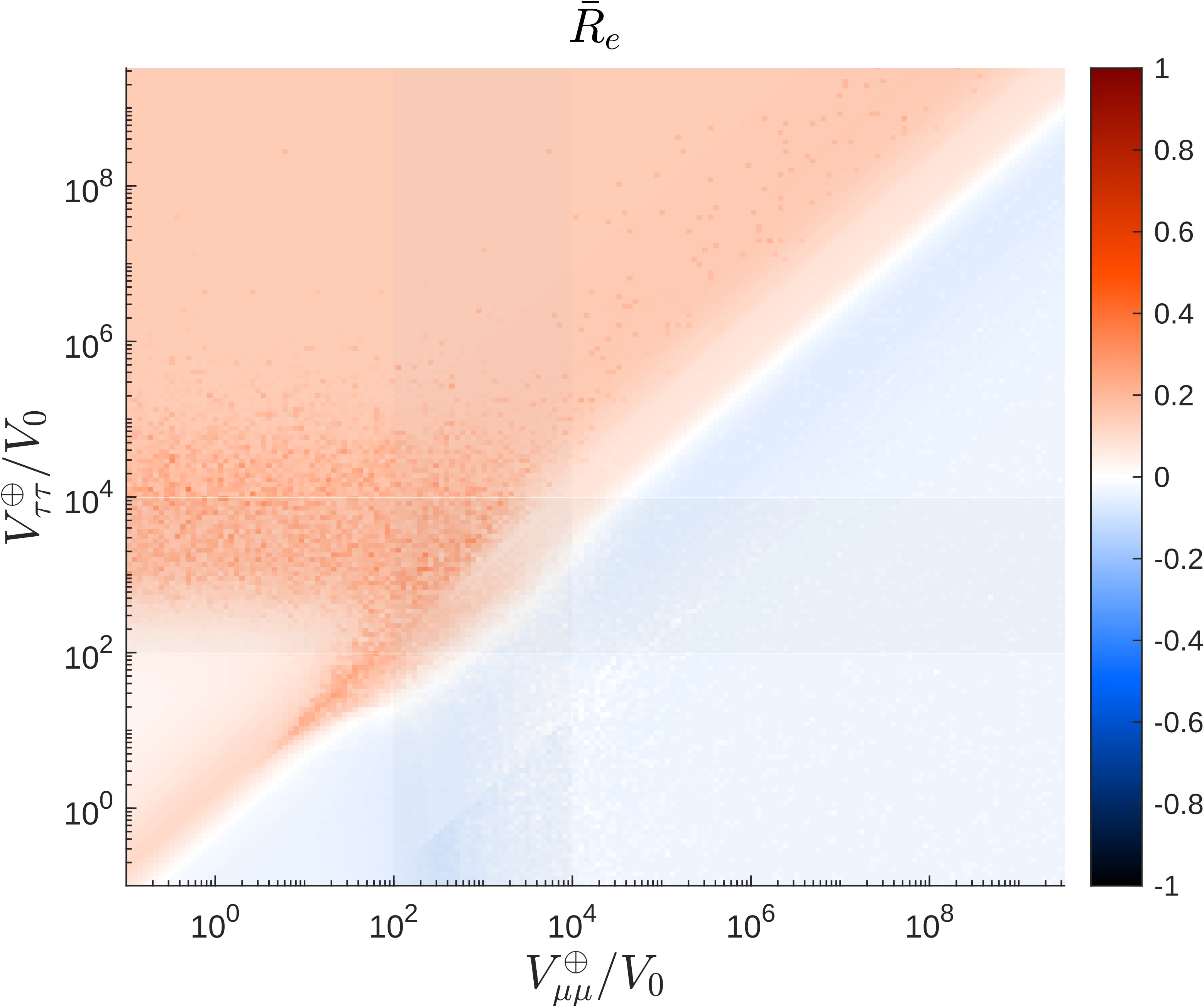}\quad
    \includegraphics[width=0.33\linewidth]{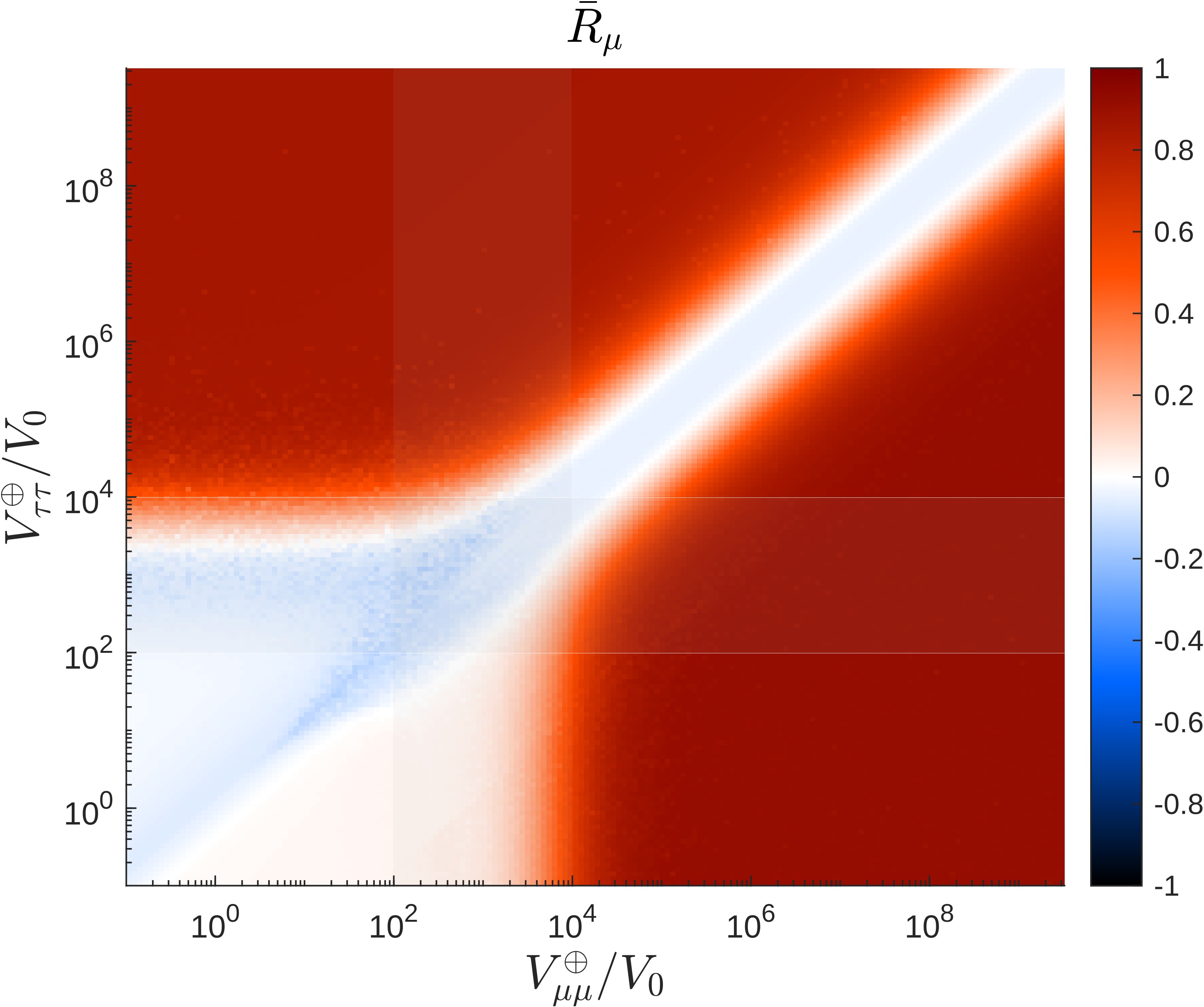}\quad
    \includegraphics[width=0.33\linewidth]{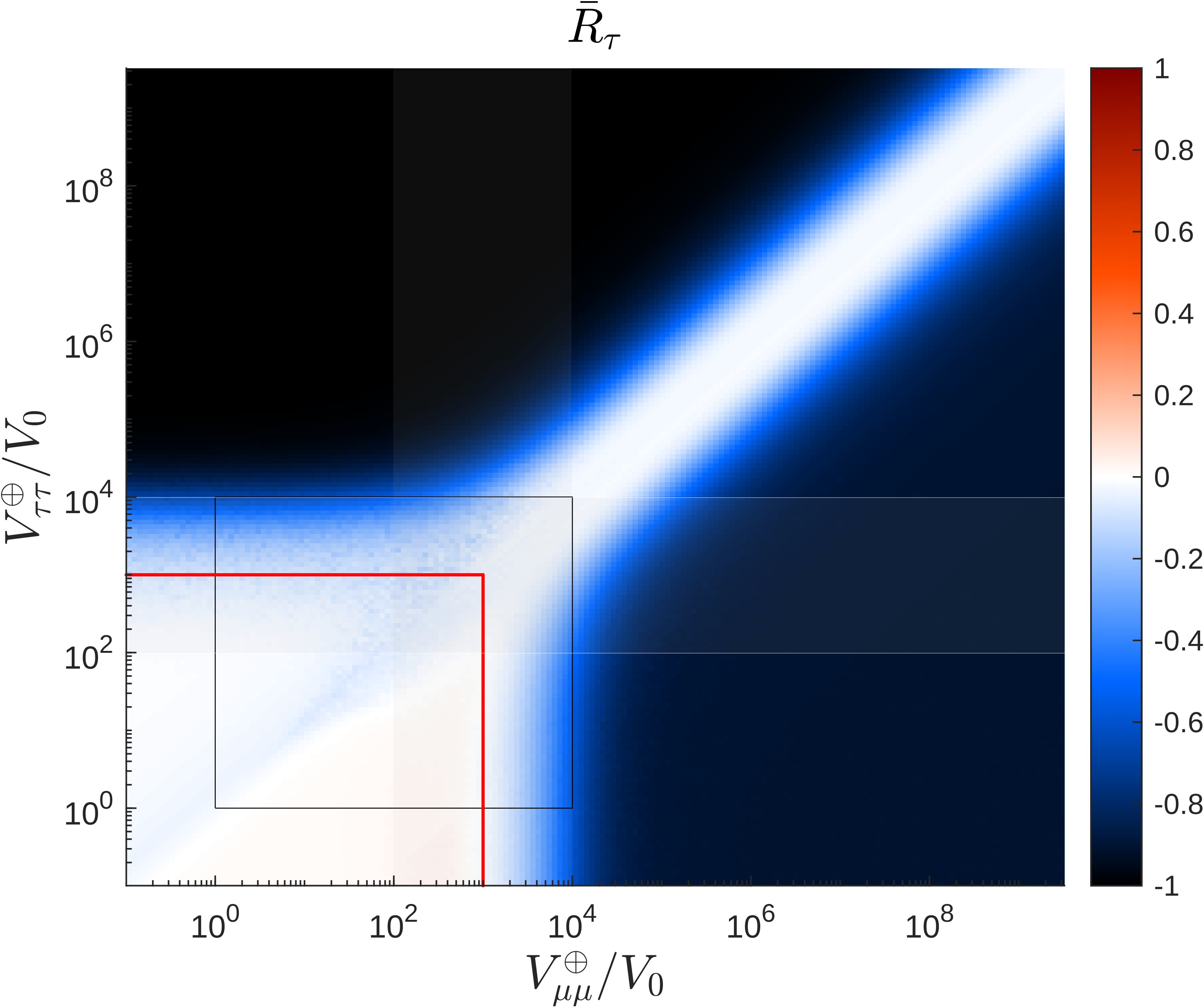}}

    \subfloat[(c1)\hspace{4.8cm} (c2)\hspace{4.8cm}(c3)]{\includegraphics[width=0.33\linewidth]{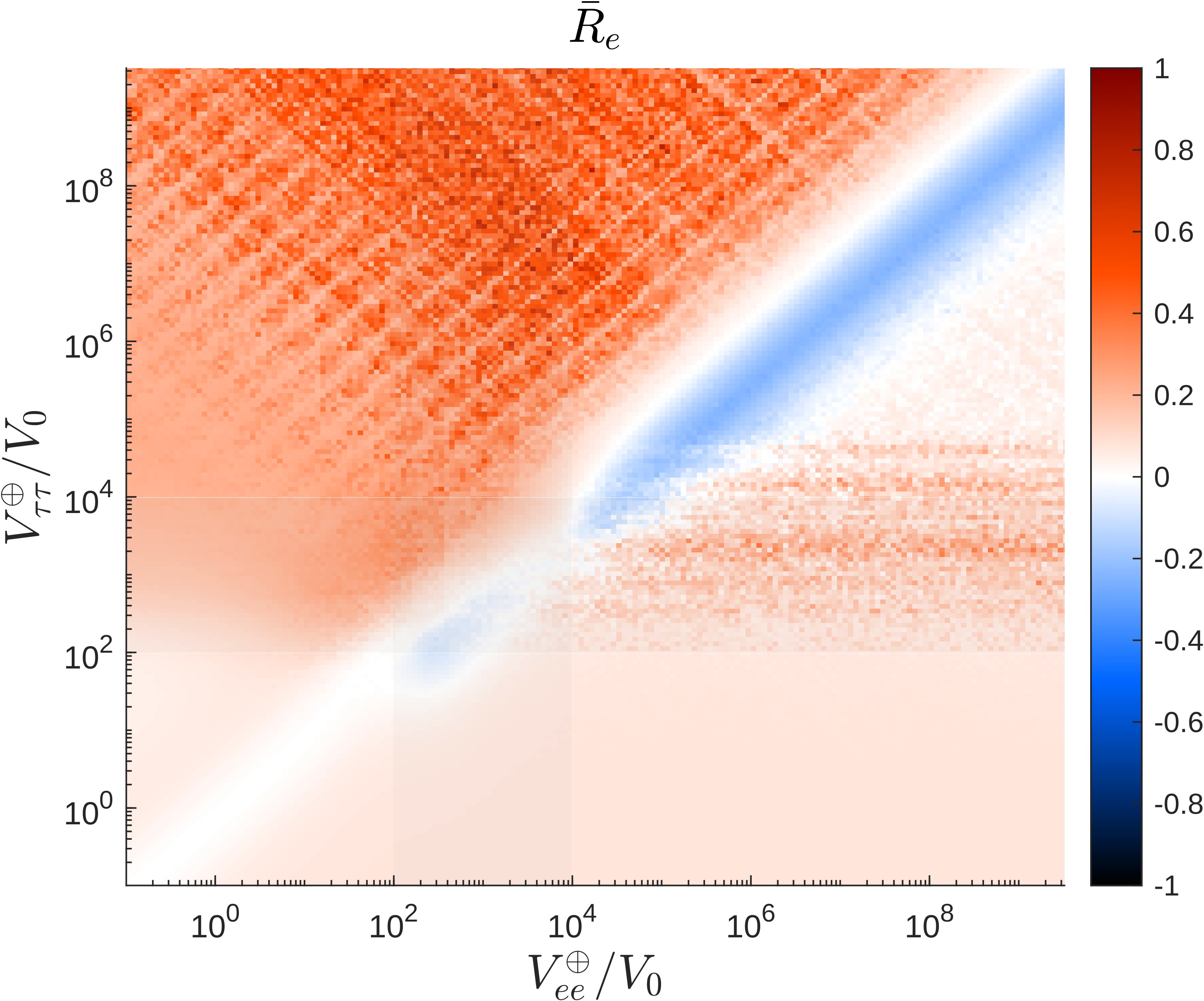}\quad
    \includegraphics[width=0.33\linewidth]{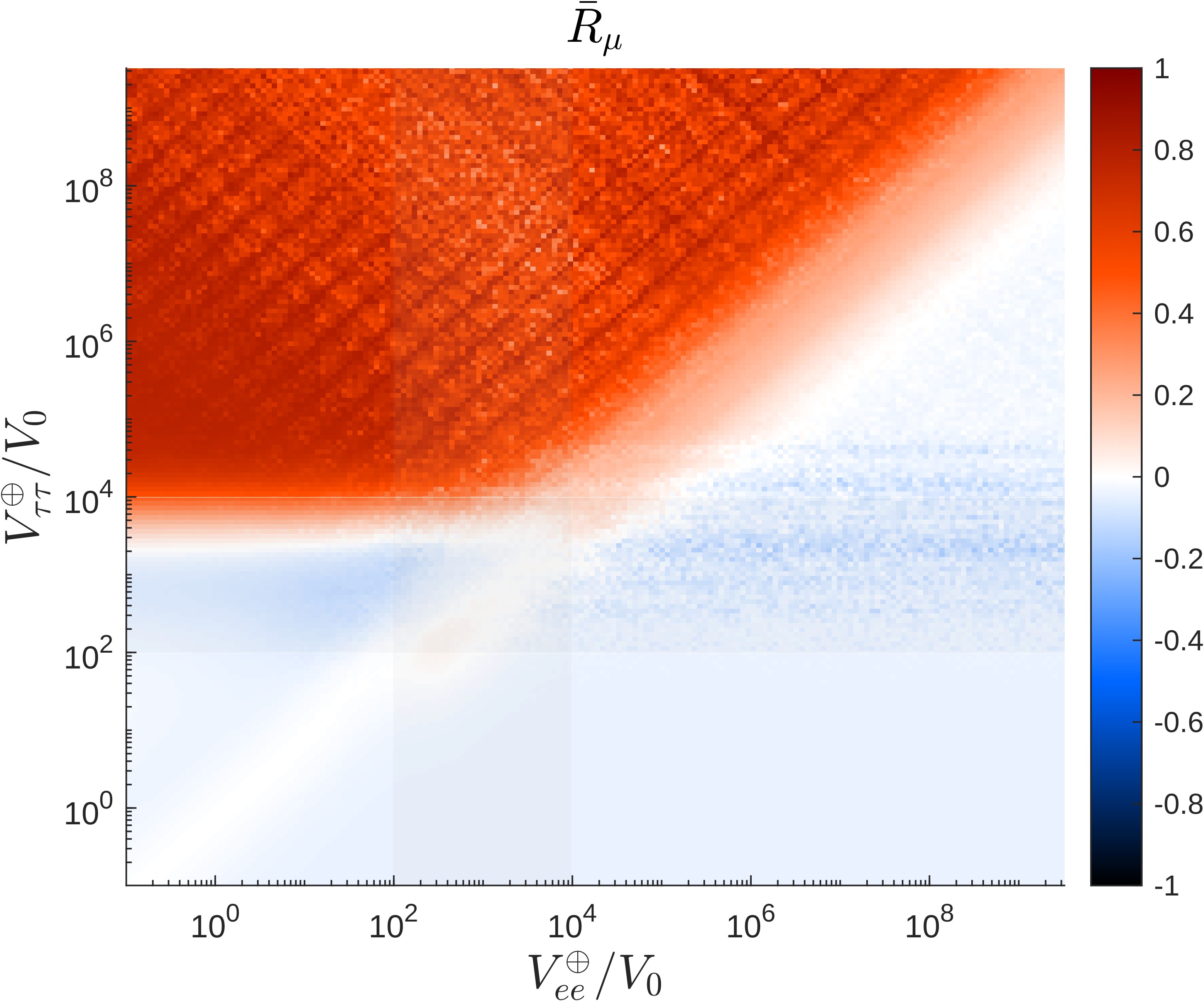}\quad
    \includegraphics[width=0.33\linewidth]{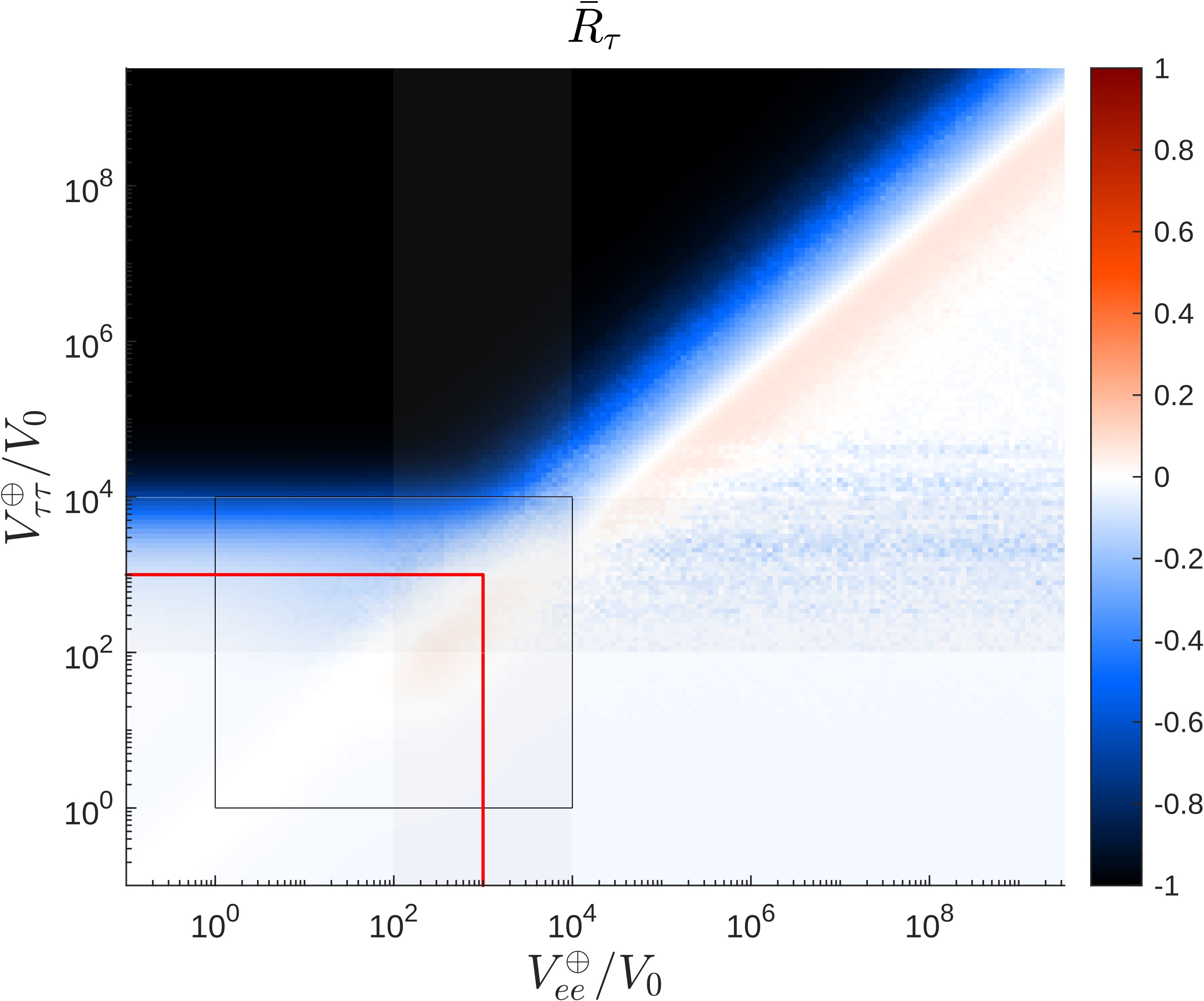}}
    \caption{The distributions of event-averaged deviation $\bar{R}_\beta$ of Fig.~\ref{fig2} for Case-(a,b,c). The red and black square boxes in panel (a3, b3, c3) represent the potential range of $m_\phi=100$ peV and $m_\phi=10$ peV, respectively.}
    \label{fig2_2}
\end{figure}

As an another illustrative example of correspondence between the event-averaged deviation plots and the flavor ratio distribution plots, we compare the $m_\phi=100\,{\rm peV}$ distributions (black) and $m_\phi=10\,{\rm peV}$ distributions (blue) of Fig.~\ref{fig2} (a1,b1,c1). The red squares in Fig.~\ref{fig2_2} (a3,b3,c3) depicts the range of $V^\oplus$ of $m_\phi=100\,{\rm peV}$ distributions. The regions enclosed by these squares are very lightly colored. Comparing to the black squares in Fig.~\ref{fig2_2} (a3,b3,c3), they contains small amount of darker-blue areas. Therefore, the $m_\phi=10\,{\rm peV}$ case of Fig.~\ref{fig2} have more points with $\bar{f}^\oplus_\tau\leq \bar{f}^\oplus_{0\tau}$ than the $m_\phi=100\,{\rm peV}$.
The black square in Fig.~\ref{fig2_2} (a3,b3,c3) is the range $V^\oplus$ of $m_\phi=10\,{\rm peV}$ case of Fig.~\ref{fig2}, and the enclosed regions also contain some darker-blue areas, similar to those in Fig.~\ref{fig3_2} (a3,b3,c3). So, the $m_\phi=10\,{\rm peV}$ distributions (blue) in Fig.~\ref{fig2} have similar shapes to those (black) in Fig.~\ref{fig3}.
Due to these features, we infer that the IceCube 95\% C.L. sensitivity may not distinguish the TXS 0506+056 signals from different range of production locations.

\newpage
\section{The Coupling Constants Excluded by Combined $\nu$ Telescopes}\label{appendixB}
\begin{figure}[H]
    \renewcommand{\thefigure}{8}
    \captionsetup[subfloat]{labelformat=empty}
    \centering

    \subfloat[(d4-NO)\hspace{4.2cm} (d5-NO)\hspace{4.2cm}(d6-NO)]{\includegraphics[width=0.33\linewidth]{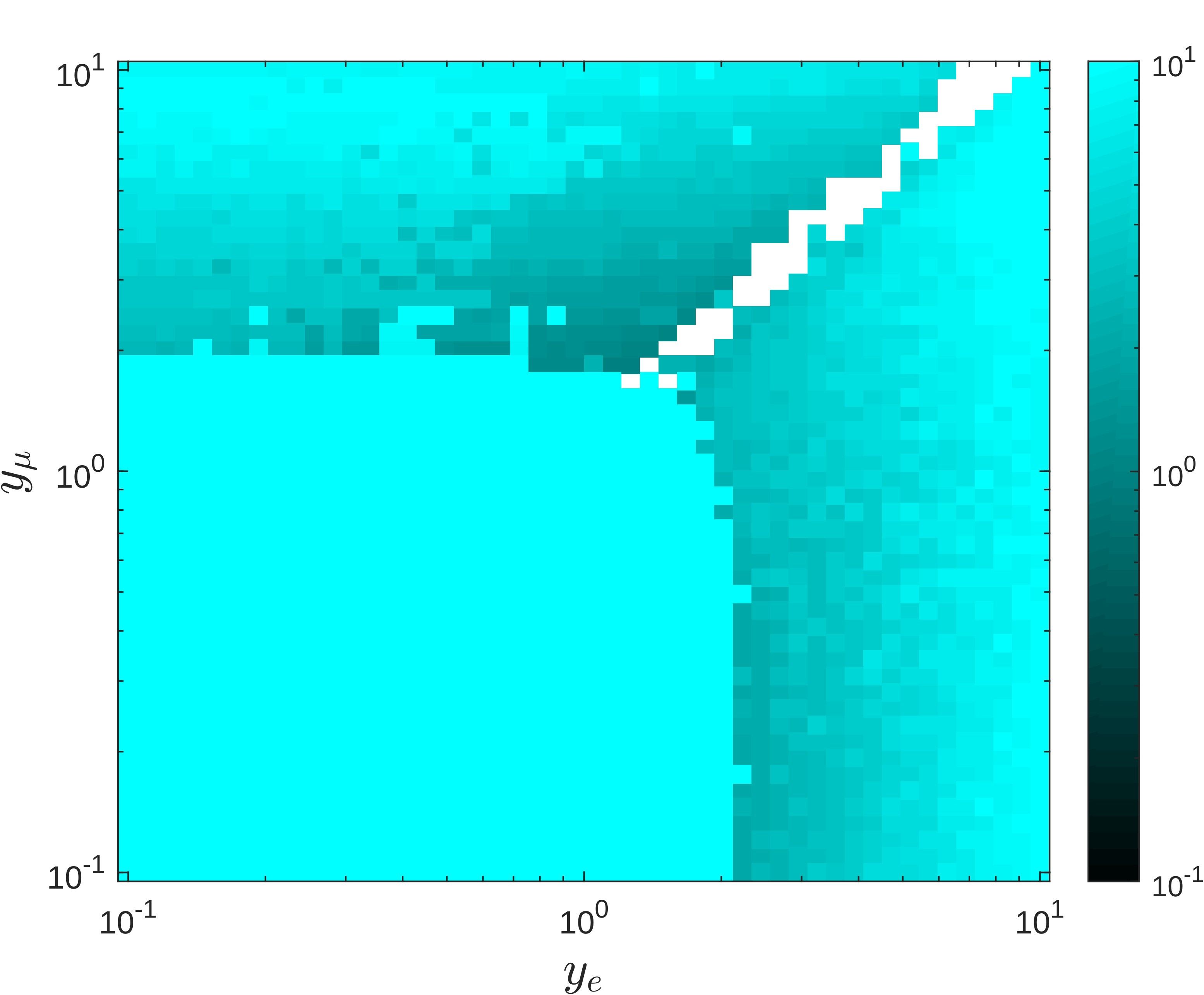}\quad
    \includegraphics[width=0.33\linewidth]{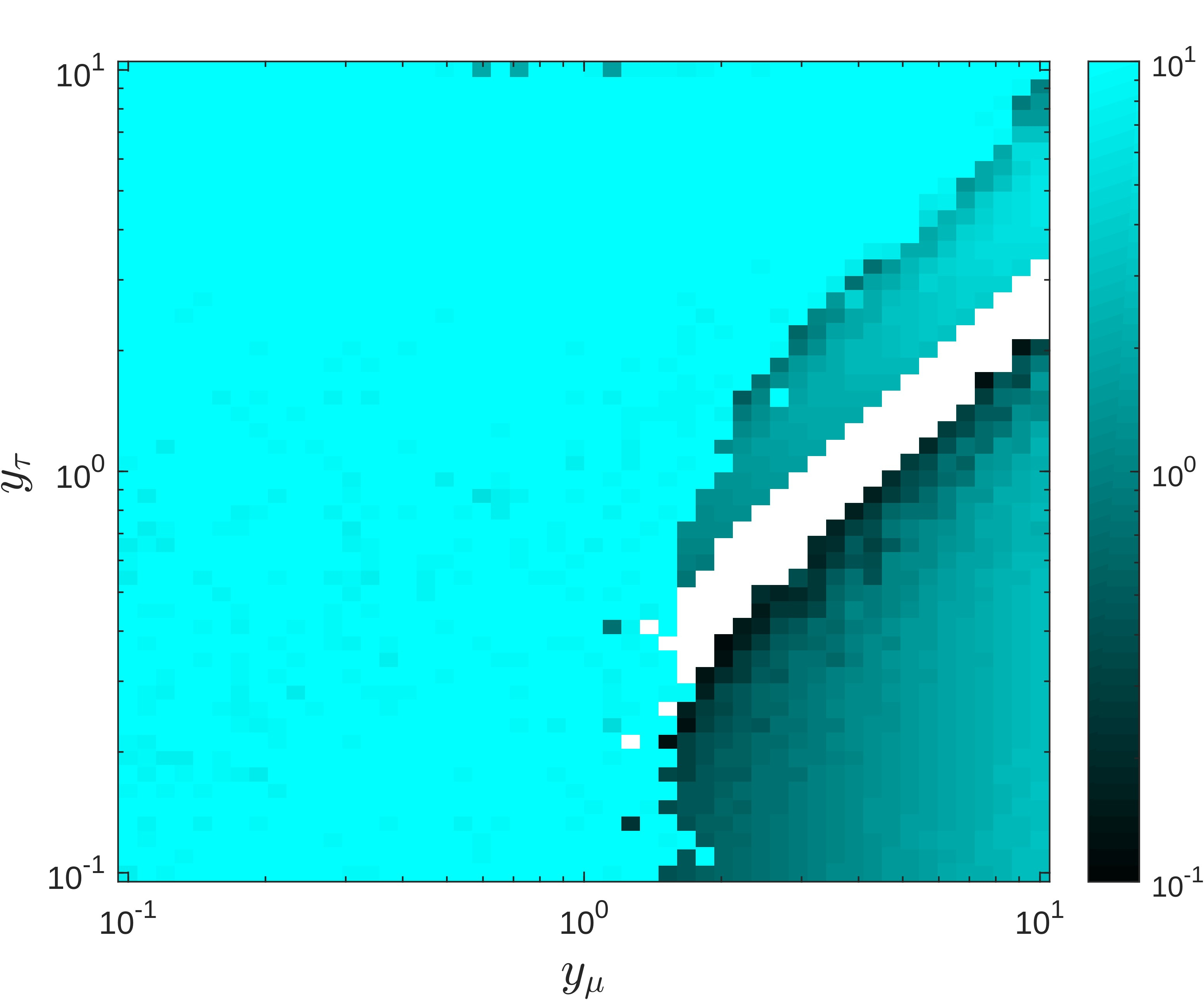}\quad
    \includegraphics[width=0.33\linewidth]{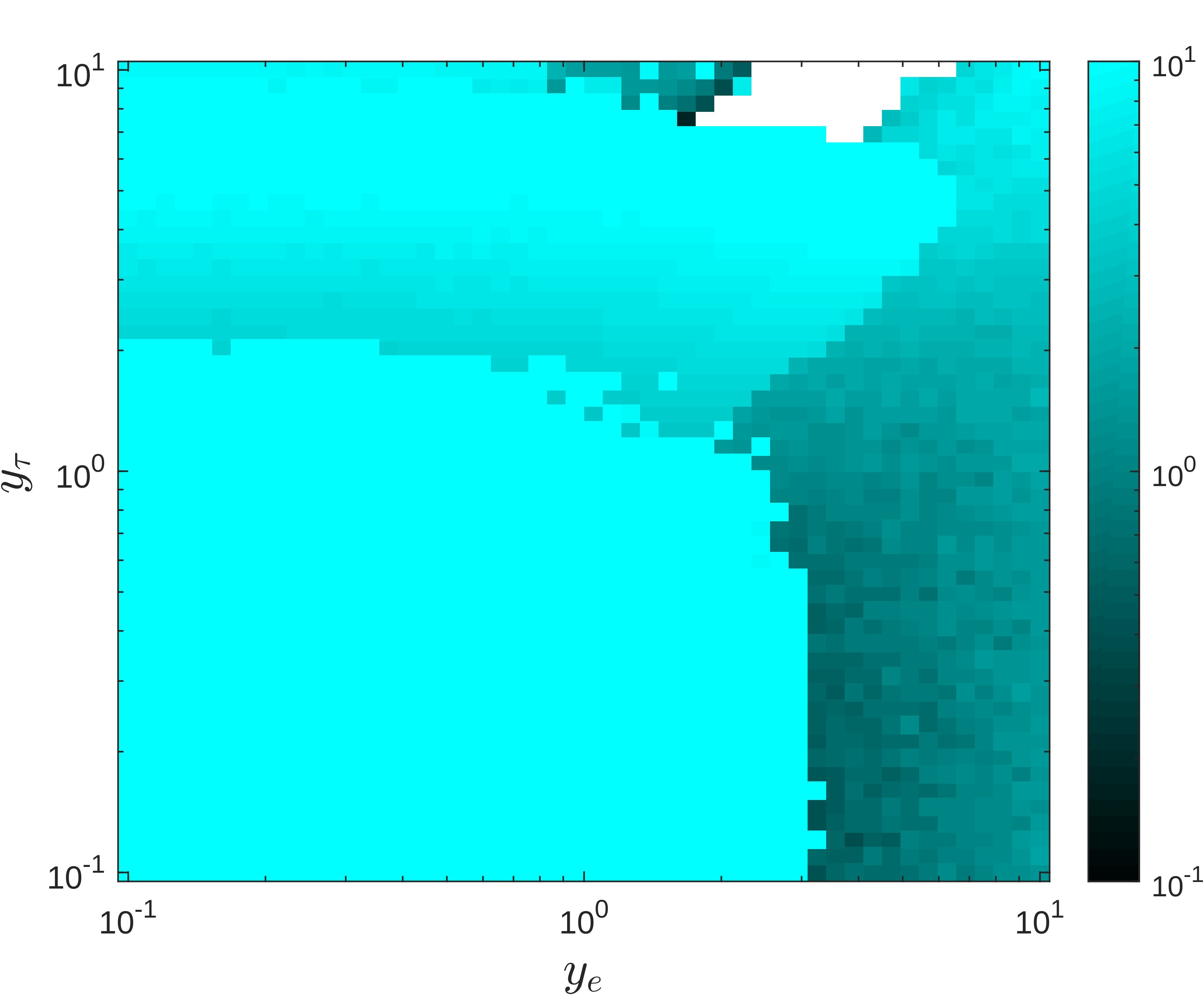}}


    \subfloat[(d4-IO)\hspace{4.2cm} (d5-IO)\hspace{4.2cm}(d6-IO)]{\includegraphics[width=0.33\linewidth]{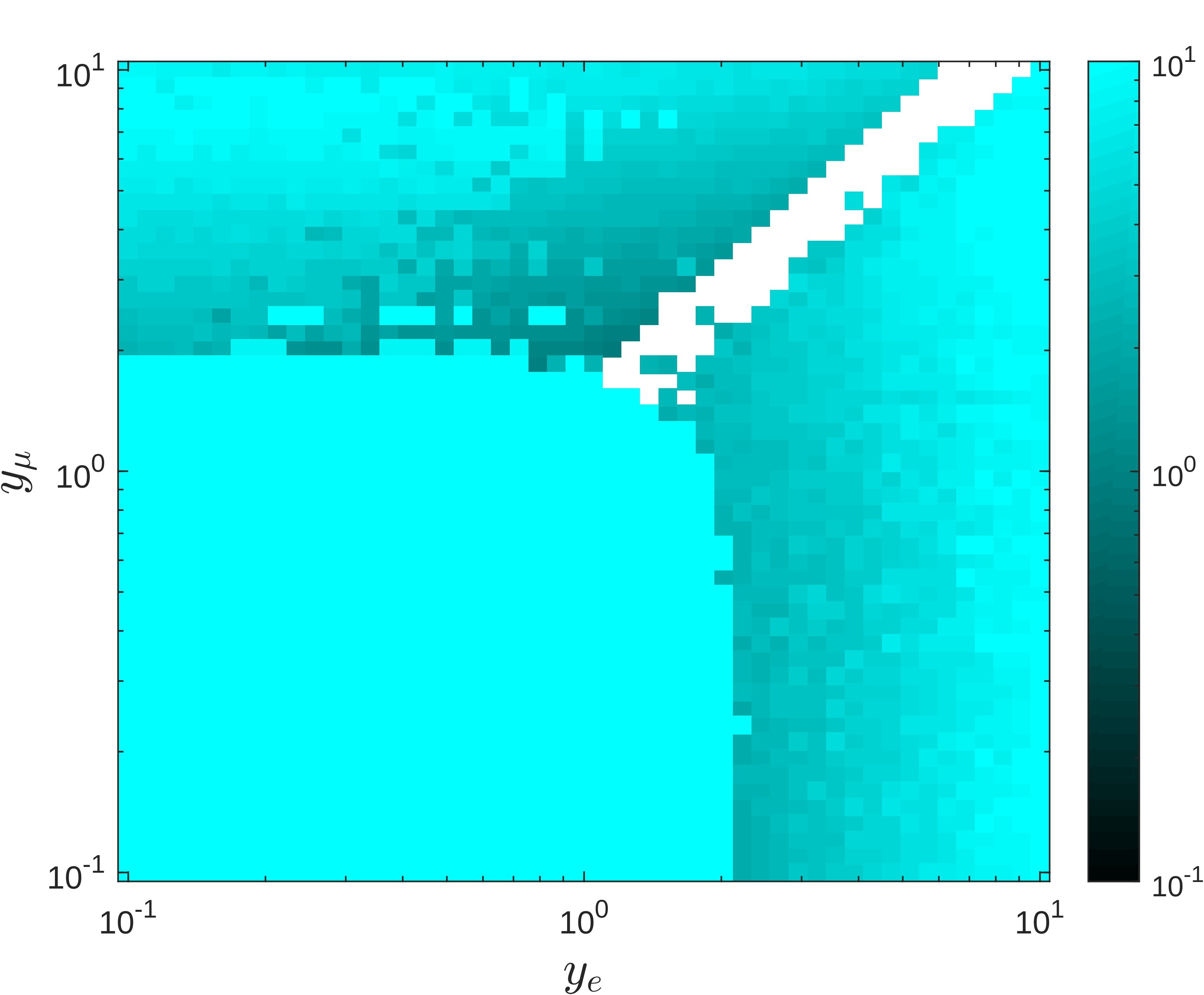}\quad
    \includegraphics[width=0.33\linewidth]{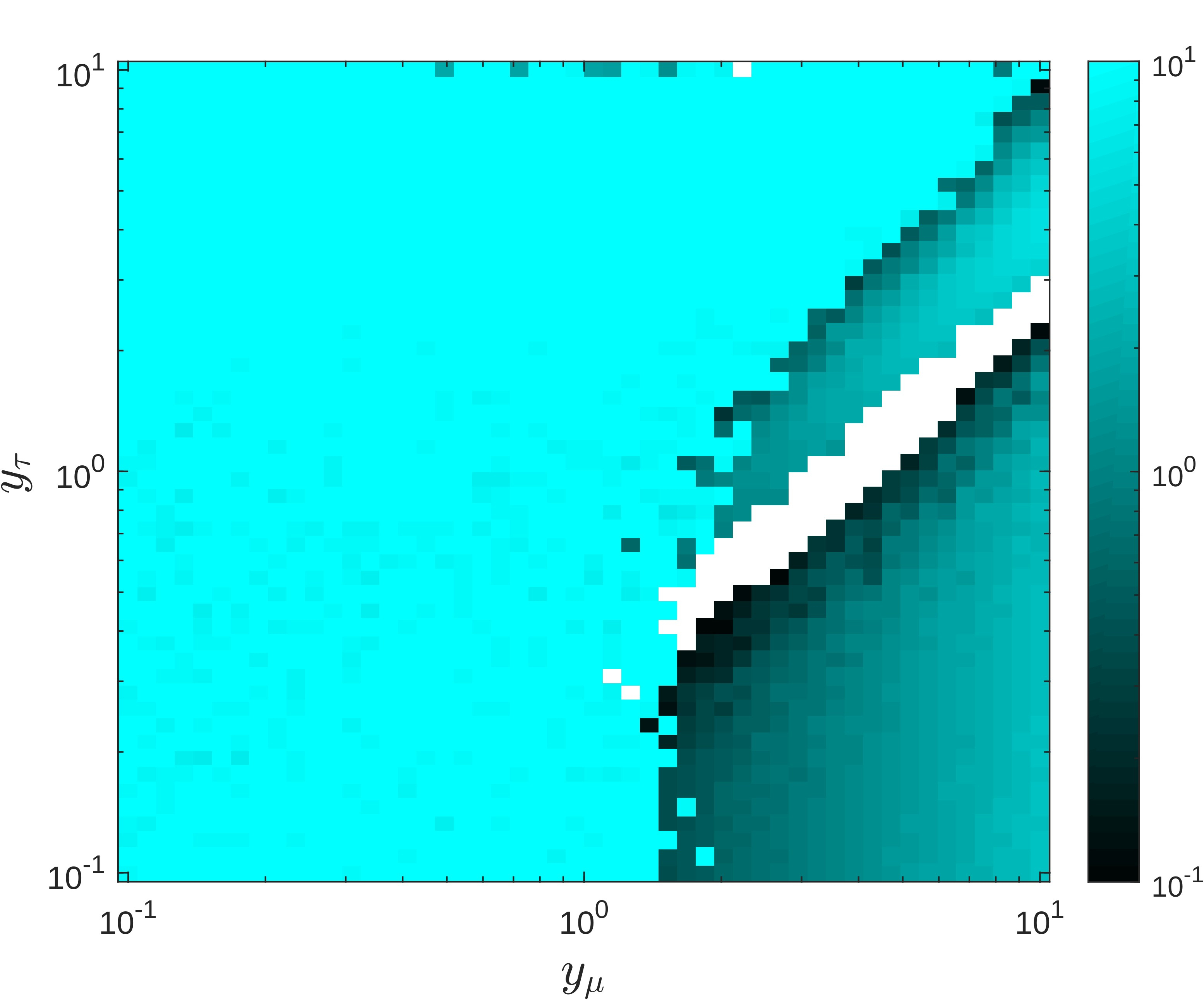}\quad
    \includegraphics[width=0.33\linewidth]{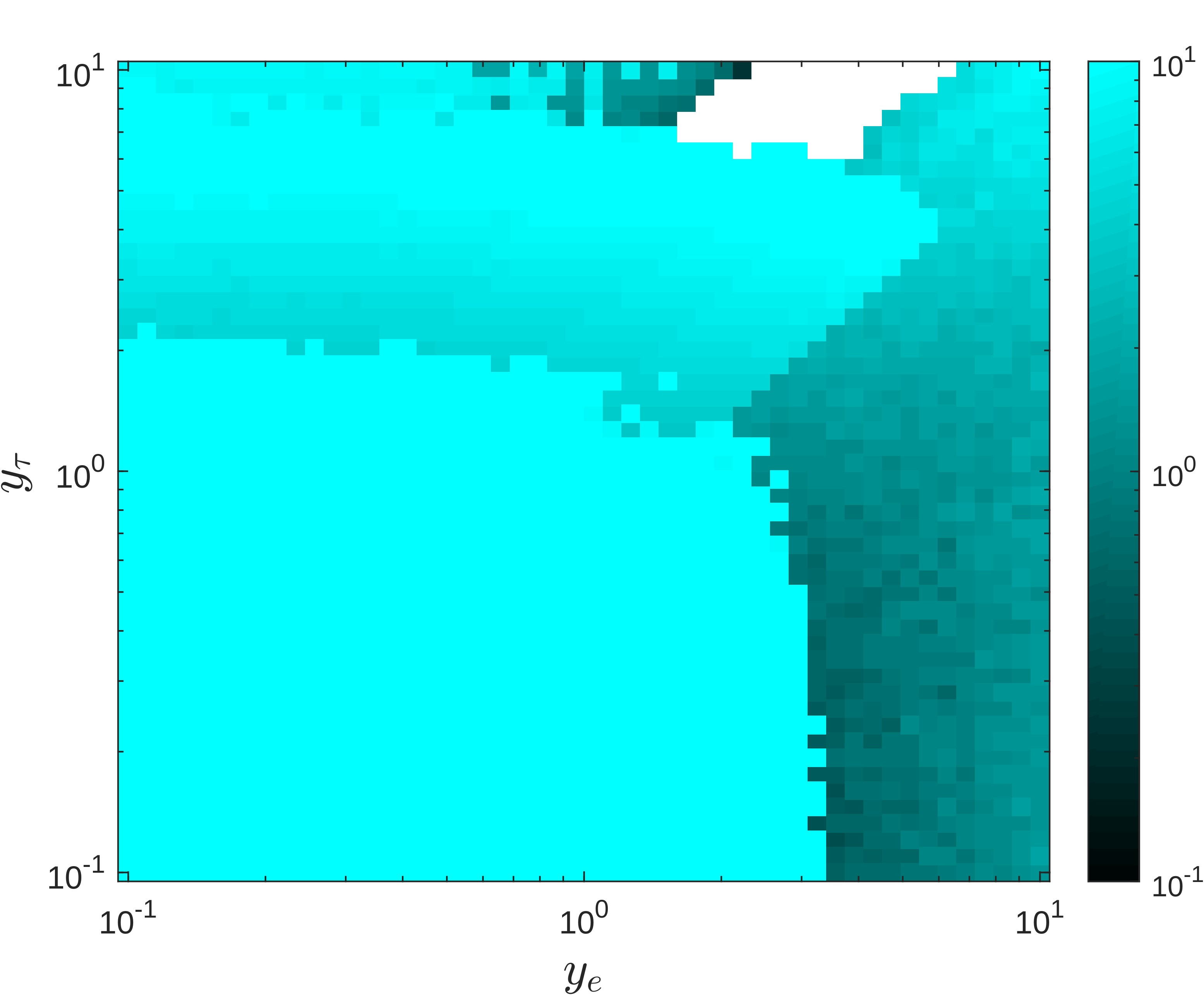}}
    \caption{The color regions represent the couplings of Case-(d) excluded by combined $\nu$ telescopes sensitives. The excluded regions with $m_\phi=1\,{\rm feV}$ by NGC 1068 are projected to two of three Yukawa couplings, and the color bar represents the maximum value of the third coupling. Upper (Lower) panels are for normal (inverted) mass ordering. 
    }
    \label{fig8}
\end{figure}

Fig.~\ref{fig8} shows the constraint on the coupling constants of Case-(d), which is obtained from the combined $\nu$ telescopes sensitivities. 
Since the contour of the combined $\nu$ telescopes sensitivities is much smaller than the IceCube MESE 95\% C.L. contour, there are more points in $(y_e,y_\mu,y_\tau)$ parameter space be excluded comparing to Fig.~\ref{fig6}.


\clearpage
\nocite{*}
\bibliographystyle{kp}

\end{document}